\documentclass[english,iop,apj]{emulateapj}
\usepackage[T1]{fontenc}
\usepackage{verbatim}
\usepackage{graphicx}
\usepackage{subfigure}
\usepackage{tabularx}
\usepackage{natbib}

\makeatletter
\slugcomment{}
\shorttitle{}
\shortauthors{}

\usepackage{graphicx}
\usepackage{txfonts}
\usepackage{cleveref}

\def\apjl{ApJL }

\def\apj{ApJ }
\def\pasp{PASP }

\def\apjs{ApJS }
\def\araa{ARAA }
\def\aap{A\&A }
\def\nat{Nature }

\def\apss{\ref@jnl{Ap\&SS}}
\def\jgr{\ref@jnl{J.~Geophys.~Res.}}

\def\mnras{MnRAS}
\def\qjras{\ref@jnl{QJRAS}}

\newcommand{\unit}[1]{\ensuremath{\, \mathrm{#1}}}
\newcommand{\ramses}{{\sc ramses}}          
\newcommand{\vapor}{{\sc vapor}}          
\newcommand{\stagger}{{\sc stagger}}        
\newcommand{\Fig}[1]{Fig.\ \ref{fig:#1}}    
\newcommand{\Figure}[1]{Figure \ref{fig:#1}}

\makeatother

\usepackage{babel}
\begin{document}

\title{Tracking the distribution of $^{26}$Al and $^{60}$Fe during the early phases of star and disk evolution}

\author{Michael Kuffmeier}

\affil{Centre for Star and Planet Formation, Niels Bohr Institute and Natural History Museum of Denmark, University of Copenhagen,
{\O}ster Voldgade 5-7, DK-1350 Copenhagen K, Denmark}

\email{<kueffmeier@nbi.ku.dk>}

\author{Troels Frostholm Mogensen}

\affil{Centre for Star and Planet Formation, Niels Bohr Institute and Natural History Museum of Denmark, University of Copenhagen,
{\O}ster Voldgade 5-7, DK-1350 Copenhagen K, Denmark}

\author{Troels Haugb{\o}lle}

\affil{Centre for Star and Planet Formation, Niels Bohr Institute and Natural History Museum of Denmark, University of Copenhagen,
{\O}ster Voldgade 5-7, DK-1350 Copenhagen K, Denmark}

\author{Martin Bizzarro}

\affil{Centre for Star and Planet Formation and Natural History Museum of Denmark, University of Copenhagen, {\O}ster Voldgade 5-7, 1350 Copenhagen, Denmark}

\author{{\AA}ke Nordlund}

\affil{Centre for Star and Planet Formation, Niels Bohr Institute and Natural History Museum of Denmark, University of Copenhagen,
{\O}ster Voldgade 5-7, DK-1350 Copenhagen K, Denmark}

\begin{abstract}

The short-lived $^{26}$Al and $^{60}$Fe radionuclides are synthesized and expelled in the interstellar medium by core-collapse supernova events. The solar system's first solids, calcium-aluminium refractory inclusions (CAIs), contain evidence for the former presence of the $^{26}$ Al nuclide defining the canonical $^{26}$Al/$^{27}$ Al ratio of $\sim5 \times10^{-5}$. A different class of objects temporally related to canonical CAIs are CAIs with fractionation and unidentified nuclear effects (FUN CAIs), which record a low initial $^{26}$Al/$^{27}$Al of $10^{-6}$. The contrasting level of $^{26}$Al between these objects is often interpreted as reflecting the admixing of the $^{26}$Al nuclide during the early formative phase of the Sun. We use giant molecular cloud (GMC) scale adaptive mesh-refinement numerical simulations to trace the abundance of $^{26}$Al and $^{60}$Fe in star-forming gas during the early stages of accretion of individual low mass protostars. We find that the $^{26}$Al/$^{27}$Al and $^{60}$Fe/$^{56}$Fe ratios of accreting gas within a vicinity of 1000 AU of the stars follow the predicted decay curves of the initial abundances at time of star formation without evidence of spatial or temporal heterogeneities for the first 100 kyr of star formation. Therefore, the observed differences in $^{26}$Al/$^{27}$Al ratios between FUN and canonical CAIs are likely not caused by admixing of supernova material during the early evolution of the proto-Sun. Selective thermal processing of dust grains is a more viable scenario to account for the heterogeneity in $^{26}$Al/$^{27}$Al ratios at the time of solar system formation.
\end{abstract}

\section{Introduction}
Giant molecular clouds (GMCs)  are the primary reservoirs of cold, star-forming  gas in the Galaxy. Astronomical observations and numerical simulations of star-forming regions suggest that GMCs have typical lifetimes of a few tens of Myr \citep{2015arXiv150904663P,2014prpl.conf....3D,2009ApJS..184....1K,1999PASJ...51..745F,1977ApJ...217..464B,1980ApJ...238..148B}, during which multiple episodes of star formation may take place. Stars more massive than eight solar masses eventually end their lives in type II supernova explosions and, during these events, pollute their environments with nucleosynthetic products. As such, the nucleosynthetic make-up of a protostellar core in a GMC is expected to reflect a mixture of an old galactically-inherited component with younger supernova-derived material, including freshly-synthesized radioactive $^{26}$Al and $^{60}$Fe, produced during the lifetime of the GMC. The $\gamma$-ray emission from radioactive $^{26}$Al and $^{60}$Fe nuclei, observable throughout the Milky Way due to the low opacity of $\gamma$-rays, have been used to determine the current average ISM abundance of $^{26}$Al and $^{60}$Fe and, hence, an estimate of the Galactic $^{60}$Fe/$^{26}$Al ratio \citep{2006Natur.439...45D,2007A&A...469.1005W}.

Meteorites and their components provide insights into the formation history of the earliest solar system, including the birthplace of the Sun. The most primitive meteorites, chondrites, contain calcium-aluminum-rich inclusions (CAIs) representing the oldest dated solar system solids, formed $4567.30 \pm 0.16$ Myr ago \citep{2012Sci...338..651C}. These sub-millimeter-to-centimeter objects are thought to have formed as fine-grained condensates from a gas of approximately solar composition, in a region with high ambient temperature (>1,300 K) and low total pressures ($10^{-4}$ bar) \citep{1987ppic.proc..333T,2000GeCoA..64..339E}, possibly during a brief (<10,000 years) \citep{2011ApJ...735L..37L}
heating event temporally associated with the very earliest phase of the proto-Sun \citep{2009GeCoA..73.4963K}. The presence in CAIs of the decay products of the short-lived radioisotope $^{10}$Be formed by solar energetic particle irradiation \citep{2000Sci...289.1334M} is further evidence that they formed in the vicinity of the proto-Sun. Importantly, CAIs typically contain evidence for an early presence of $^{26}$Al, defining a canonical initial $^{26}$Al/$^{27}$Al ratio of $\sim 5\times10^{-5}$ \citep{1995Metic..30..365M,2008E&PSL.272..353J,2011ApJ...735L..37L}. This initial abundance is approximately 10 times higher than that expected from the galactic background abundance, apparently requiring a late-stage seeding of the protosolar molecular cloud core from a nearby supernova. However, numerical simulations of the production, transport, and admixing of freshly synthesized $^{26}$Al in star-forming regions within GMCs \citep{2013ApJ...769L...8V} indicate that, under typical star formation conditions, the levels of $^{26}$Al in most star-forming regions are comparable to that deduced from CAIs. Thus, the presence of short-lived radionuclides (SLRs) such as $^{26}$Al in the early solar system does not require special circumstance but rather represent a generic feature of the chemical evolution of GMCs.

However, a class of refractory grains and inclusions believed to be temporally related to the formation of canonical CAIs record much lower levels of $^{26}$Al corresponding to initial $^{26}$Al/$^{27}$Al of $<5\times10^{-6}$ \citep{1987Metic..22..377F}. Of interest are the coarse-grained refractory inclusions with fractionation and unidentified nuclear effects (FUN CAIs, \citet{1977GeoRL...4..299W}), which, in addition to their low initial abundance of $^{26}$Al, are characterized by large mass-dependent fractionation effects and nucleosynthetic anomalies in several elements. Moreover, the abundance of rare earth elements and the oxygen isotope composition of FUN CAIs indicate that their precursors formed as condensates from a solar gas \citep{Holst28052013}. Collectively, these observations are often interpreted as reflecting formation of FUN CAIs prior to the admixing of stellar-derived $^{26}$Al to the CAI forming gas \citep{1998ApJ...509L.137S,2008ApJ...680L.141T,2011ApJ...733L..31M,2012ApJ...756..102P,2012ApJ...756L...9B,2013ApJ...770...51B,2014ApJ...788...20B,2015ApJ...809..103B}. If  this interpretation is correct, these objects can be used to track the timing of addition of $^{26}$Al to the forming protoplanetary disk. Alternatively, the contrasting initial $^{26}$Al abundance of canonical and FUN CAIs may reflect unmixing of two distinct dust components by thermal processing \citep{2009Sci...324..374T,2013ApJ...763L..40P,2015E&PSL.420...45S}, namely an old, galactically-inherited homogeneous dust component and a new, supernova-derived dust component formed shortly prior to the collapse of the protosolar molecular cloud core. Such a bi-modal dust distribution can either be achieved by having separate populations of grains, or having old grains being covered with newly synthesised gas condensates, resulting in a multilayered grain-structure. Distinguishing between these two interpretations is critical for understanding the origin and distribution of SLRs in the early solar system.

In this paper, we use GMC-scale adaptive mesh-refinement numerical simulations to trace the abundance of $^{26}$Al and $^{60}$Fe in star-forming gas during the early stages of accretion of individual low mass protostars. We first model the star formation process on the time scale of an evolving GMC structure, and use additional adaptive mesh refinement to zoom in on individual stars, allowing us to study the accretion dynamics of individual stars and their disks down to scales of a few astronomical units. This approach allows us, for the first time, to evaluate the level of $^{26}$Al and $^{60}$Fe heterogeneity during the early evolutionary stages of individual protostars that may result from the variable contributions of different supernova sources during the lifetime of the GMC structure.
More than 200 stars with masses of at least 0.2 M$_{\odot}$ form during our simulation, of which we select ten stars that end up having about 1-2 solar masses and one of about 7 solar masses for detailed high-resolution zoom-in investigations. Our models indicate a homogeneous level of $^{26}$Al in the accreting gas for all systems during the first 100 kyr of their formation, although some level of heterogeneity is possible in the later evolutionary stages. Therefore, the contrasting initial $^{26}$Al/$^{27}$Al ratios recorded by canonical and FUN CAIs cannot easily be understood as a result of heterogeneous accretion processes.

\section{Methods}

The simulations were carried out with the magnetohydrodynamic
(MHD) adaptive mesh refinement (AMR) code \ramses\
 \citep{2002A&A...385..337T, 2006A&A...457..371F}. We
solve the equations of magnetohydrodynamics using a MUSCL Godunov
method with a constrained transport HLLD solver \citep{2005JCoPh.208..315M} using
a multi-dimensional MonCen slope limiter. To maintain numerical stability in super-sonic
flows, and ensure a reasonable time-step, cells where the combined advection
and fast-mode speed -- the total signal velocity -- is above about $150 \unit{km}\unit{s}^{-1}$ are evolved with a
more diffusive local Lax-Friedrichs solver.
In \ramses\ the adaptive mesh is described with a fully threaded
oct-tree, and a cell refined to level $n+1$ has half the linear size compared to a cell refined to level $n$. Refinement
can be done according to a variety of criteria. The basic criterium used
in this paper is a Truelove density-based refinement with a factor of 4 increase
in threshold density for each level of refinement, resulting in a constant minimum
number of cells per Jeans-length \citep{1997ApJ...489L.179T}. This is
complemented with a number of refinement criteria based on gradients in density,
pressure, and magnetic fields, as described below.

To model a star forming region, we include self-gravity, cooling parameterized with
a table lookup based on \citet{2012ApJS..202...13G}, heating from cosmic rays,
and photo-ionization \citet{2006agna.book.....O} with a density dependent exponentially cut-off of 500 cm$^{-3}$
\cite{1986PASP...98.1076F}. For a more detailed description of the thermodynamics see also \cite{2015arXiv150904663P}.
When the gas reaches a density where a Jeans length at the
highest level of refinement is resolved with only a few cells,
and several other criteria are fulfilled (see below),
sink particles are inserted that interact with the gas through accretion
\citep[cf.][]{2012ApJ...759L..27P, 2014ApJ...797...32P}.
Sink particles more massive than 8 M$_{\odot}$ eventually explode as supernovae, with
a delay time (stellar life time) given by a mass-dependent lookup table \citep{1992A&AS...96..269S}.
Fresh SLR material is admixed into the supernova ejecta
according to the mass-dependent yields given in \citep{2006ApJ...647..483L}.

\subsection{Setup and initial evolution for a GMC}

The current simulation is a partial rerun, with much higher numerical fidelity, of
the model in \citet{2013ApJ...769L...8V}, which used
a $(40\unit{pc})^3$ periodic box with a total mass of $9\times10^{4}$M$_{\odot}$,
and a mean magnetic field of 3.5 $\mu$G.
The initial evolution was simulated using the unigrid {\sc stagger-code} \citep{2011ApJ...737...13K,2011ApJ...730...40P}.
That model was started up by driving turbulence with a typical root mean square velocity
of 6-7 km/s \citep{2011ApJ...730...40P}, consistent with Larson's velocity dispersion-size relation
$\sigma(\unit{km}\unit{s})^{-1}\propto L_{\unit{pc}}^{0.38}$ \citep{1969MNRAS.145..271L, 1981MNRAS.194..809L}.
Self-gravity was then turned on, and subsequently, sufficiently dense gas was converted to a distributions of
sink particles, sampled according to a Salpeter initial mass function (IMF) \citep{2002ApJ...576..870P}.
When the kinetic energy feedback due to supernova explosions
became significant, the turbulence driving was turned off, and the evolution was continued with
the {\ramses} code, with a root grid of $128^3$ and 16 levels of refinement relative to the box
size; i.e. with a minimum cell size of about 126 AU.  Essentially the same star formation recipe as described
below was used to self-consistently follow the formation of massive stars, which gradually took
over driving from the generation of massive stars originating from IMF sampling.
The result is a realistic, evolving GMC model, polluted through supernova explosions
with SLRs, and having a mature and diverse  population of massive stars \citep[cf.][for details]{2013ApJ...769L...8V}.
Here we repeat parts of that run with 4-6 additional levels of refinement, using an updated version of {\ramses} that also provides higher fidelity at a given grid size than the version used by \citet{2013ApJ...769L...8V}, by allowing more aggressive choices of Riemann solver and slope limiter.
In the current paper we define t=0 as the time of birth of the first massive star formed in the
\citet{2013ApJ...769L...8V} \ramses\ run.

\subsection{Sink particle creation and accretion}

Sink particles are used as a subgrid model for stars. Sink particles form from cold gas on the
highest refinement level, when it exceeds a certain
threshold density, has a convergent velocity flow, is at a local minimum of
the potential, and is at least 30 cells
($\approx 3777$ AU) from any already formed sink particle.
In addition, the temperature of the gas has to be below 2000 K.
Over time, it is evident that stars form clustered in filaments of
high density, in contradiction to the classical model of stars forming
isolated due to gravitational collapse \citep{1977ApJ...214..488S}. This
is consistent with numerical simulations by other groups
(e.g. \cite{2013A&A...556A.153H, 2014MNRAS.439.3420M}, \cite{2015arXiv150903436B})
and as seen in observations \citep{2003ARA&A..41...57L,2010MNRAS.409L..54B}.
It demonstrates the necessity of using our more complex large-scale zoom-in
model, instead of using idealized spherical core collapse models.
The sink particles move through the molecular cloud, accreting
gas from surrounding cells within a radius of 8 cells from the sink if the
total energy in the gas in a nearby cell is negative. The rate
of accretion increases gradually from zero at the edge to a fraction of
$\approx 0.01$  per orbital time near
the sink particle, similar to the prescription given in \citep{2014ApJ...797...32P}.
Galilean transformations to the rest frame of a single sink particle make it possible to turn on the
built-in \ramses\ geometric refinement and keep the particle centered, allowing
us to explicitly zoom to the environment around the sinks of interest.  The
geometric refinement in \ramses\ does not force refinement.
Instead, it constrains potential refinement only to cells that are located close enough to the sink in order to avoid unnecessary computational costs.
We set up geometric of refinement in such a way that with decreasing distance from the sink the allowed maximum level of refinement gradually increases.
Here we allow refinement in concentric spherical regions with
radii of at least 40 cells at each refined level.

Self-gravity in the simulation is accounted for in three steps: first,
we compute the potential from only the gas, using it to compute the
gravitational force from the gas on the sink particles. Second, we deposit the sink
particle masses to the grid using a triangular shaped cloud (TSC) method and use the combined potential from
the gas and sink particles to compute the gravitational forces on the gas. The gravitational forces
between sink particles are accounted for by explicitly using Newton's law, with a smoothed gravitational
potential using a piece-wise polynomial with a softening length of $0.3 \Delta x$ \citep{2010ApJ...713..269F}.
Particles are evolved with a symplectic kick-drift-kick leap-frog integrator, identical to the one used
for dark matter particles in {\ramses}. This ensures that close encounters between particles
are properly accounted for, and close binaries settle in orbits of $\sim \Delta x$.
To make sure sink particles move in stable orbits
we have added two courant conditions related to the max velocity and acceleration
of the sink particles
\begin{eqnarray}
\Delta t_v & = C_{\Delta t} \frac{\min(r_{ss'},\Delta x)}{v} \\
\Delta t_a & = \left[C_{\Delta t} 2 \frac{\min(r_{ss'},\Delta x)}{a}\right]^{1/2}\,,
\end{eqnarray}
where $C_{\Delta t}$ is the courant number (typically 0.5), $r_{ss'}$ is the minimum distance between
two sink particles, $\Delta x$ is the cell size, $v$ is
the speed of each sink and $a$ the size of the acceleration.

\subsection{Zoom-in on individual stars}
To follow the accretion history of a number of individual stars in detail, while
simultaneously retaining realistic initial and boundary conditions of the surrounding
medium in the model, we use the method of zoom-in introduced in \citet{2014IAUS..299..131N},
proceeding in two steps:
First, we simulate the evolution of the entire box for approximately 4 Myr with
a maximum resolution of 126 AU. During this step, several hundred stars of different 
masses are formed, from which eleven stars are selected. Ten of these stars accrete
to from 1 to 2 M$_{\odot}$ and one (star 4) accretes to a final mass of about 2.8 M$_{\odot}$. 
In table \ref{run-overview}, we provide an overview of the different sinks. The selected
stars are formed --- at different points in time and in different environments --- from
collapsing pre-stellar cores generally located in filamentary structures of the GMC.
Our choice of selecting stars that accrete to more than 1 M$_{\odot}$ is motivated
by the fact that young stars eject a fraction of the accreting mass in strong
outflows, which we do not resolve with a resolution of 126 AU, but (partly) resolve
when zooming in. Consequently, in order to model the formation and evolution of
what becomes solar mass stars, we need to select stars that accrete more than
1 M$_{\odot}$ in the first step, i.e. the low resolution run.

In the second step, we rerun the simulation with higher resolution
around the selected stars, to follow the accretion onto these individual stars
in as much detail as we can afford.  A compromise between resolution and
time coverage allows using up to 20 to 22 AMR levels of refinement relative 
to the box size, instead of the 16 levels used in the first step.
This corresponds to a minimum cell size of either 8 AU (20 levels) 
or 2 AU (22 levels), which still allows following the accretion over  
about 100 kyr (during which we use output file cadences between 0.2 to 1 kyr).
This is sufficient to cover the periods of time when most of the mass of the 
stars accretes.
To ensure proper coverage of the early phase
of star formation, we start most of the second step simulations more than ten thousand
years before formation of the selected star, while imposing a `geometric refinement' 
zoom-in region centered on the pre-stellar core of the star\footnote{In RAMSES, `geometric 
refinement' is a technique where successively larger regions disallows refinement, one
level at a time.  Within each region, refinement is allowed, but is not imposed.}.
Simultaneously, we also insert about $10$ million tracer particles in a cubic region 
of about $1.28\times10^{5}$ AU (0.62 pc) in diameter for eight of the zoom-in runs, 
namely for star 1, 5, 6, 7, 8, 9, 10 and 11.
The tracer particles are distributed with a probability density proportional to mass 
density, and are passively advected with the gas motion.

\begin{table}[]
\centering
\resizebox{\columnwidth}{!}{
\begin{tabular}{r|r|r|r|r|r}


\begin{tabular}{@{}c@{}} \# of \\ star \end{tabular} & \begin{tabular}{@{}c@{}} $\Delta x_{\rm min}$ \\ in AU \end{tabular}& \begin{tabular}{@{}c@{}} $t_{\rm birth}$ \\ in kyr \end{tabular} & \begin{tabular}{@{}c@{}} x \\ in pc \end{tabular} & \begin{tabular}{@{}c@{}} y \\ in pc \end{tabular} & \begin{tabular}{@{}c@{}} z \\ in pc \end{tabular} \\ \hline
1                   & 2                              & 631                           & 33.2             & 30.8             & 7.8             \\
2                   & 2                              & 667                           & 13.5             & 27.4             & 25.6             \\
3                   & 2                              & 1743                          & 11.1             & 10.9             & 0.1           \\
4                   & 2                              & 2055                          & 11.9             & 9.5             & 27.5             \\
5                   & 8                              & 2212                          & 37.9             & 27.3             & 33.0             \\
6                   & 2                              & 2471                          & 3.2             & 9.2             & 3.2             \\
7                   & 2                              & 2576                          & 3.5             & 8.9             &2.6             \\
8                   & 2                              & 2653                          & 10.2             & 12.3             & 3.4             \\
9                   & 2                              & 3157                          & 9.3             & 12.0            & 32.3             \\
10                   & 8                              & 3271                          & 26.1             & 29.3             & 2.6             \\
11                   & 2                              & 3389                          & 3.3             & 4.6             & 2.2             \\
\end{tabular}}
\caption{Overview of the eleven stars selected for zoom-in. First column: number of star, second column: cell size at the highest resolution, third column: time of formation of the star in the parental run, fourth to sixth column: $x$, $y$ and $z$-coordinate of the star at the time of formation.}
\label{run-overview}
\end{table}

\section{Distribution of SLRs in space and time on different scales }

In this Section, we present the distribution of $^{26}$Al and $^{60}$Fe
abundance in a GMC, as obtained in our simulations.
We present and discuss the evolution and distribution of the SLRs in
the entire GMC of (40 pc)$^3$ according to our AMR simulation with a
maximum resolution of 126 AU, which corresponds to a maximum level of
refinement of 16 powers of 2 (i.e. $\frac{40 \unit{pc}}{2^{16}}=126 \unit{AU}$).
In the following, we refer to this run as our parental run. First, we
analyze the distribution of the SLRs in the gas phase of the entire GMC
and discuss how it affects the abundance in and around the stars.
Afterwards, we elaborate in more detail on the SLR abundance around
particular stars, by zooming in with a maximum resolution of 2 AU
(22 levels of refinement) on nine stars, and with a maximum resolution
of 8 AU (20 levels of refinement) around 2 stars. We distinguish between
the early (first $\sim$100 kyr after star formation) and late phase (times
later than $\sim$100 kyr).

\subsection{Distribution of SLRs in the gas phase of the GMC}

\begin{figure*}
\centering
\subfigure{\includegraphics[scale=0.44,bb=28bp 262bp 602bp 530bp,clip]{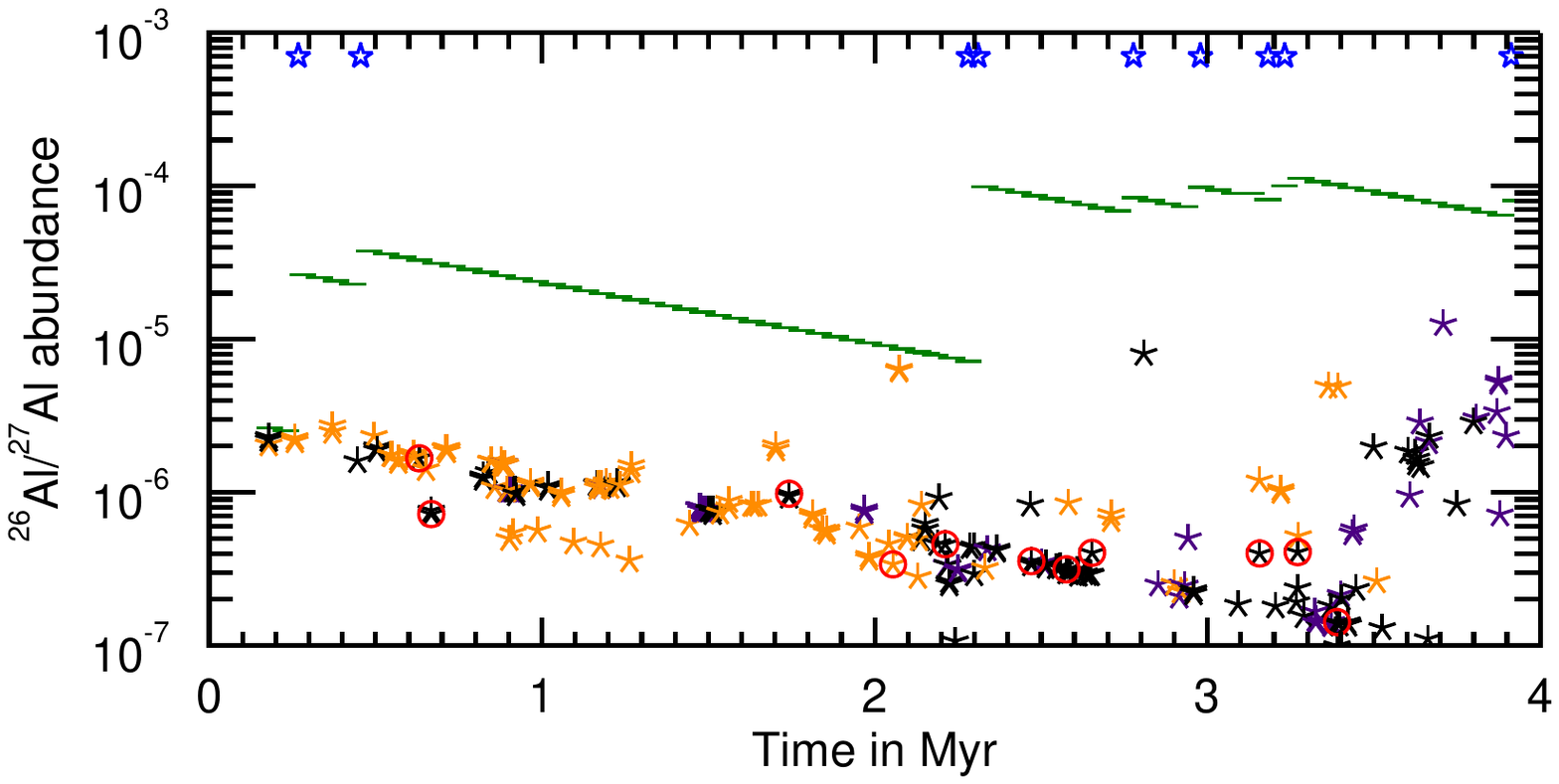}}
\subfigure{\includegraphics[scale=0.44,bb=28bp 262bp 602bp 530bp,clip]{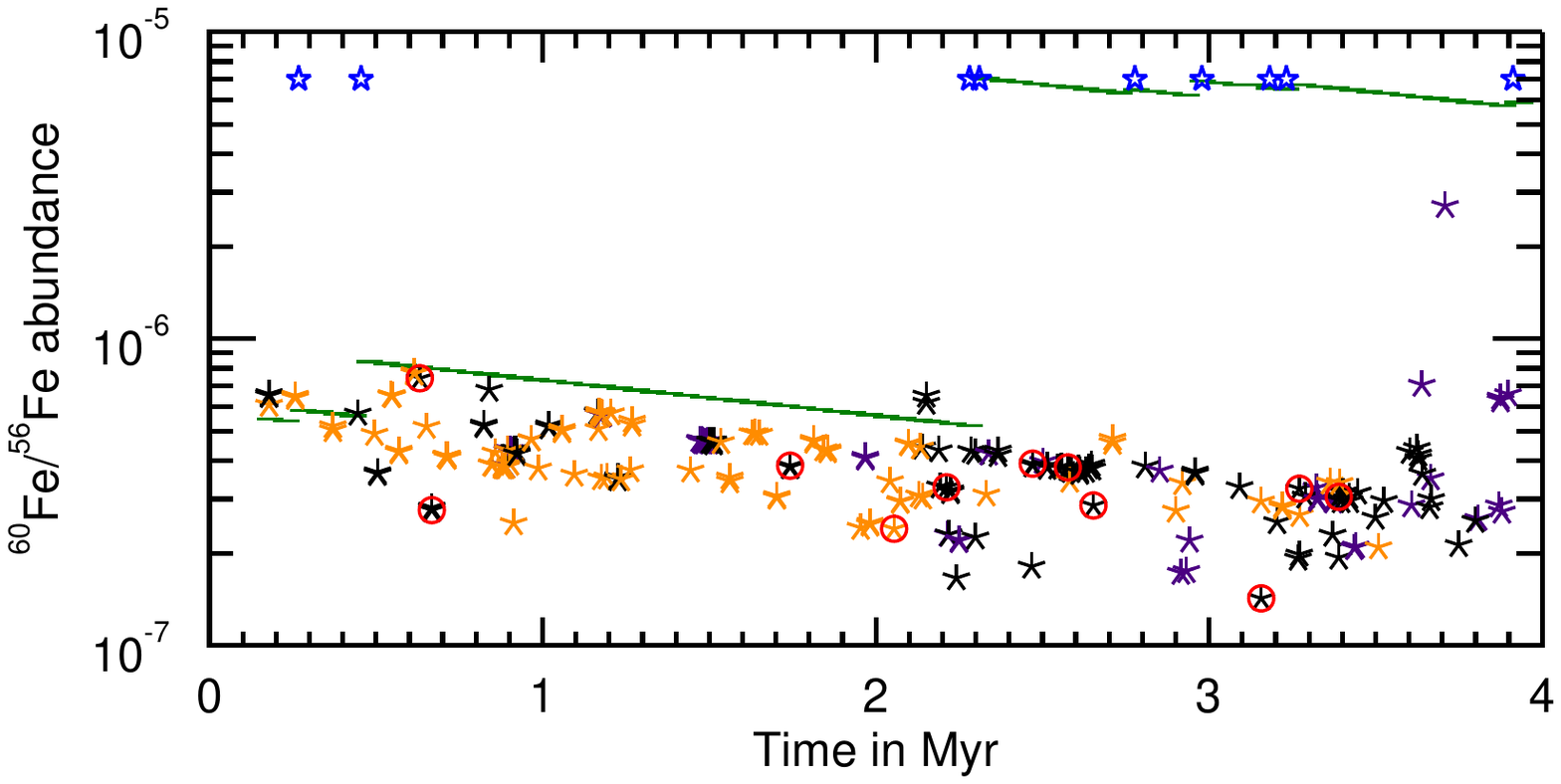}}
\protect\caption{\label{fig:SLR_sinks} The two panels illustrate the SLR-ratios for all the sinks in the mass range $0.2$ M$_{\odot}$ to $0.5$ M$_{\odot}$ (purple asterisks), $0.5$ M$_{\odot}$ to $2.5$ M$_{\odot}$ (black asterisks), $2.5$ M$_{\odot}$ to $8$ M$_{\odot}$ (orange asterisks) and the average value of all the gas (green horizontal line) vs. time
of formation in the \ramses\ simulation. The red circles mark the 11 stars that are selected for zoom-ins. The blue stars in the upper part of both plots indicate times of supernova explosions -- they do not reflect the SLR abundance/enrichment of the supernovae. Left panel: $^{26}$Al/$^{27}$Al;
right panel $^{60}$Fe/$^{56}$Fe}
\end{figure*}

During the roughly 4 Myr of GMC evolution considered for this paper, nine of the massive stars adopted from the previous {\stagger} and {\ramses} runs explode as supernovae after their mass dependent life times and admix $^{26}$Al as well as $^{60}$Fe at different locations in the GMC.

In \Fig{SLR_sinks}, we show the average mass-weighted abundance of $^{26}$Al (left panel) and $^{60}$Fe (right panel) as green horizontal lines together with the abundances of the individual stars of masses from 0.2 to 0.5 M$_{\odot}$ (purple asterisks), 0.5 to 2.5 M$_{\odot}$ (black asterisks) and 2.5 to 8 M$_{\odot}$ (yellow asterisks) at their times of formation. The initial abundance of $^{26}$Al and $^{60}$Fe in the cloud prior to the first supernova explosion reflect contributions from earlier supernova events that occurred prior to our t=0. It is clear from Fig. 1 that the average $^{26}$Al and $^{60}$Fe abundances in the cloud are highly modulated by supernova events (illustrated by the blue stars on top of the plots), followed by a gradual decrease due to radioactive decay. The first supernovae corresponds to a star of 13 M$_{\odot}$ and result in a significant enhancement of the $^{26}$Al abundance relative to $^{60}$Fe. There are two reasons for this. First, $^{26}$Al ($\tau_{\rm 1/2,^{26} Al} \approx 717$ kyr) decays about three times faster than $^{60}$Fe ($\tau_{\rm 1/2,^{60}Fe} \approx 2.6$ Myr \citep{2009PhRvL.103g2502R}) and, hence, the $^{60}$Fe abundance is depleted less than $^{26}$Al before the first supernova event. Second, the first supernova event is not a very massive star, which produces less $^{60}$Fe per unit mass relative to more massive supernovae. The second supernova explodes with a mass of $22$ M$_{\odot}$ less than 200 kyr later and significantly enriches the cloud in $^{60}$Fe as it is more efficient in producing $^{60}$Fe than the first supernova. During the time until the next supernova explosion, one can clearly recognize the characteristic decay of both SLRs before the box gets efficiently enriched in SLRs by the third supernova. This supernova is the most massive one during the entire evolution of the GMC with $75$ M$_{\odot}$ and is thus particularly responsible for the enhancement in $^{60}$Fe. The subsequent supernovae, with masses of $15$ M$_{\odot}$, $29$ M$_{\odot}$, $40$ M$_{\odot}$, $13$ M$_{\odot}$, $22$ M$_{\odot}$ and $29$ M$_{\odot}$ do not enhance the average abundance as much, partly because enrichments of the already enhanced GMC appear less significant on the logarithmic scale. In general, we can see an overall increase of SLR abundances (ranging from about $2.5\times 10^{-6}$ to about $1\times 10^{-4}$ for $^{26}$Al/$^{27}$Al and from about $5\times10^{-7}$ to about $7\times10^{-6}$ in $^{60}$Fe) over time, consistent with earlier results of \cite{2013ApJ...769L...8V}.

\begin{figure*}
\subfigure{\includegraphics[scale=0.22]{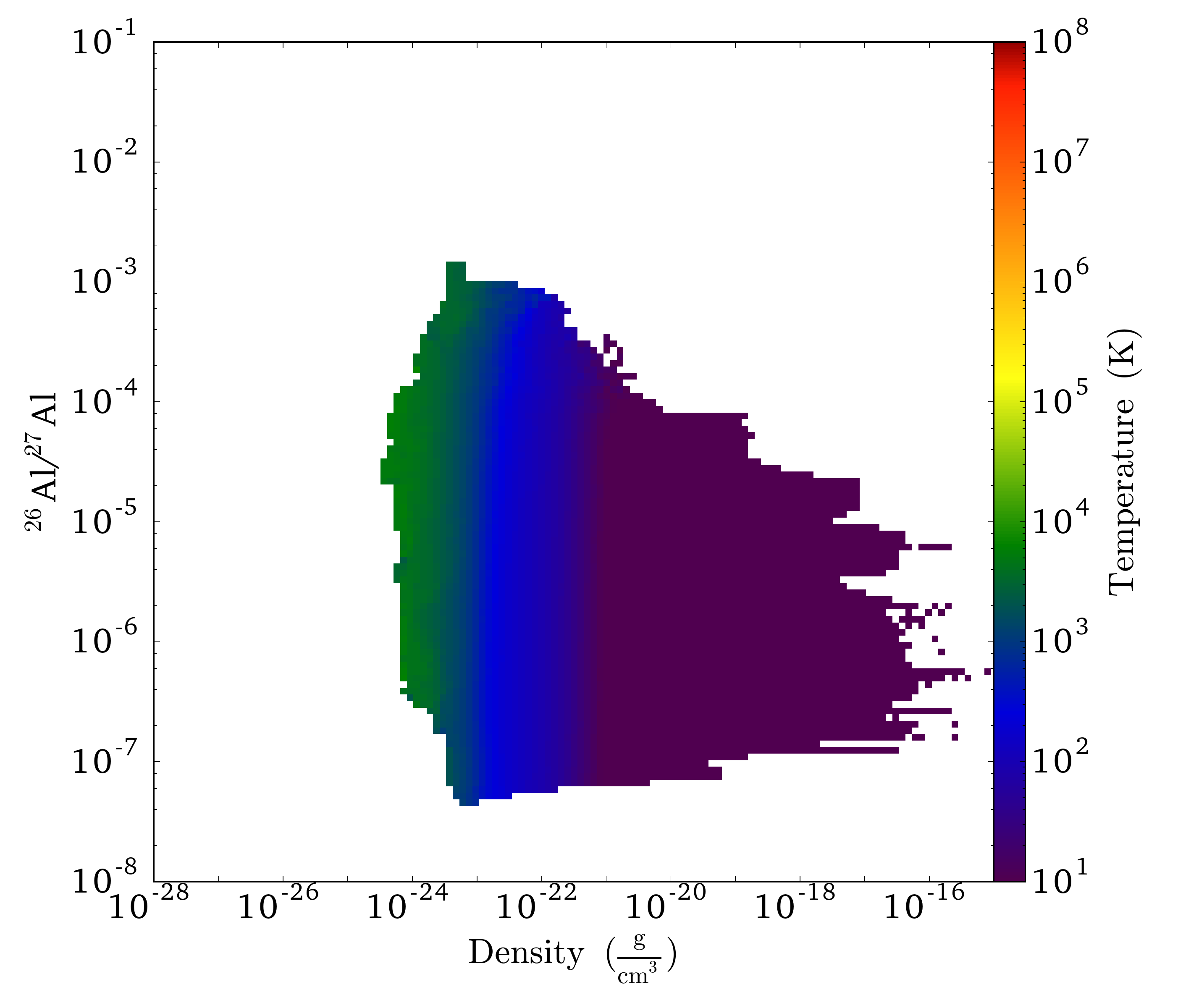}
\includegraphics[scale=0.22]{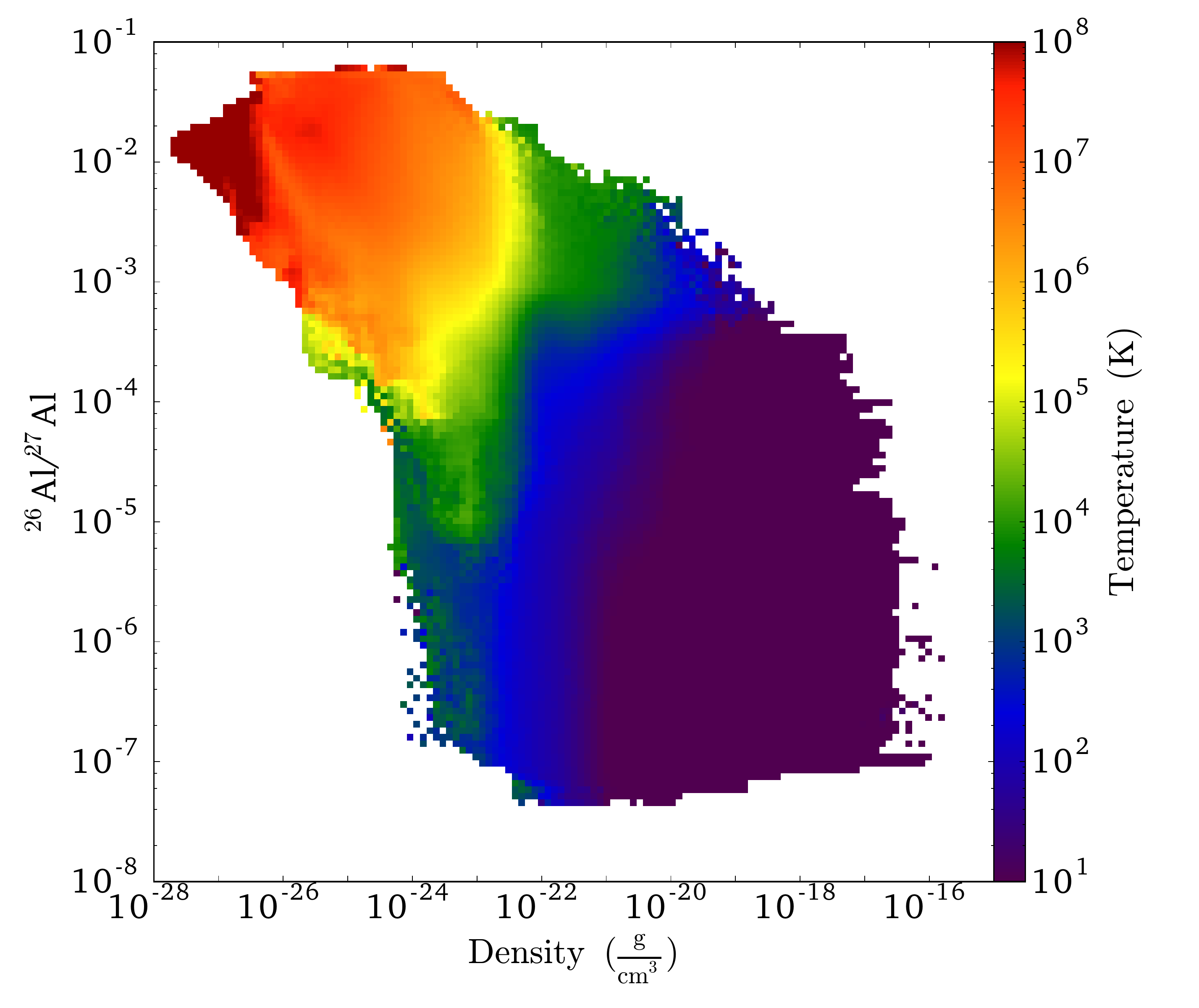}
\includegraphics[scale=0.27]{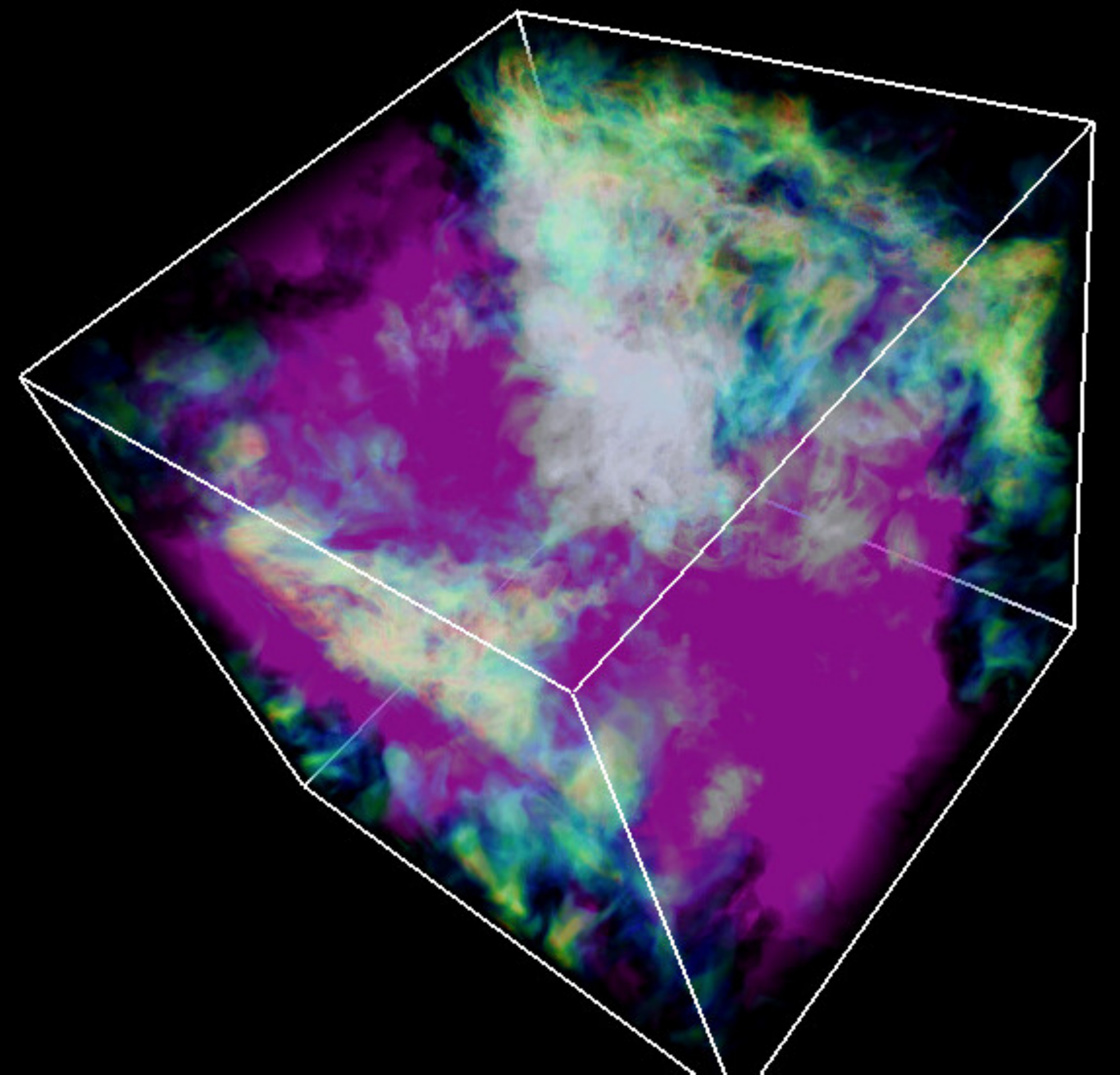} }

\protect\caption{\label{fig:phase_space_box}Distribution of $^{26}$Al/$^{27}$Al abundance
inside the entire box of all the cells just at the end of the supernova quiet period at $t=2.2$ Myr(left panel) and at the end of the simulation $t=3.9$ Myr(middle panel)
in dependence of their density. The colors in the diagram represent the temperature gas temperature from cold (purple) to warm (red). The right panel illustrates how $^{26}$Al/$^{27}$Al is distributed in the GMC at the end of the simulation. White color represents low abundances ($10^{-7}$), while purple represents high abundances ($10^{-2}$). }
\end{figure*}

\begin{figure*}
\centering
\subfigure{\includegraphics[scale=0.44,bb=28bp 262bp 602bp 530bp,clip]{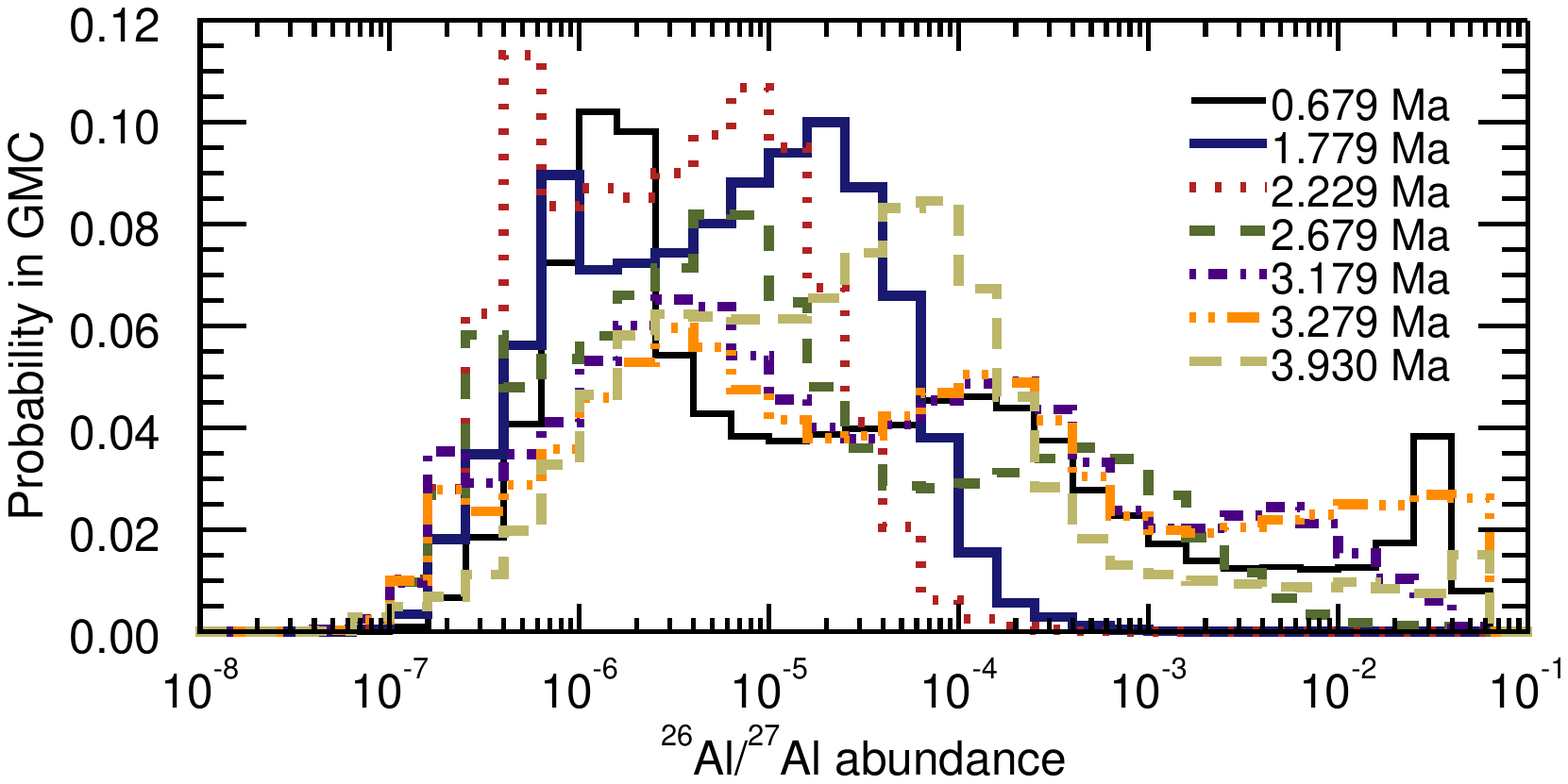}}
\subfigure{\includegraphics[scale=0.44,bb=28bp 262bp 602bp 530bp,clip]{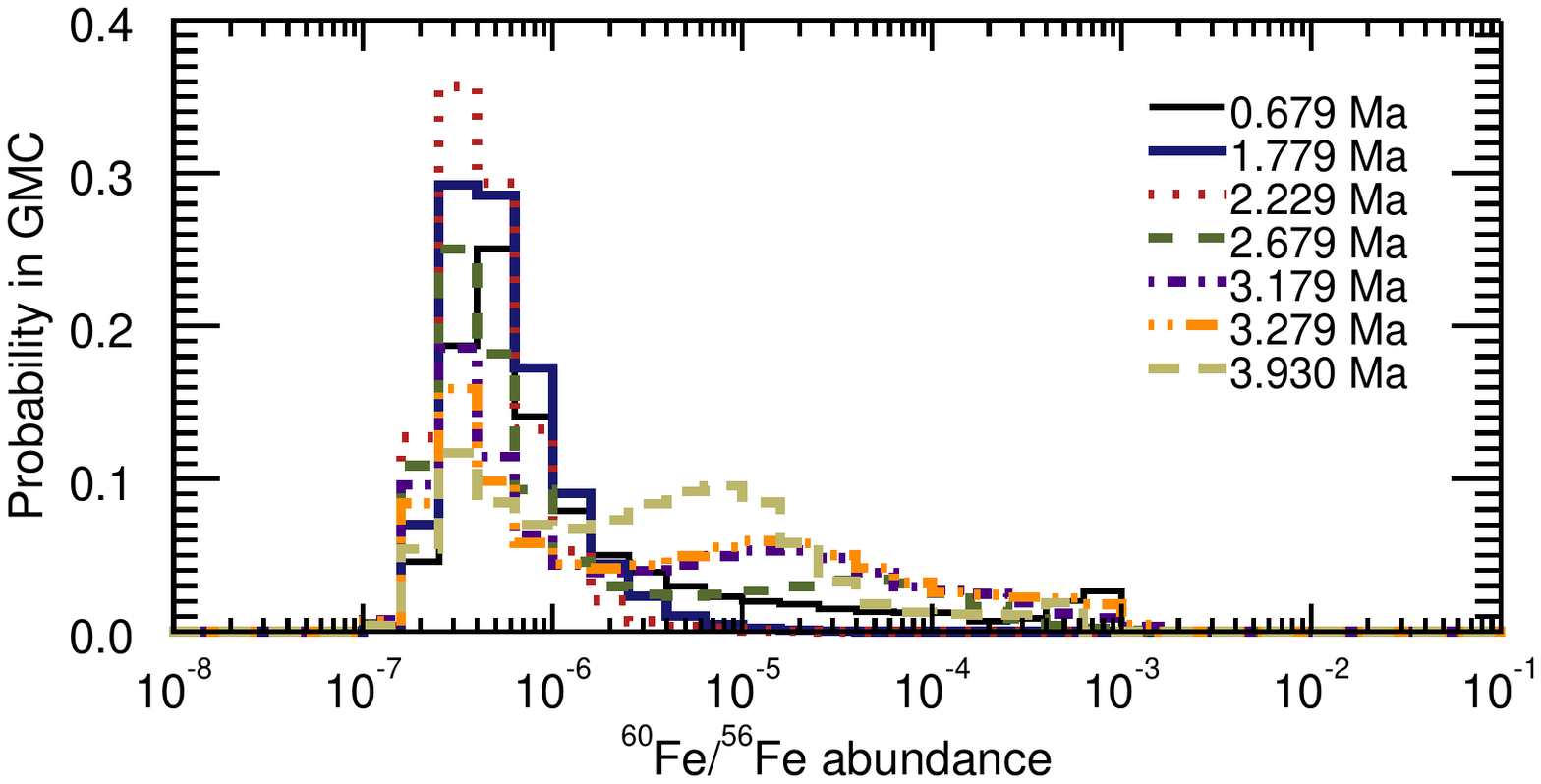}}
\protect\caption{\label{fig:SLR_histo}The two panels illustrate the probability distribution of SLR-values in logarithmic steps of 0.1 at times briefly after the formation of star 1 and 2, star 3, star 5, star 6, star 7, star 8 and 9 and at the end of the simulation. We have omitted star 4 in these histograms for clarity. Note that the histograms for star 4 is similar to that of star 3 and 5. Left panel: $^{26}$Al/$^{27}$Al;
right panel $^{60}$Fe/$^{56}$Fe}
\end{figure*}

Enrichment in SLR abundances is also reflected in the spatial distributions of SLRs at different times.
The two left panels of \Fig{phase_space_box} show the distribution of $^{26}$Al/$^{27}$Al in the cloud for all cells with respect to density and temperature for two different times. The left panel illustrates the distribution just at the end of the quiescent period at $t=2.2$ Myr, while the middle panel corresponds to the end of the simulation. In both diagrams one can see that the temperature of the gas decreases with increasing density. Also, the spread in $^{26}$Al/$^{27}$Al is wider for low density gas than for high density gas. Both diagrams reveal that the highest abundances occur for highest temperatures, but due to several recent supernova enrichments, this property is much more evident at the end of the simulation (middle panel) than after the supernova quiet phase (left panel). The significant amount of cells with high temperatures and low density reveals the admixing of enriched gas from supernova explosions into the GMC. To illustrate the spatial distribution of the SLRs, in particular $^{26}$Al/$^{27}$Al, we present the distribution of $^{26}$Al/$^{27}$Al inside our entire box with the visualization software \vapor\ \citep{clyne2007interactive}.
The color-scheme represents $^{26}$Al/$^{27}$Al ratios from a lowest value of about $3.9\times10^{-8}$
to highest values of about $5.6\times10^{-2}$. For clarity, we set the floor values to $10^{-2}$ ($10^{-8}$) and colored all values above (below) this ratio in white (violet). However, because the high $^{26}$Al/$^{27}$Al values are typically associated with hot and, hence, very low density gas, the gas enriched in SLR does not contribute significantly to the overall mass distribution in the GMC. The variability in $^{26}$Al/$^{27}$Al ratios present in dense and cold star-forming gas is much more limited relative to that of the entire GMC.

To quantify the distribution of SLR abundances in our GMC at different times more acurately, we plot the $^{26}$Al/$^{27}$Al and $^{60}$Fe/$^{56}$Fe distribution in the form of histograms in \Fig{SLR_histo}. The distributions clearly show that large spatial heterogeneities in $^{26}$Al/$^{27}$Al abundance exist throughout the entire GMC due to supernova enrichments. The SLR ratios cover a range of up to 6 orders of magnitude for $^{26}$Al/$^{27}$Al and up to about 4 orders of magnitude for $^{60}$Fe/$^{56}$Fe at times not too long after recent supernova events. Furthermore, the distribution is narrower and the GMC lacks very high values of SLRs at the end of more quiescent periods (blue solid line, red dotted line). Comparing the histogram for $t=2.229$ Myr, corresponding to the end of a quiescent period, with the distribution shortly after the two first supernova enrichments at $t=0.679$ Myr shows that the maximum SLR values are about two orders of magnitude lower as seen in \Fig{phase_space_box}. This significant decrease in $^{26}$Al and $^{60}$Fe abundance observed at the end of the quiescent period cannot be explained by radioactive decay and, instead, must reflect progressive admixing of the SLR-enriched high density gas with older, lower density gas present in the GMC.
Considering the range of abundances found in our model,
both the canonical value measured in bulk CV CAIs as well as lower
values measured in FUN CAIs are well represented within the range found in our simulation, although the majority of star forming gas is of lower abundance.

\begin{figure*}
\centering
\subfigure{\includegraphics[scale=0.44,bb=28bp 262bp 602bp 530bp,clip]{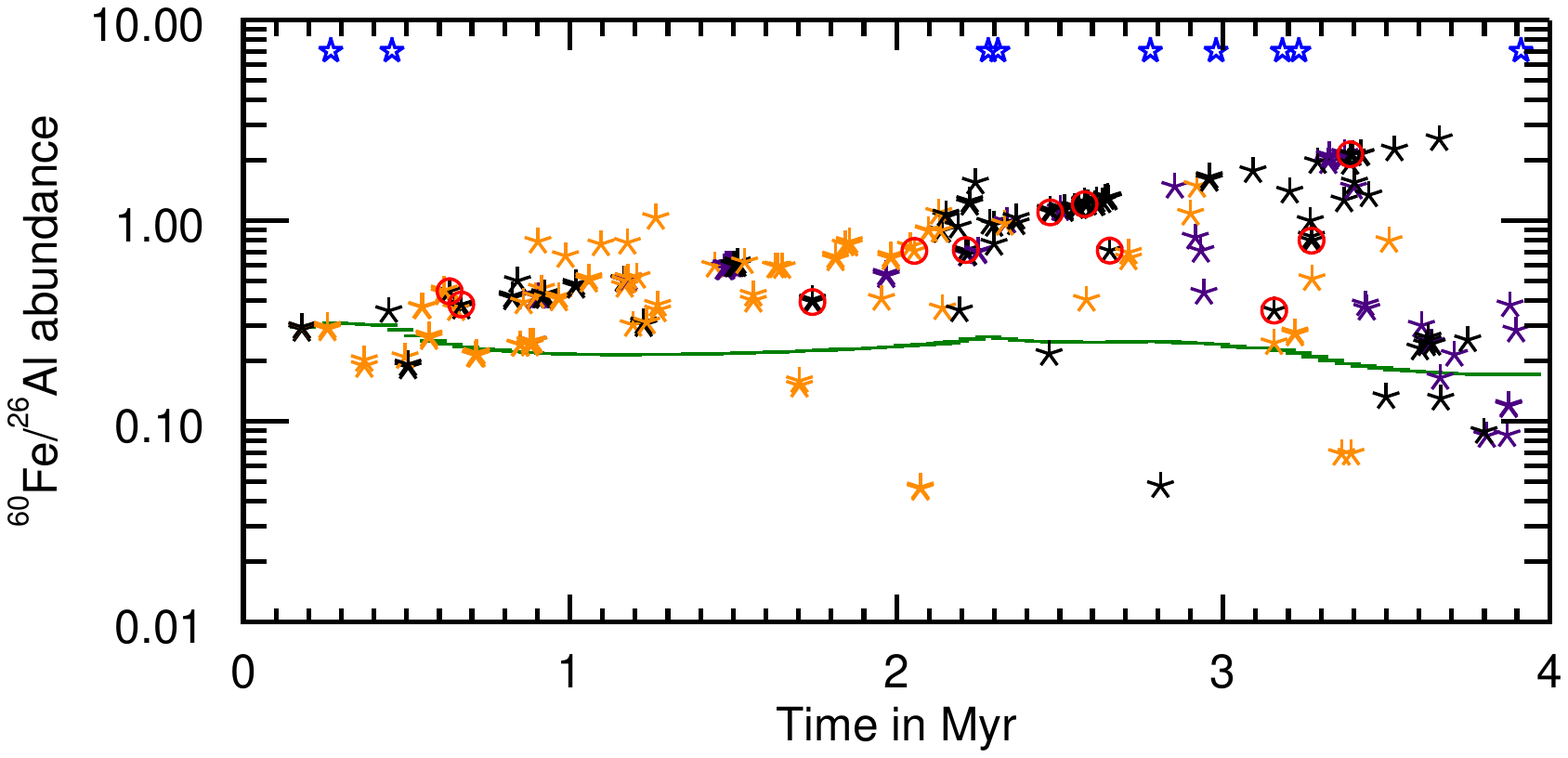}}
\subfigure{\includegraphics[scale=0.44,bb=28bp 262bp 602bp 530bp,clip]{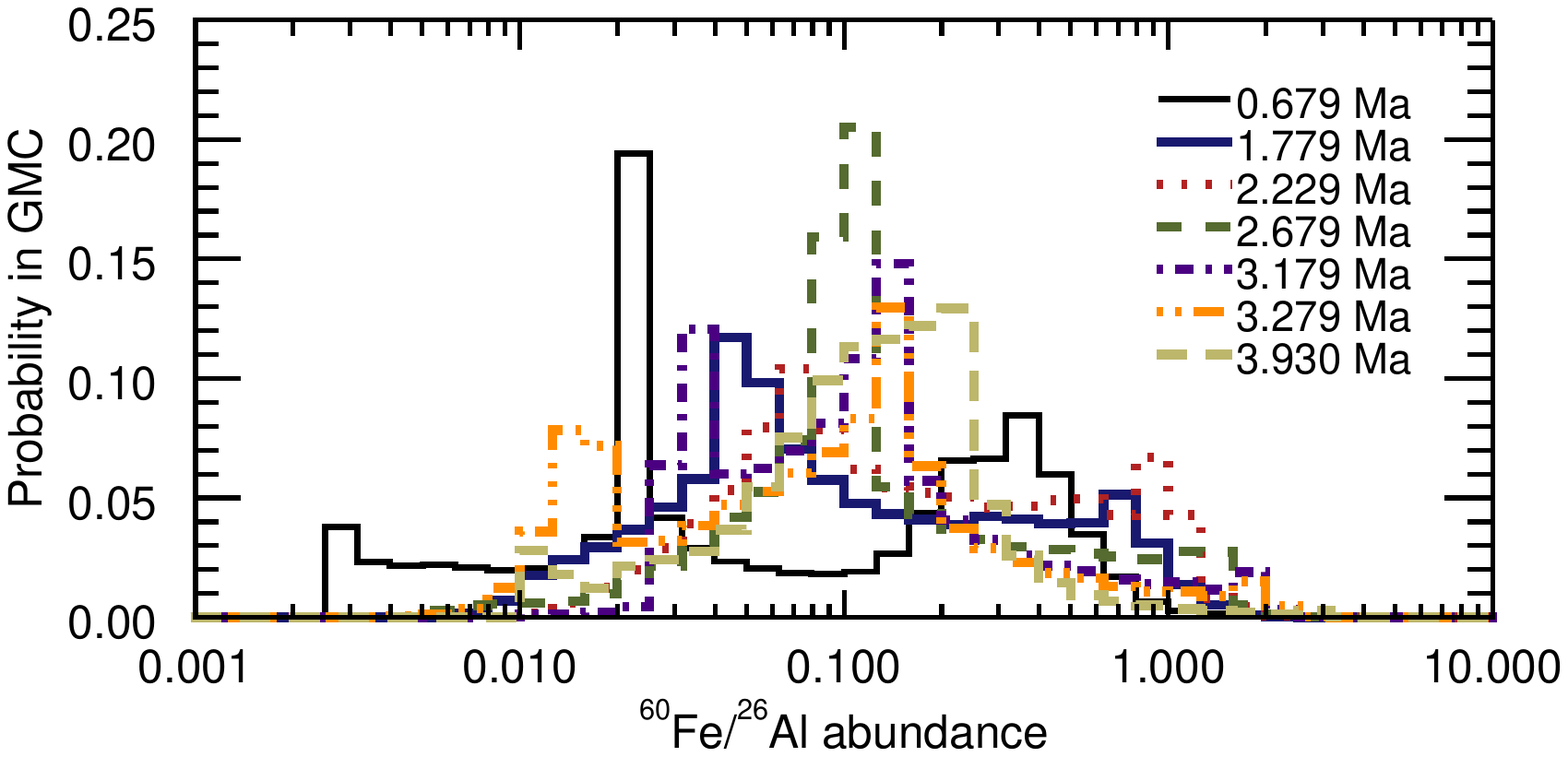}}
\protect\caption{\label{fig:FeAl_sinks} The plot illustrate the evolution of the average $^{60}$Fe/$^{26}$Al in the gas phase of \ramses\ simulation (green horizontal lines) together with the ratios inside the stars (black asterisks). The blue stars in the upper part of both plots only indicate times of supernova explosions, but no SLR abundance of the supernovae.}
\end{figure*}

Similarly to \Fig{SLR_sinks}, we illustrate in \Fig{FeAl_sinks} (left panel) the evolution of the average $^{60}$Fe/$^{26}$Al ratio in the GMC together with the abundances of the individual stars at their time of formation (asterisks) and the exploding supernovae. Although the average ratios vary due to the different decay times of $^{60}$Fe and $^{26}$Al as well as the different supernova enrichments during the evolution of the GMC, the value generally decreases from about 0.3 to about 0.16 at the end of the simulation due to the larger amount of supernovae that admix more $^{26}$Al than $^{60}$Fe into the GMC. Again, this value is in agreement with \cite{2013ApJ...769L...8V}, who found an average value of about 0.2. Furthermore, the value at later times is also consistent with the galactic value of $0.15\pm0.06$ \citep{2006Natur.439...45D,2007A&A...469.1005W}. Our average value is higher than the galactic value throughout the entire evolution of about 4 Myr, but could eventually have become lower, if we had continued the simulation for a longer time. Despite of natural fluctuations of the $^{60}$Fe/$^{26}$Al value, the overall trend of a decrease is expected, since supernovae generally admix less Fe relative to Al for decreasing masses of the progenitor, and therefore the $^{60}$Fe/$^{26}$Al in GMCs will tend to decrease over time.
This argument is also supported by the changing distribution of $^{60}$Fe/$^{26}$Al ratios inside our cloud \Fig{FeAl_sinks} (right panel). Since our GMC has already evolved long enough before the start of our simulation, the less massive supernova occurs already at the beginning of our simulation. Although somewhat counter-intuitive, the associated distribution (black solid line) reveals that the lowest $^{60}$Fe/$^{26}$Al ratios occur after a low-mass supernova. In general, however, the early SLR abundances are predominantly dominated by enrichments of short-lived heavy supernovae, while longer-lived low mass supernovae mostly occur at later times and cause new enrichment.
At this point, we emphasize the difficulties of measuring and estimating one single $^{60}$Fe/$^{26}$Al value for a GMC considering the large range of $^{60}$Fe/$^{26}$Al ratios reflected in \Fig{SLR_histo} and \Fig{FeAl_sinks} (right panel) present throughout the entire GMC. Although mixing occurs inside the GMC, fluctuations are still significant indicating that the process of mixing occurs on longer timescales relative to that depicted in our simulations.

To better understand the heterogeneities inside the GMC, we compare the time scales that are relevant for mixing of the SLRs.
The dynamical crossing time is
\begin{equation}
t_{\rm cross}= \frac{l}{\mathcal{M} c_s},
\end{equation}
where $l = L_{\rm box} / 2$ is a characteristic length and $\mathcal{M} c_s = 6-7$ $\unit{km s}^{-1}$ is the typical turbulent RMS speed.
This gives a mixing time of $t_{\rm cross}\sim 3 \unit{Myr}$. Another relevant time-scale is the cooling time of
the hot medium
\begin{equation}
t_{\rm cool} = \frac{3k_bT}{2n\Lambda}
\end{equation}
where $\Lambda$ is the cooling rate. If we assume approximate pressure equilibrium between different phases, then hot
gas has a vastly different density compared to cold gas, and it has to be cooled down and compact to efficiently mix with
the cold gas. In our case the cooling time can be calculated to be $\sim1 \unit{Myr}$ for $10^6$ K gas.
The average time between different supernovae $t_{\rm \Delta SN}$ provides the time between supernova enrichments.
Given that nine supernova explosions occurred in roughly 4 Myr during our simulation,
we set $t_{\rm \Delta SN}$ to 450 kyr. This is significantly lower than either the crossing time or the cooling time, which
explains the heterogeneous SLR abundance in the gas of the GMC.

\subsection{Abundance of SLRs in stars}
After having shown that SLR abundances of the gas are heterogeneous, we investigate to what extent the heterogeneity is present in the stars (represented by the asterisks in \Fig{SLR_sinks}.
Altogether 252 stars of masses higher than $0.2$ M$_{\odot}$ form in between $t=0$ and the end of the simulation of which
46 evolve to masses higher than 8 M$_{\odot}$ and will end their lives in supernova explosions. As mentioned earlier, none of these stars has exploded in a supernova event by the end of the simulations. The GMC also contains lower mass stars, but we exclude stars of masses lower than 0.2 M$_{\odot}$ because the minimum cell size of 126 AU is not sufficient to properly sample the tail of the turbulence and resolve the cores lower mass stars. As we are mostly interested in the evolution of solar mass stars and due to the lack of radiative transfer, we also exclude the high mass stars for our analysis and only focus on the 206 stars in the range of 0.2 M$_{\odot}$ to 8 M$_{\odot}$.
In agreement with \cite{2013ApJ...769L...8V} and \cite{2009ApJ...694L...1G}, the stars show different relative abundances
in $^{26}$Al (varying from about $1\times10^{-7}$ up to about $1\times10^{-5}$),
as well as in $^{60}$Fe (varying from about $1\times10^{-7}$ up to $3\times10^{-6}$) at their time of formation. Moreover, similar to the distribution of all the gas in the cloud, the initial abundances are on average lower and show a narrower spread in abundance than seen at later times.

However, there are significant differences in the overall distribution of SLR abundances between the gas and the stars. The stars show a smaller range of abundances than the gas and the stars have abundances that always lie below the average abundance in the gas at that time. Moreover, a significant amount of stars show abundances that follow the decay curve of the average value of the gas at the very beginning of the simulation (the barely visible small green lines). We interpret this result such that these stars formed from a first rather old gas reservoir. The parental run does not include tracer particles, which would have allowed us to track the history of the gas from at least one supernova enrichment during the evolution of the GMC. Nevertheless, we can use the decay time of the SLRs as a clock to draw some qualitative conclusions about the origin of the gas in stars. In agreement with the delay of enriched SLR abundances for the stars in our box, we suggest that although highly SLR enriched gas from supernovae is present in the GMC, it does not contribute to the formation of stars until at least several 100 kyr later.
This is in agreement with results from \cite{2013ApJ...769L...8V}, who followed the motion of gas injected by supernovae by using tracer particles and estimated that it takes of the order of $1 \unit{Myr}$ until such gas is incorporated in star forming cores. Observations show that stars form in regions of cold, dense gas. Hence it is obvious that the gas in SN ejecta needs time to cool before it can take part in star formation. This is in agreement with the results seen in \Fig{phase_space_box} (left and middle panel), as well as in \Fig{SLR_histo}. In general, the gas covers a large range of ratios and densities. However, $^{26}$Al/$^{27}$Al abundances higher than $10^{-3}$ only occur for densities that are lower than $10^{-13} \unit{g}\unit{cm}^{-3}$ and, thus, can not contribute to star formation yet. The plot shows that only a small faction of the cells have densities higher than $10^{-16} \unit{g}\unit{cm}^{-3}$. These cells correspond to potential star forming cores, and that gas does not show such large spreads in $^{26}$Al/$^{27}$Al ratio as for lower densities. Considering that the densities of the star forming cores are several orders of magnitude higher than the SLR-enriched gas in the vicinity of recent supernovae indicates the difficulty to contaminate the star forming cores with new gas of different abundance. Given that high abundances must be associated with recent supernova activities, we conclude that the gas needs time to cool sufficiently before it is able to clump and subsequently to form stars only from the gas with lower SLR ratio.

\subsection{SLR distribution in vicinity of stars at early times}
\begin{figure}
\subfigure{\includegraphics[width=\linewidth,bb=20bp 260bp 602bp 530bp,clip]{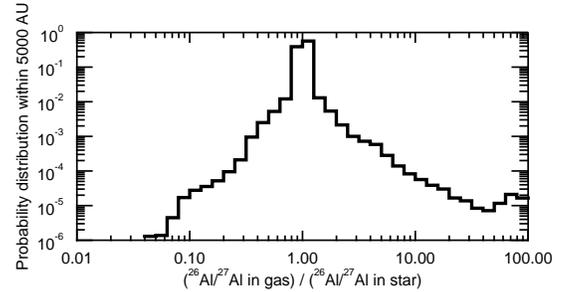} } \quad
\protect\caption{\label{fig:SLR_coarse_diff} Histogram showing the probability distribution for relative differences of $^{26}$Al/$^{27}$Al in the gas within a radius of 5000 AU compared to their host star.}
\end{figure}

With respect to the measured differences in $^{26}$Al/$^{27}$Al between canonical CAIs and FUN CAIs of more than one order of magnitude, it is of particular interest to investigate whether such differences occur during the accretion process.
It is generally accepted that the formation of CAIs (both CV and FUN types) is restricted to the very early phase of star formation and very close to the star, probably only to the first few ten thousand years and the inner AUs \citep{2009GeCoA..73.4963K,Holst28052013}.
Since the resolution around the stars is 126 AU and the time between snapshots in the parental run already is 50 kyr, we cannot resolve the surrounding to this level for all the stars in our simulation. However, late-stage contamination of an accreting star by freshly synthesized supernova material requires that the differences in abundance originate at distances far beyond the sizes of star forming cores.
Therefore, we test, whether significant differences in abundance occur within 5000 AU of the star 50 kyr after its formation. In \Fig{SLR_coarse_diff}, we show the mass-weighted distribution around the stars in the mass range of 0.2 M$_{\odot}$ to 8 M$_{\odot}$ that show ten times higher $^{26}$Al/$^{27}$Al ratios within a distance of 5000 AU and less than 50 kyr after the birth of the star with respect to the abundance in the corresponding star.
Altogether only nine of the 206 stars show contaminations of more than a factor of ten, among them only two stars with masses higher than 1 M$_{\odot}$. We emphasize that all of the cells with very different abundance are at least 1500 AU away from the star and belong to times already up to 50 kyr after the star has formed. Hence, we do not see any contaminations at early times relevant for CAI formation ($t<10$ kyr) that can account for the measured differences of $^{26}$Al/$^{27}$Al in FUN and canonical CAIs.

\begin{figure*}
\subfigure{\includegraphics[width=0.45\textwidth,bb=20bp 270bp 602bp 530bp,clip]{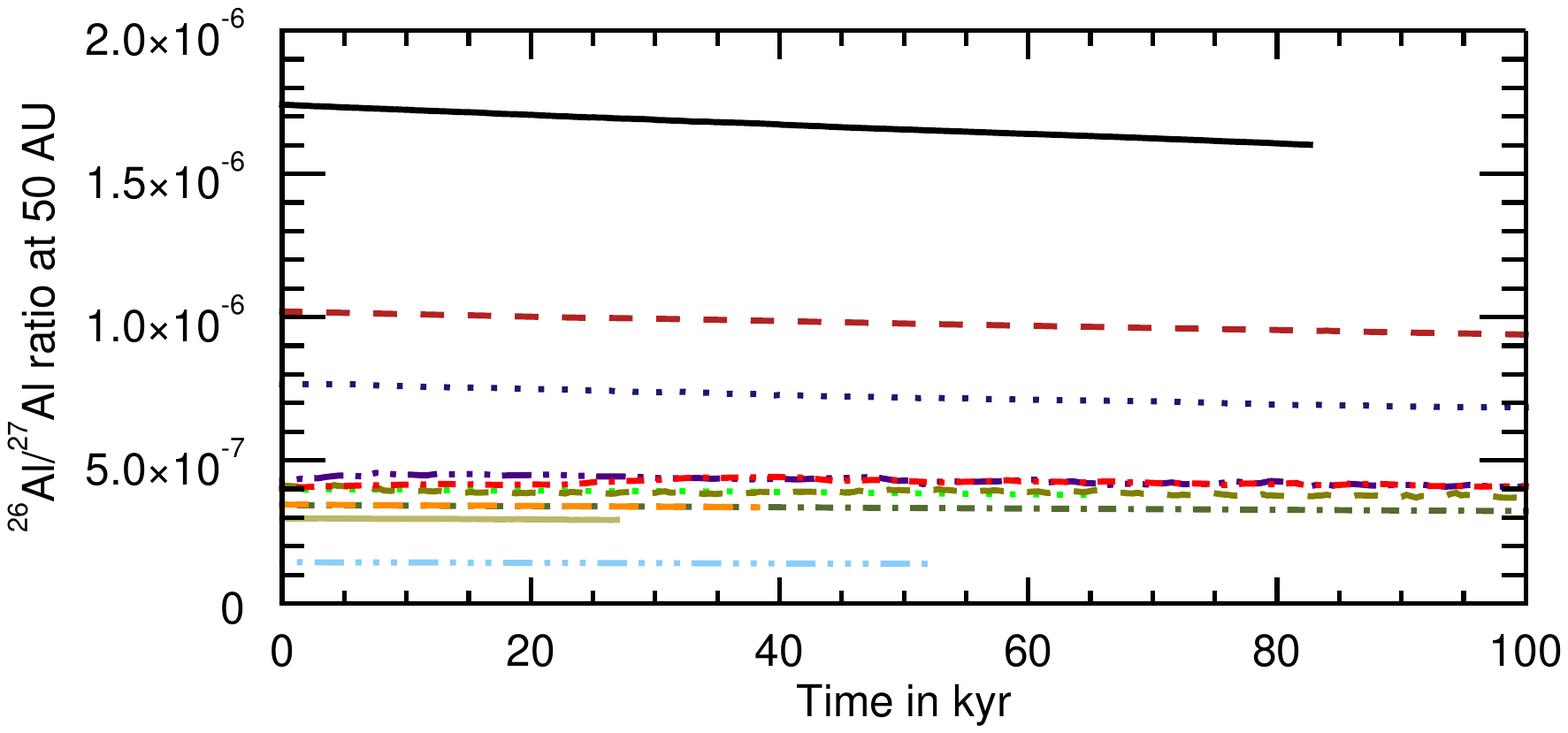} }
\subfigure{\includegraphics[width=0.45\textwidth,bb=20bp 270bp 602bp 530bp,clip]{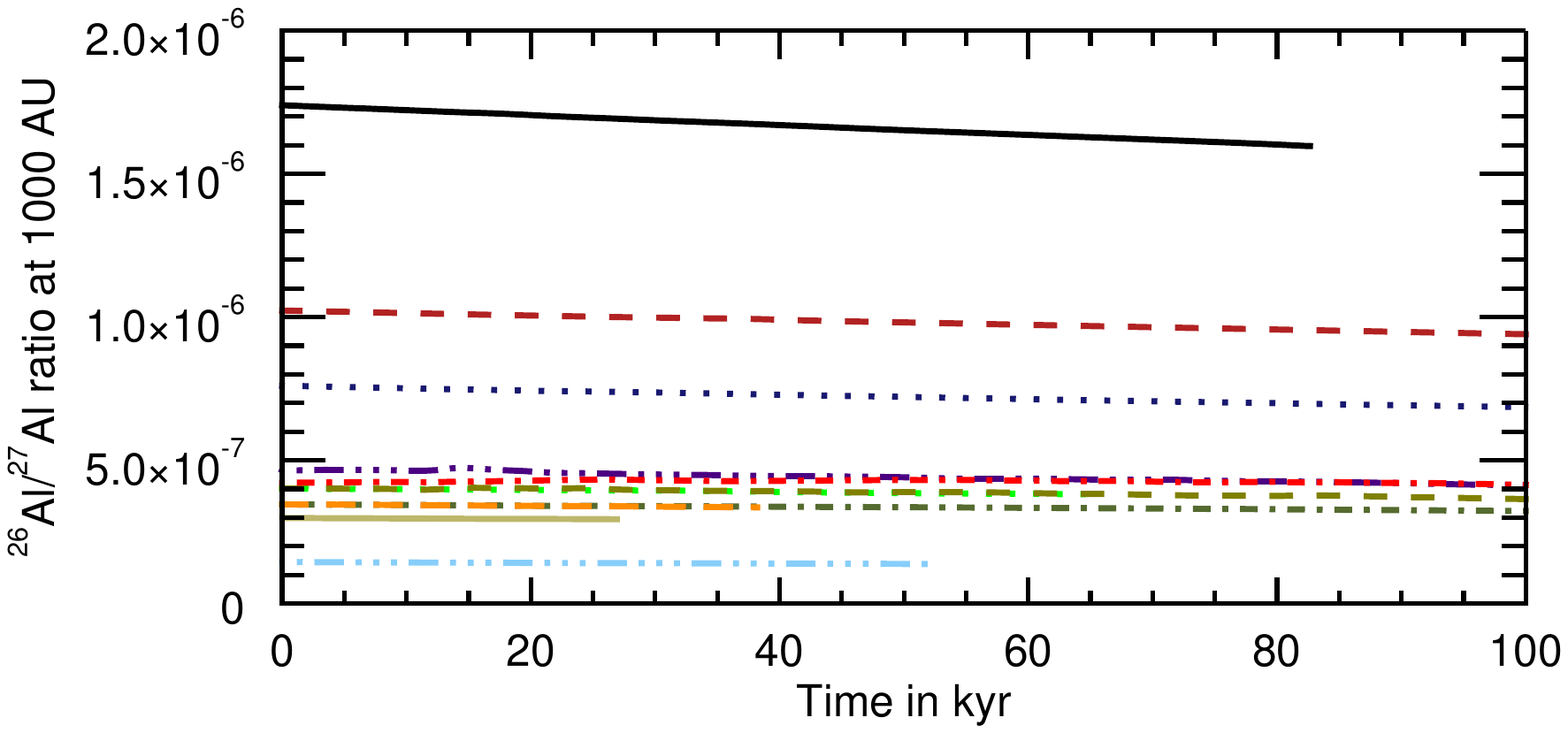} } \\
\subfigure{\includegraphics[width=0.45\textwidth,bb=20bp 270bp 602bp 530bp,clip]{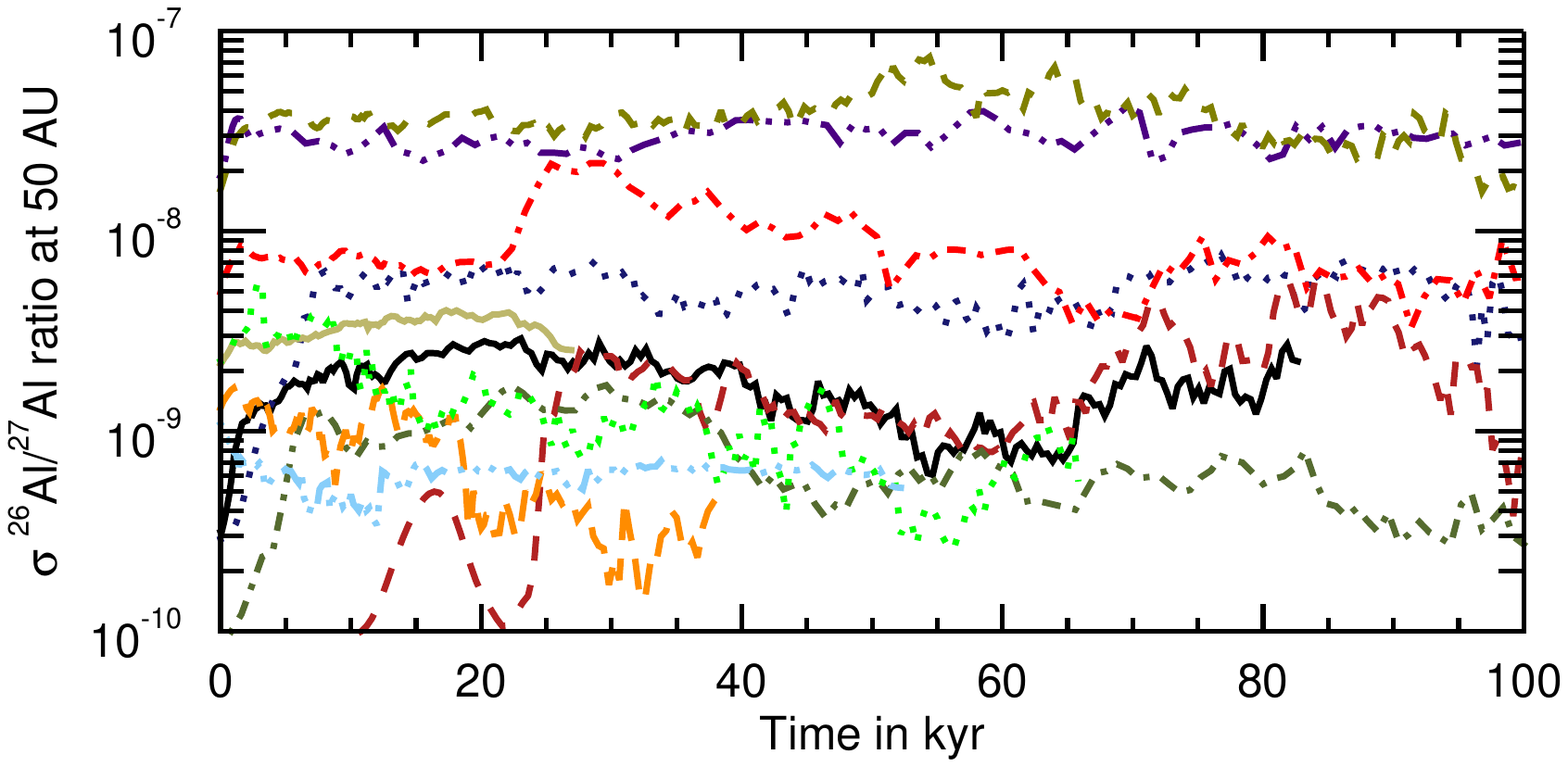} }
\subfigure{\includegraphics[width=0.45\textwidth,bb=20bp 270bp 602bp 530bp,clip]{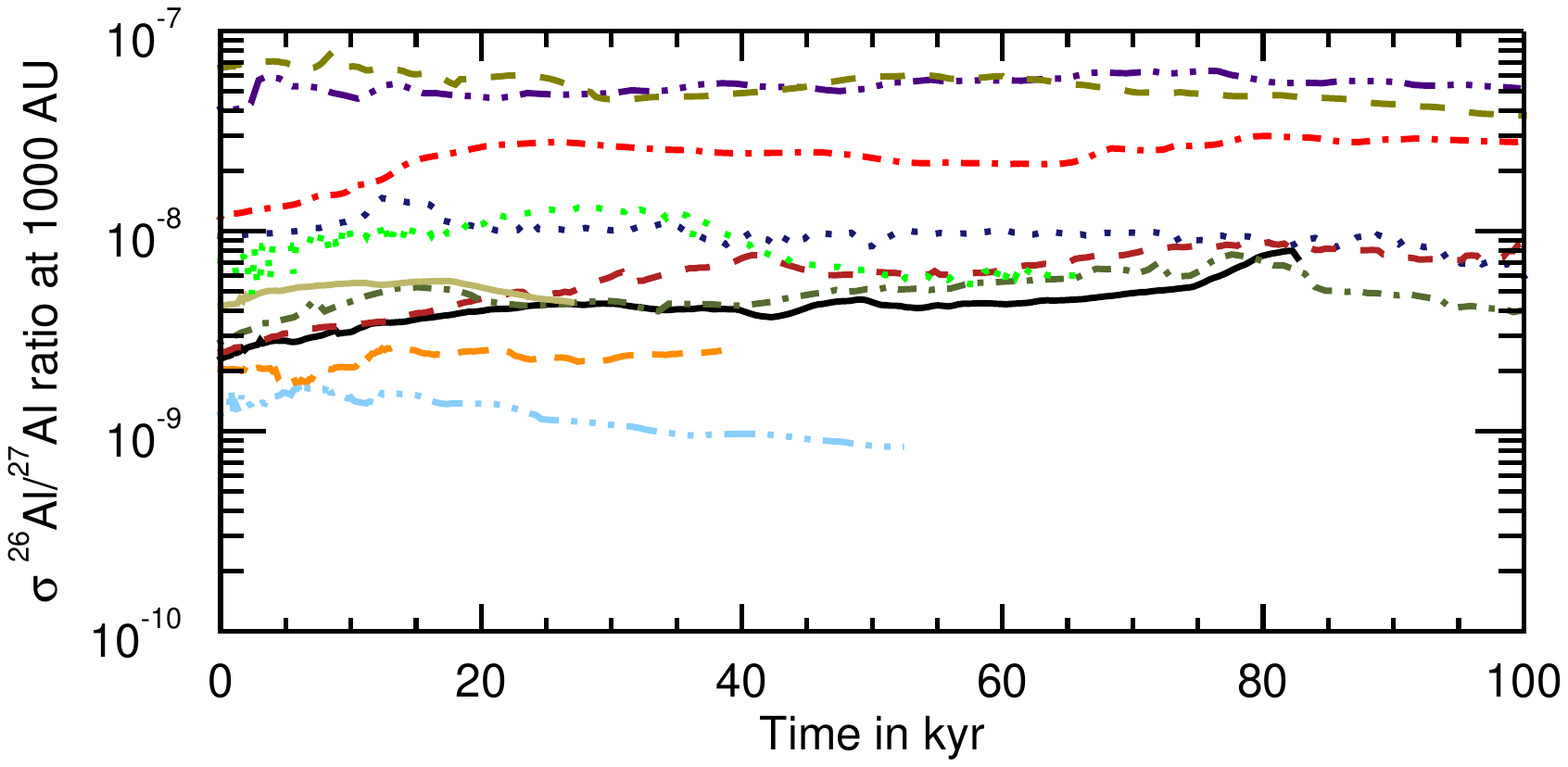} }
\protect\caption{\label{fig:SLR_t_indi}Temporal evolution of the $^{26}$Al/$^{27}$Al ratio (upper panels)
and of their standard deviation (lower panels)
around the different selected stars in spherical shells at distances of 50 AU (left)
and 1000 AU (right) from the corresponding star. Black solid corresponds to
star 1, blue dot to star 2, red dash to star 3, dark green dash-dot
to star 4, purple dot-dot-dash to star 5, orange dash to star
6, kaki solid to star 7, lime-green dot to star 8, olive green
dash to star 9, bright red dash-dot to star 10 and light blue dot-dot-dash to star 11.}
\end{figure*}

\begin{figure*}
\subfigure{\includegraphics[width=0.45\textwidth,bb=20bp 270bp 602bp 530bp,clip]{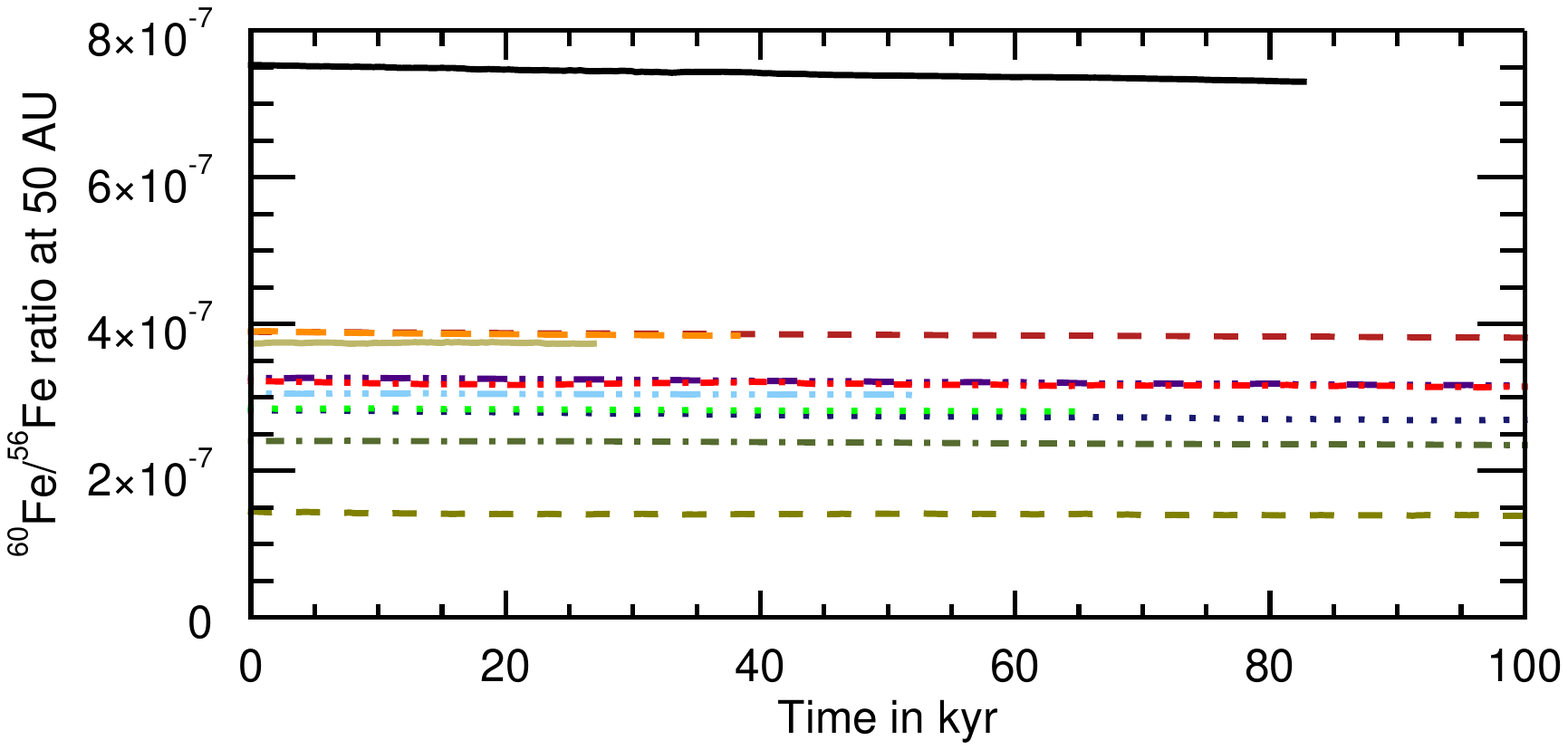} } \quad
\subfigure{\includegraphics[width=0.45\textwidth,bb=20bp 270bp 602bp 530bp,clip]{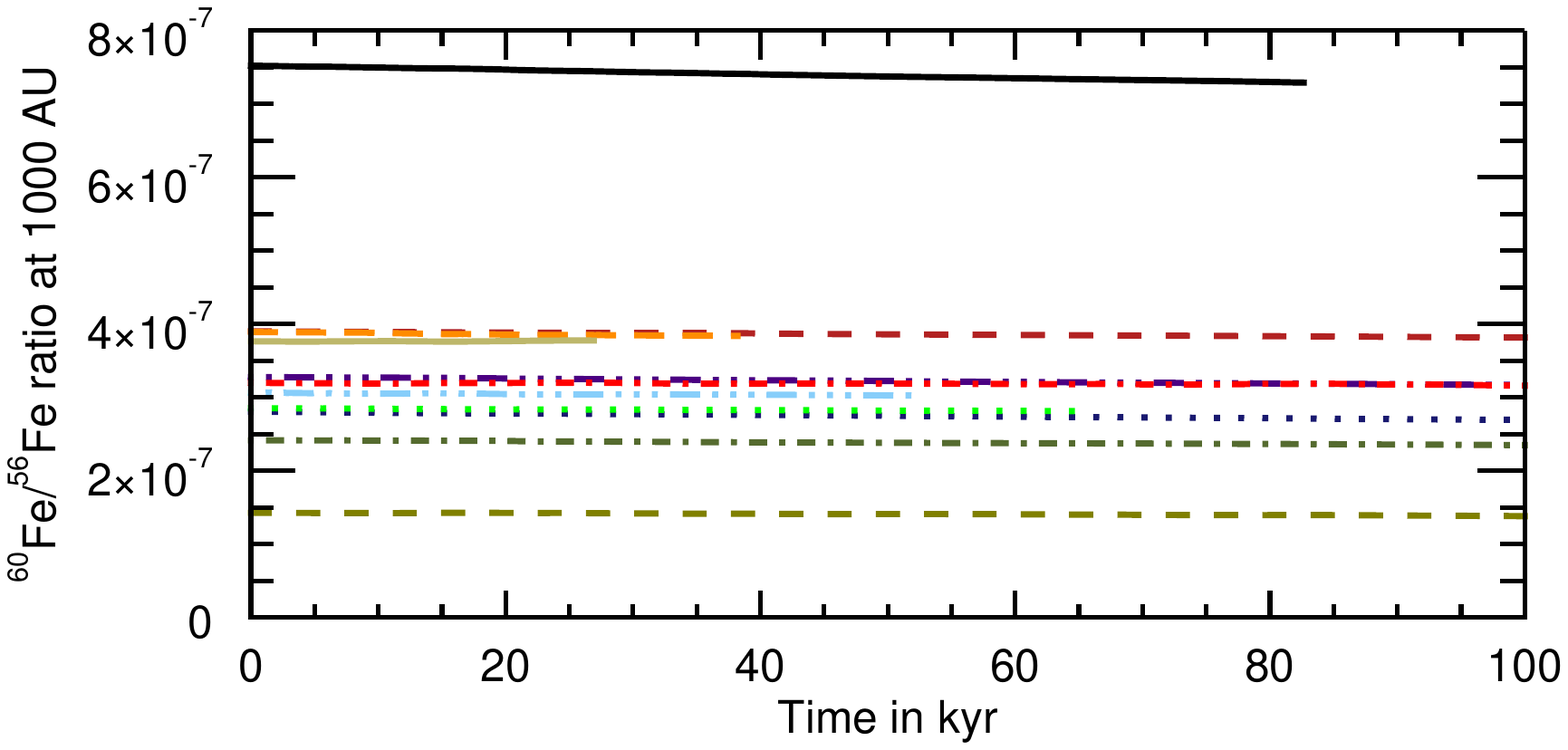} } \\
\subfigure{\includegraphics[width=0.45\textwidth,bb=20bp 270bp 602bp 530bp,clip]{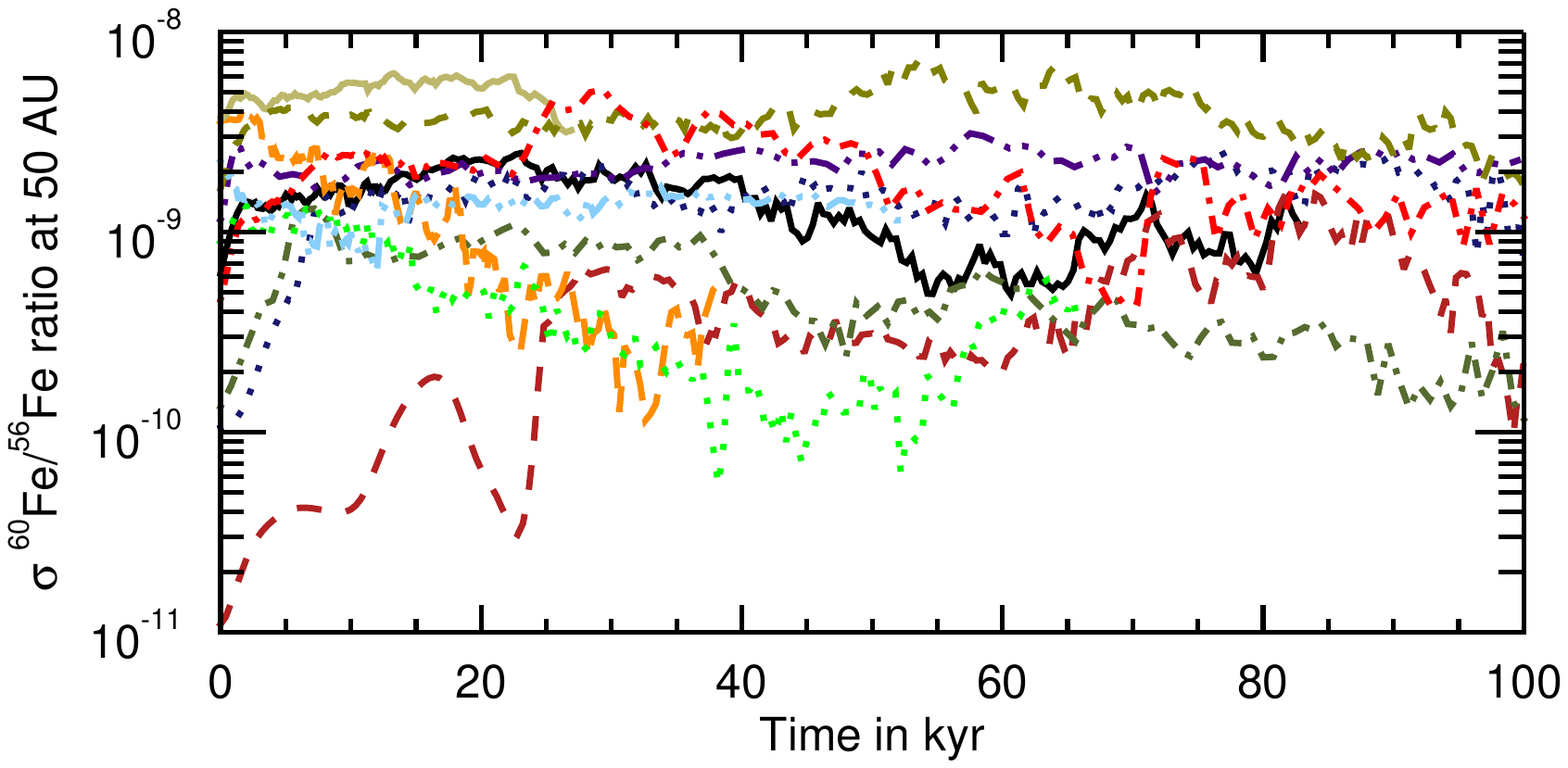} } \quad
\subfigure{\includegraphics[width=0.45\textwidth,bb=20bp 270bp 602bp 530bp,clip]{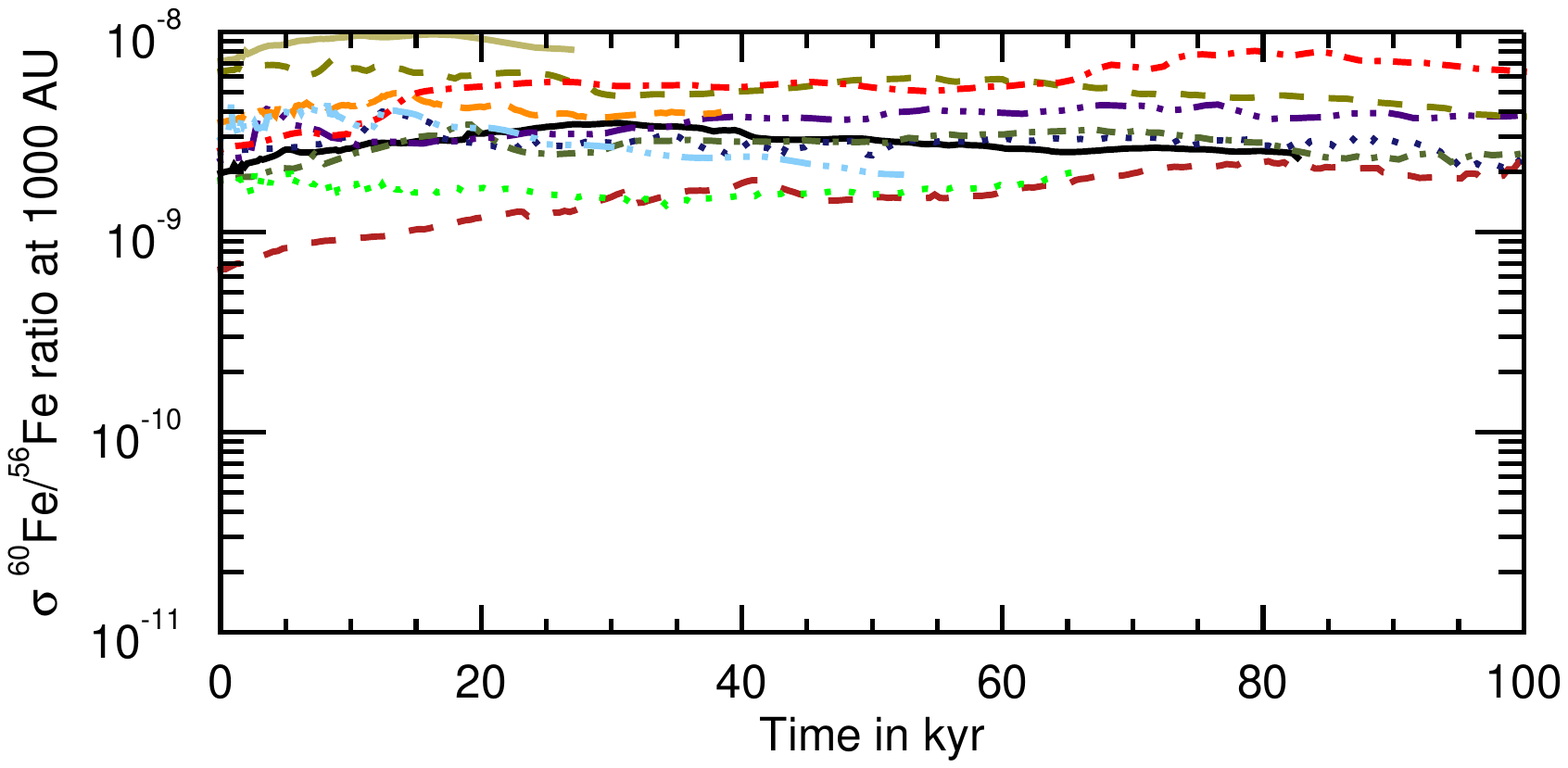} }
\protect\caption{\label{fig:Fe_t_indi}Temporal evolution of the $^{60}$Fe/$^{56}$Fe ratio (upper panels)
and for their standard deviation (lower panels) around the different selected stars in spherical shells at distances of 50 AU (left)
and 1000 AU (right) from the corresponding star. Black solid corresponds to
star 1, blue dot to star 2, red dash to star 3, dark green dash-dot
to star 4, purple dot-dot-dash to star 5, orange dash to star
6, kaki solid to star 7, lime-green dot to star 8, olive green
dash to star 9, bright red dash-dot to star 10 and light blue dot-dot-dash to star 11.}
\end{figure*}

To ensure that this result is robust, we selected a few stars to follow their formation phase with higher resolution.
The eleven stars selected for zoom-in are marked with red circles in \Fig{SLR_sinks}. Ten of the eleven stars accreted between 1 and 2 M$_{\odot}$. Additionally, we also modelled one star that accreted to more than 2.8 M$_{\odot}$.
\Figure{SLR_t_indi} and \Fig{Fe_t_indi} show the
temporal evolution of the average $^{26}$Al/$^{27}$Al ratio and $^{60}$Fe/$^{56}$Fe
in spherical shells at distances of 10 AU and 1000 AU as well as their standard deviations for the selected
stars. These have different relative abundances of $^{26}$Al/$^{27}$Al
between $10^{-7}$ and $2\times10^{-6}$, while $^{60}$Fe/$^{56}$Fe ratios
are in between $10^{-7}$ and $8\times10^{-7}$.
Although the stars selected for zoom-ins show abundances below the canonical value of $5 \times 10^{-5}$,
we emphasize that our selection is valid to test the hypothesis of $^{26}$Al enrichment
in the solar system through supernovae. Considering that bulk CV CAIs are supposed to reflect
$^{26}$Al enrichment after solar birth, the initial abundances in our selected stars
are consistent with values less than $3 \times 10^{-6}$ as measured in FUN CAIs.
Since these values are considered to reflect the original abundance in the collapsing pre-solar core, and the fact that higher abundances are available in the GMC ,
our selection provides an adequate sample to test the enrichment hypothesis.

As expected from the results obtained using the parental run, the average $^{26}$Al/$^{27}$Al and $^{60}$Fe/$^{56}$Fe ratios are almost indistinguishable at different distances from the parent stars (left and right upper panel of Fig. 6 and 7). This suggest a spatially homogeneous distribution of $^{26}$Al and $^{60}$Fe within 1000 AUs during the accretion process for the first 100 kyr of evolution. We note that the apparent time integrated variability is consistent with the typical decay curve for $^{26}$Al and $^{60}$Fe.
In principle, inflow of gas with different SLR abundances at two different locations and identical in-fall speed could nevertheless cause spatial heterogeneities without affecting the average SLR value, which would be reflected in large deviations from the mean value. However, the ratios deviate only marginally from the average values in the shells as illustrated by plotting the standard deviation from the mean value at distances of 50 AU and 1000 AU in the lower panels of \Fig{SLR_t_indi} and \Fig{Fe_t_indi}.
Generally, the different stars all have deviations that range from less than 1 \textperthousand \ to at most 20 $\%$ of the mean value. Importantly, the fluctuation observed in the star with the largest fluctuation in $^{26}$Al/$^{27}$Al ratios (star 9) is still lower by more than one order of magnitude relative to the difference between canonical and FUN CAIs.

\begin{figure*}
\subfigure{\includegraphics[bb=0bp 250bp 612bp 550bp,clip,scale=0.4]{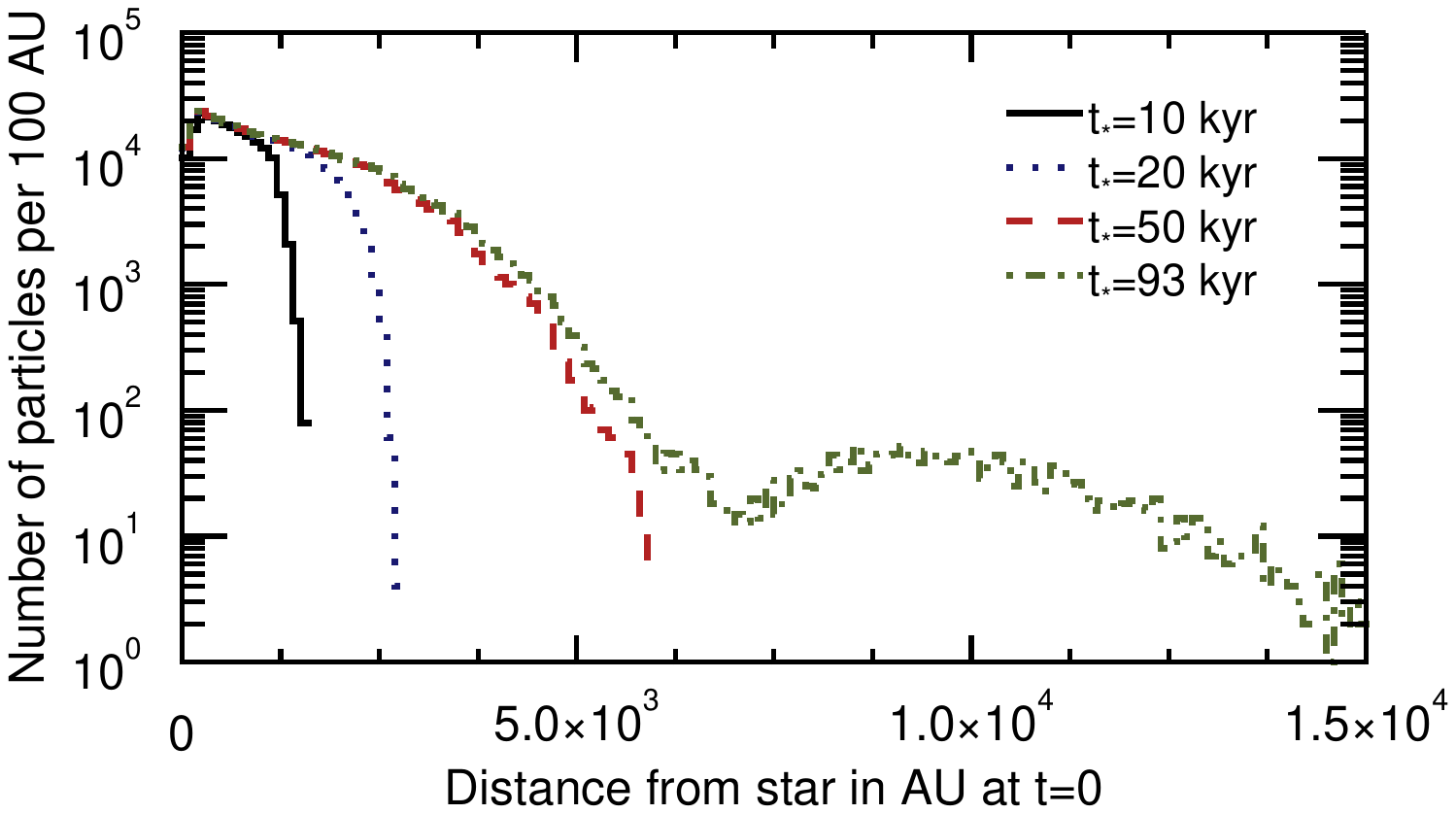} \includegraphics[scale=0.2]{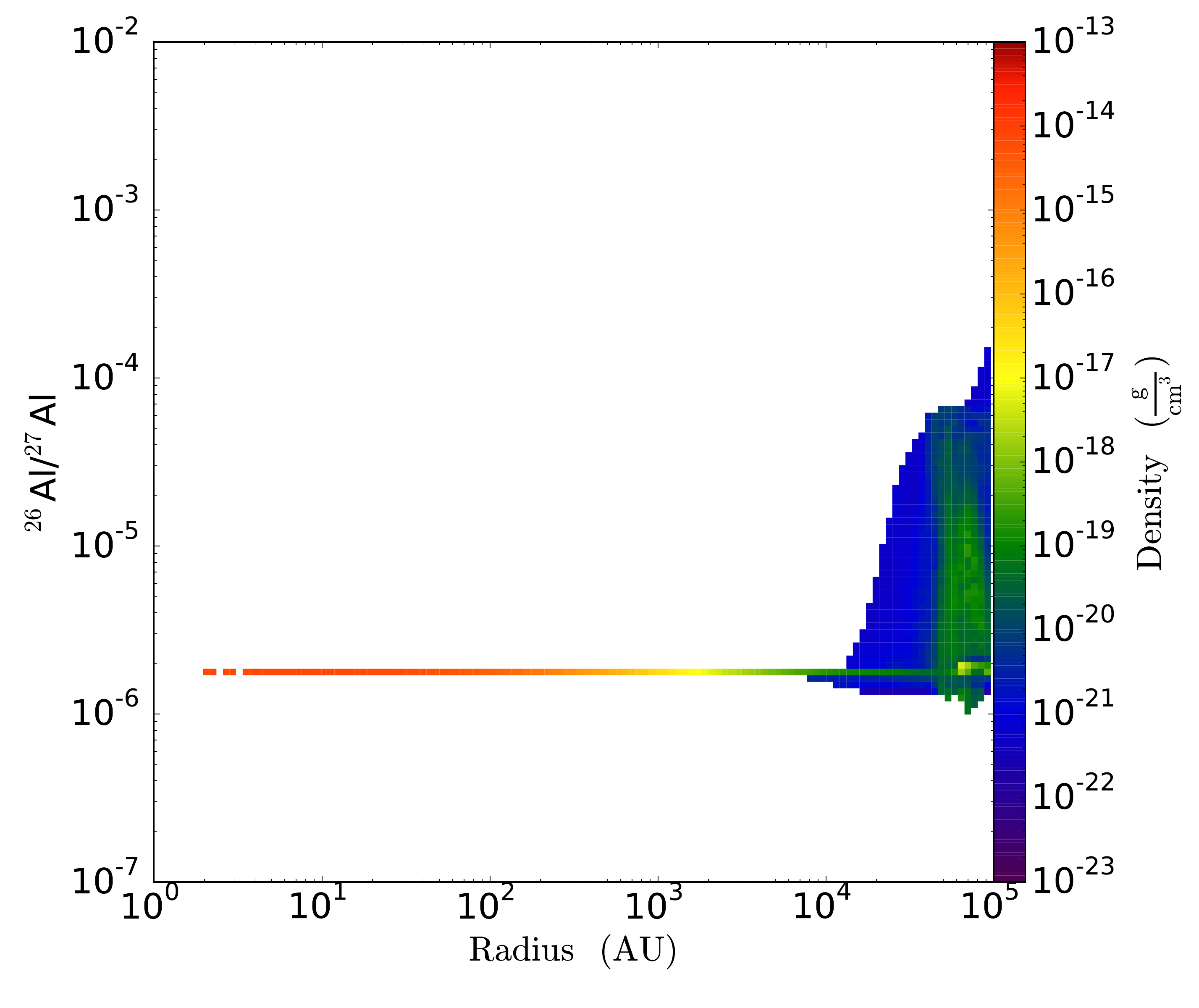} } \quad
\subfigure{\includegraphics[bb=0bp 250bp 612bp 550bp,clip,scale=0.4]{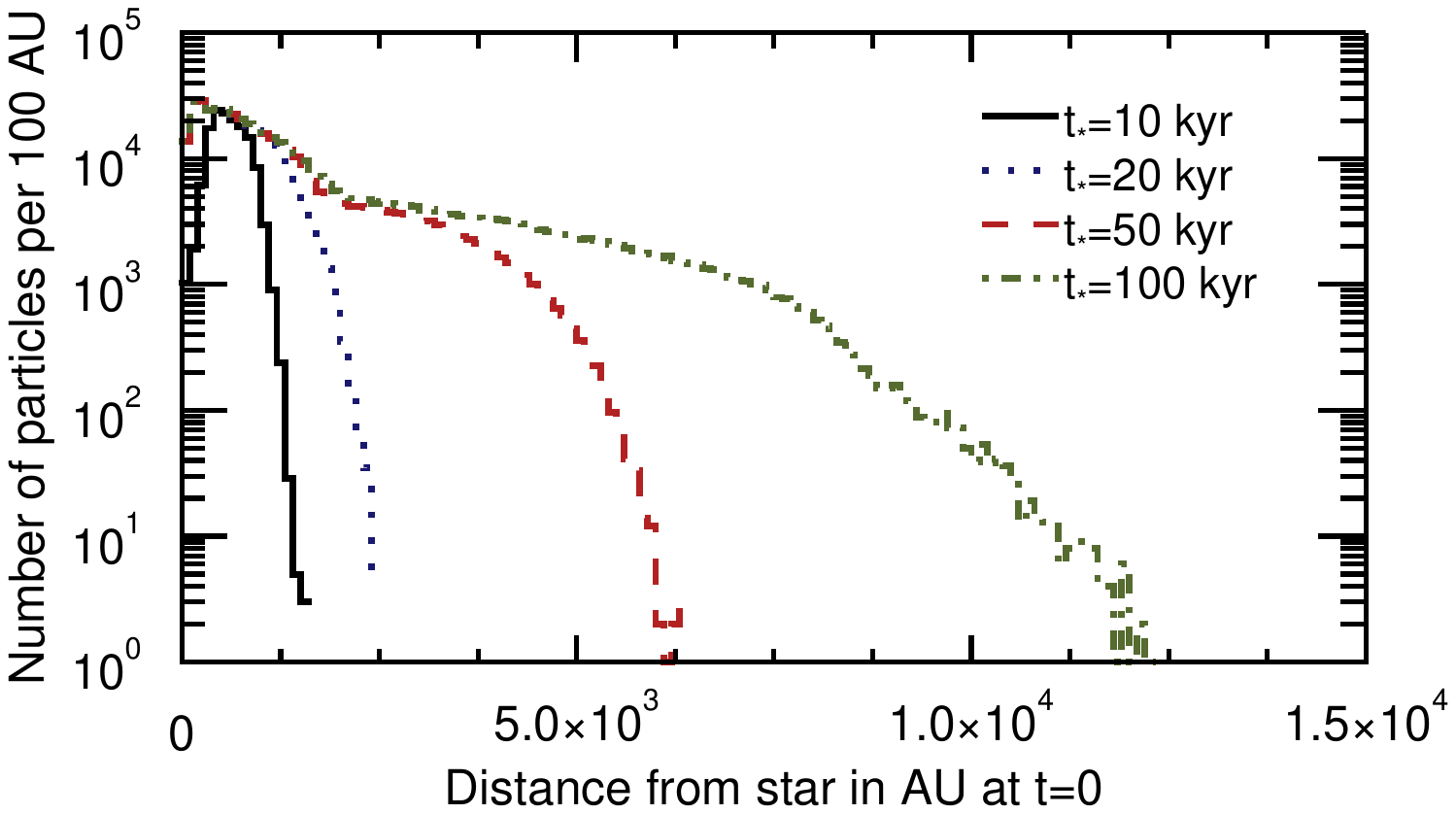} \includegraphics[scale=0.2]{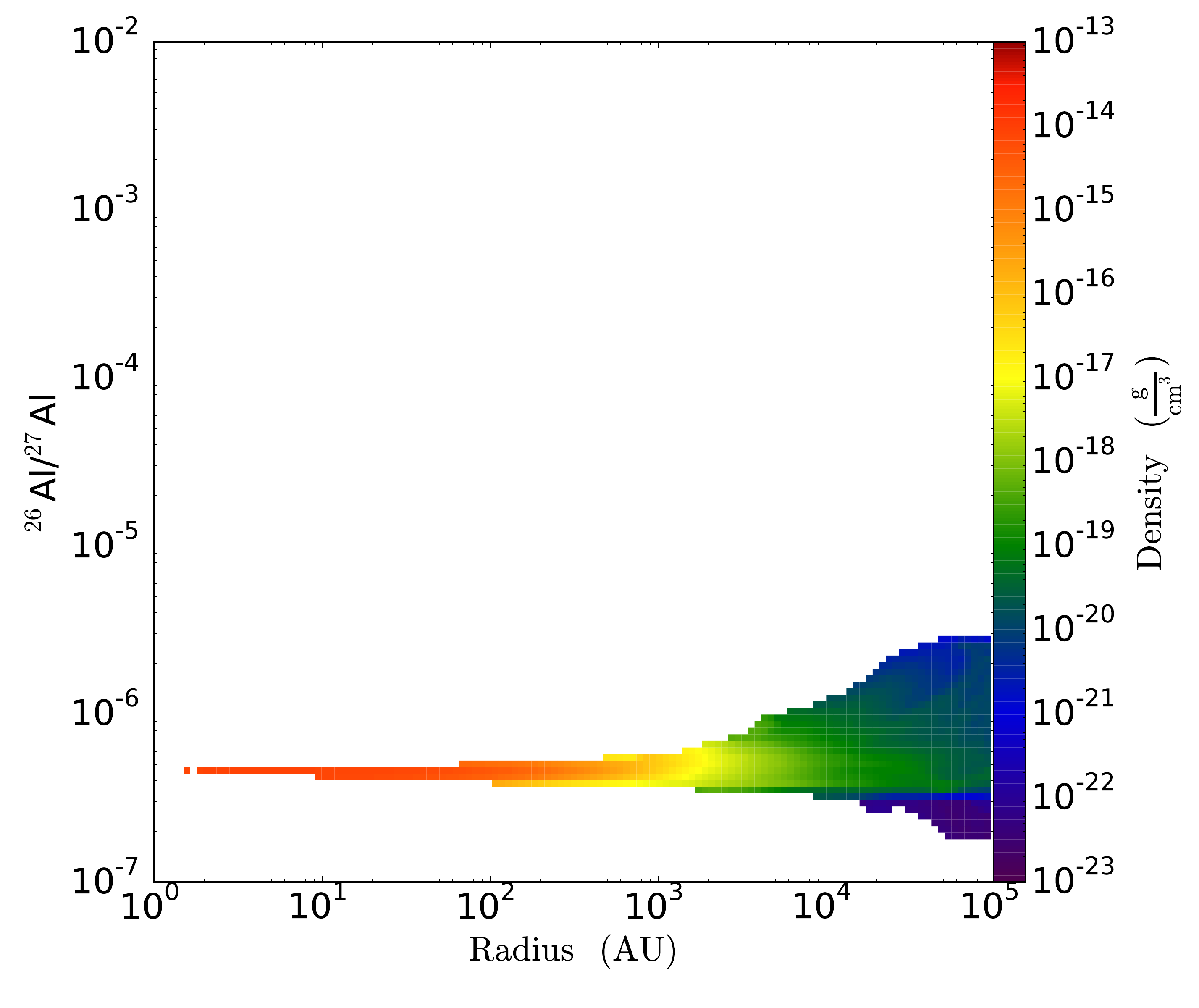} } \quad
\subfigure{\includegraphics[bb=0bp 250bp 612bp 550bp,clip,scale=0.4]{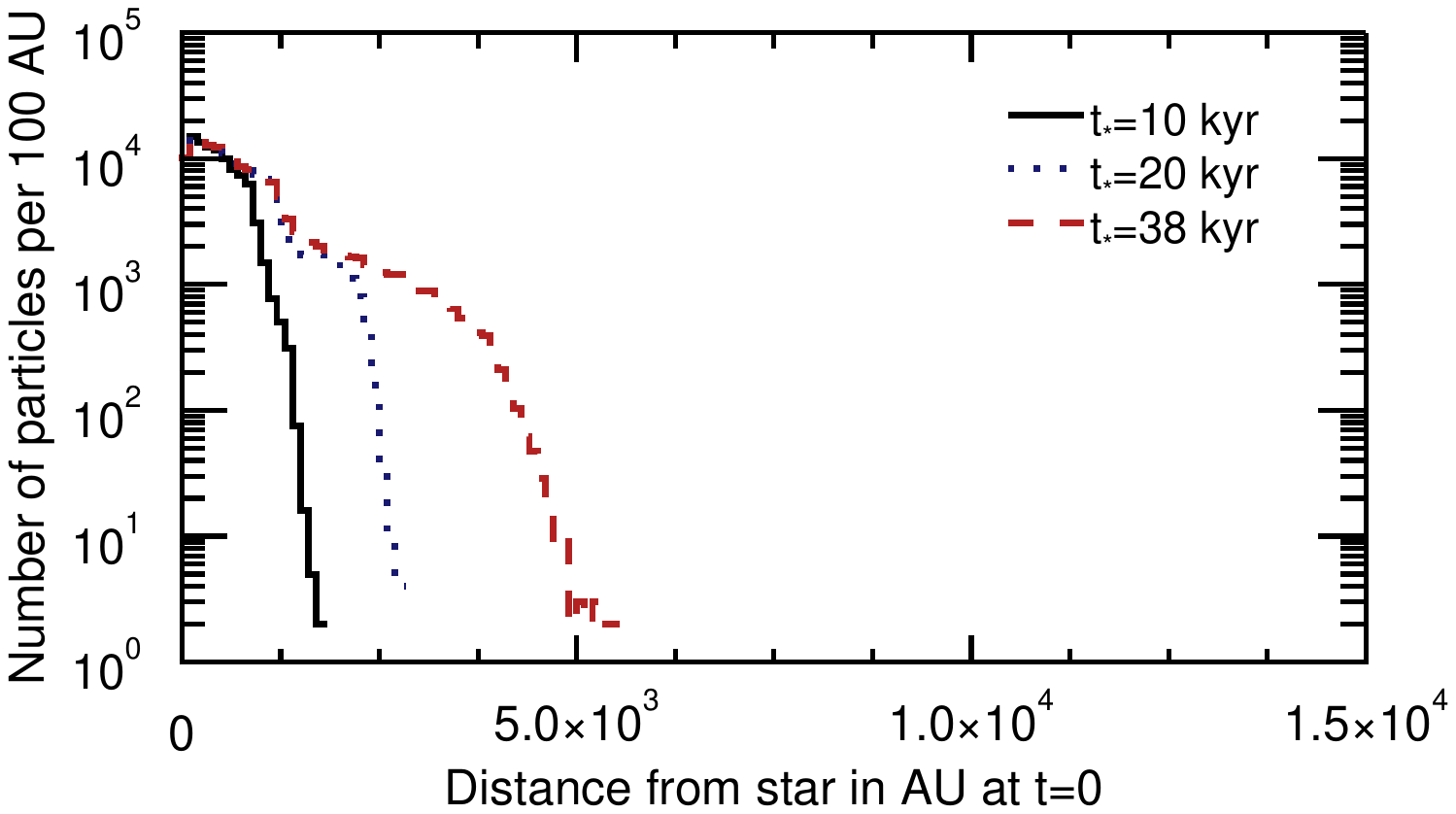} \includegraphics[scale=0.2]{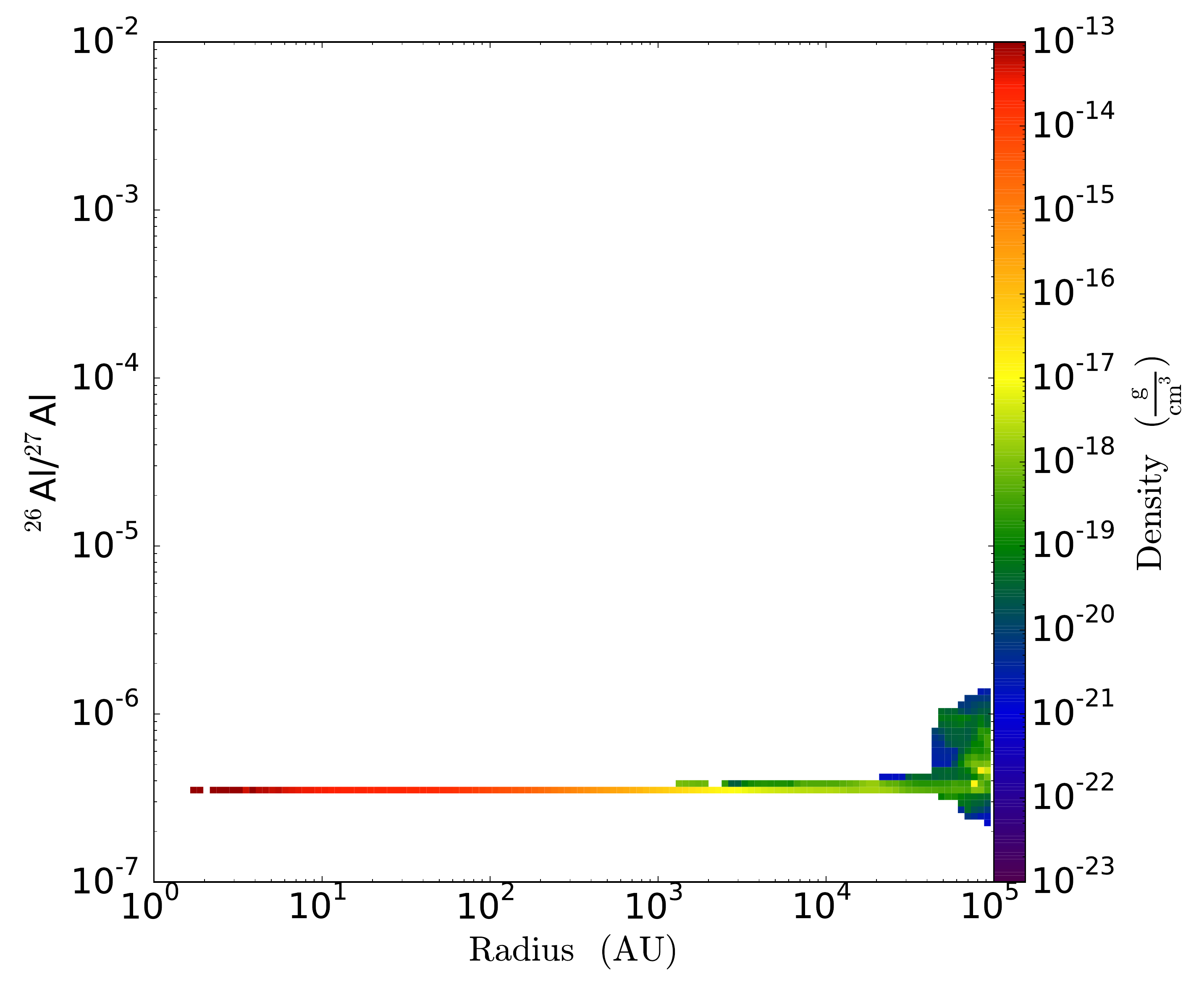} } \quad
\subfigure{\includegraphics[bb=0bp 250bp 612bp 550bp,clip,scale=0.4]{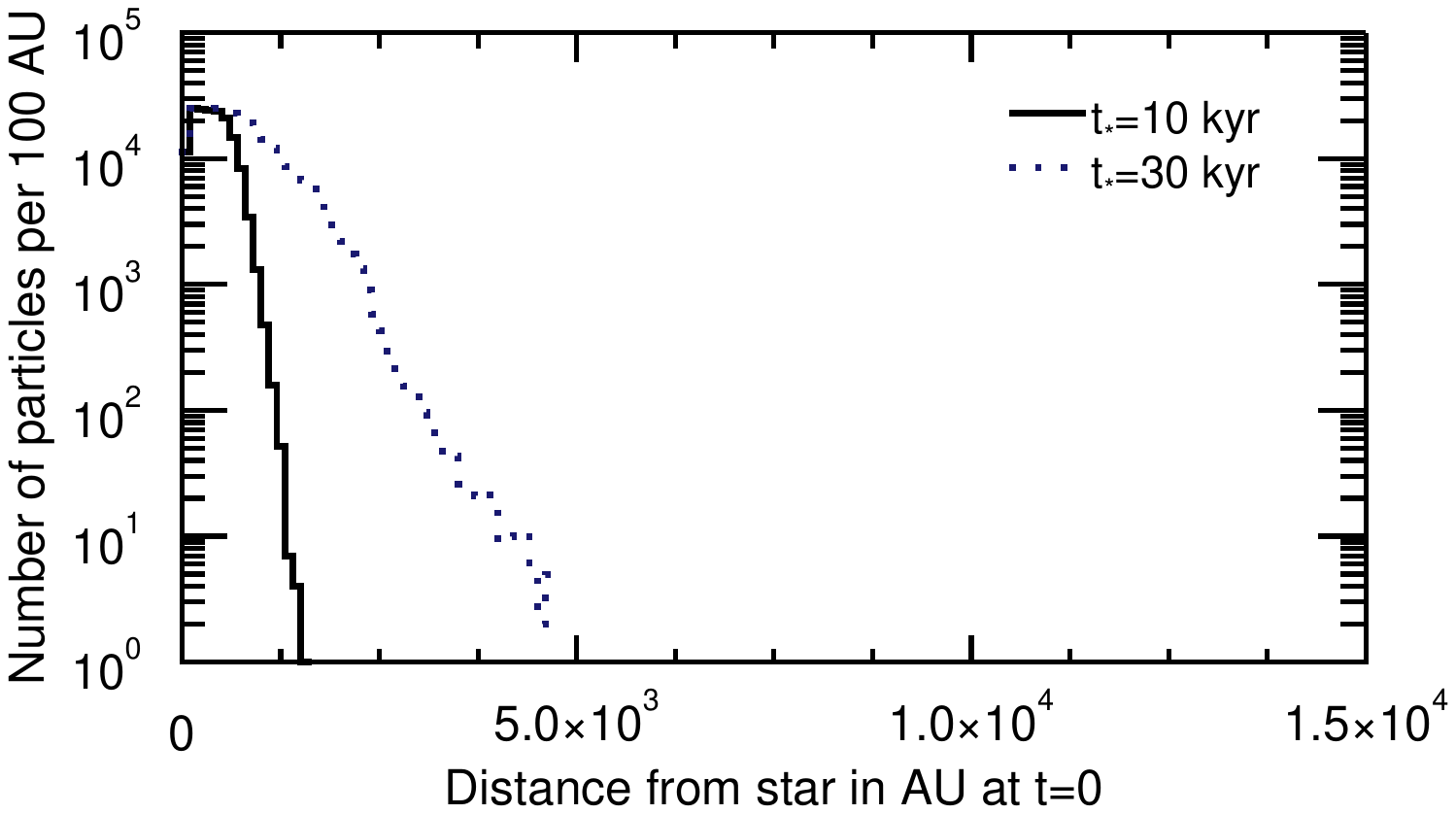} \includegraphics[scale=0.2]{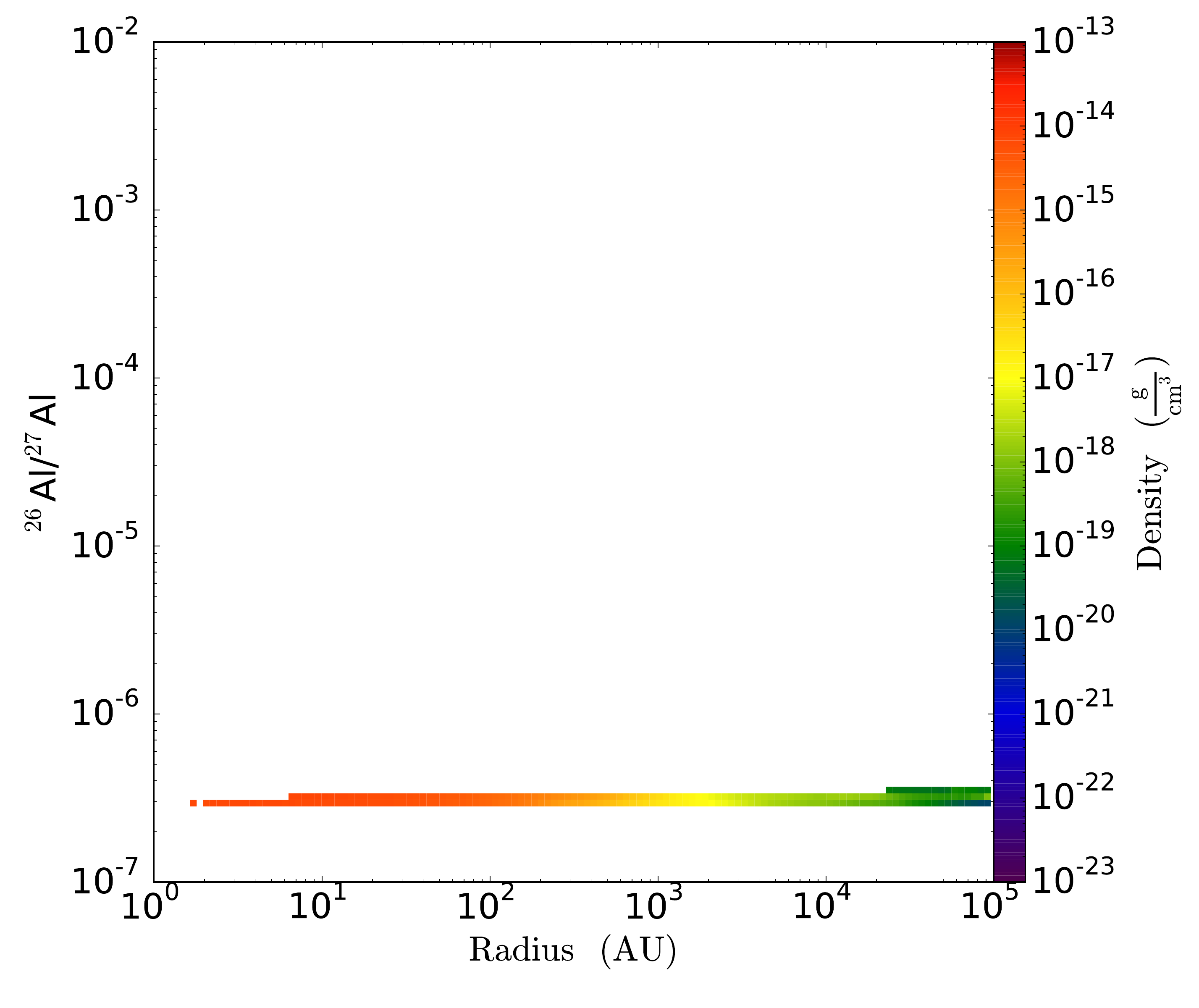} } \quad
\protect\caption{\label{fig:Origin-of-gas1}Original location of gas at time of stellar formation that is located within a distance
of 100 AU from star 1 (top left), star 5 (second row left), star 6 (third row left), star 7 (fourth row left) at times indicated in the plots. The distances
on the x-axis refer to the positions of the gas at the time of star formation. The right panels illustrate the $^{26}$Al/$^{27}$Al distribution in the gas around the star at the time of its birth.
}
\end{figure*}

\begin{figure*}
\subfigure{\includegraphics[bb=0bp 250bp 612bp 550bp,clip,scale=0.4]{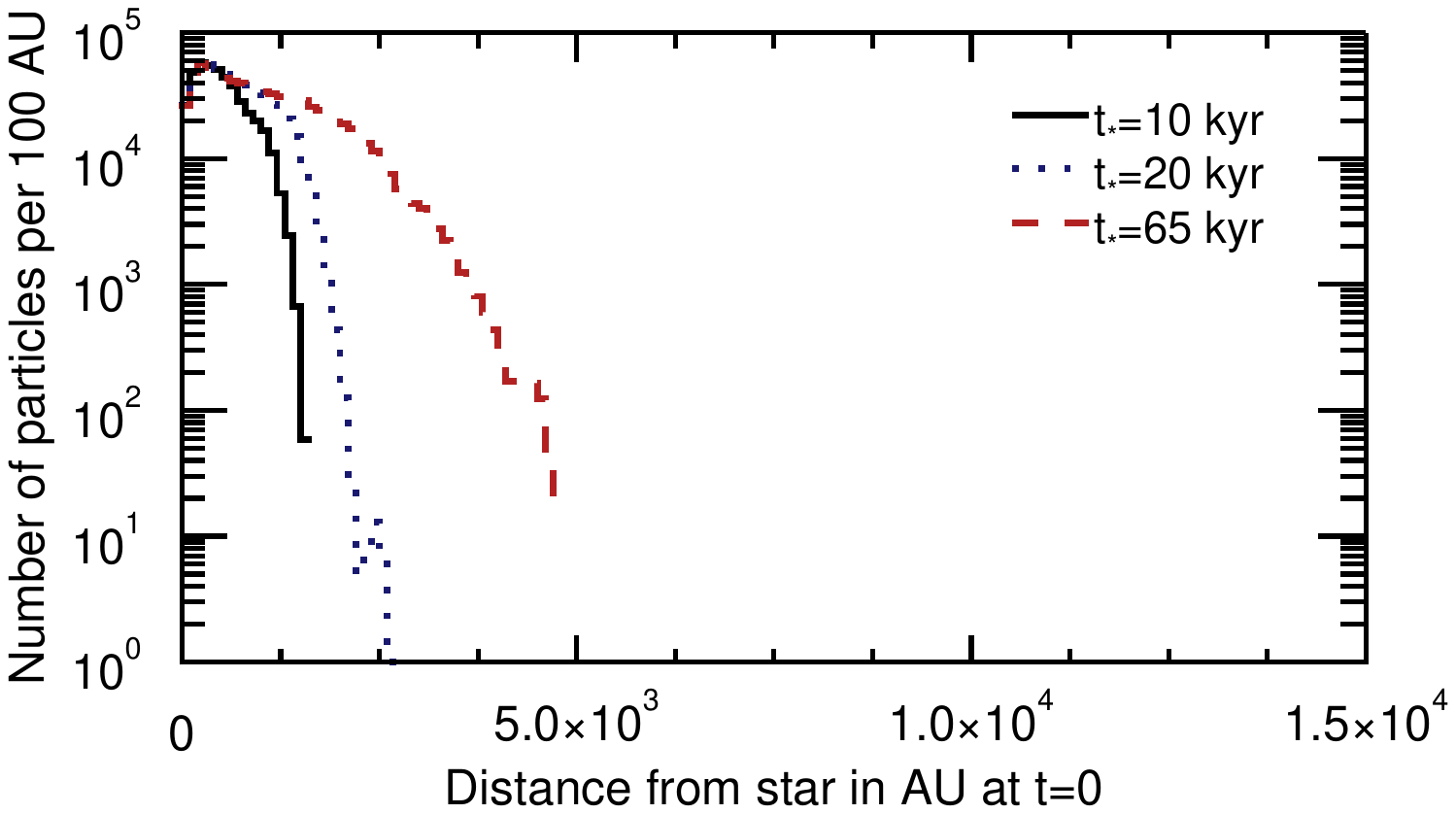} \includegraphics[scale=0.2]{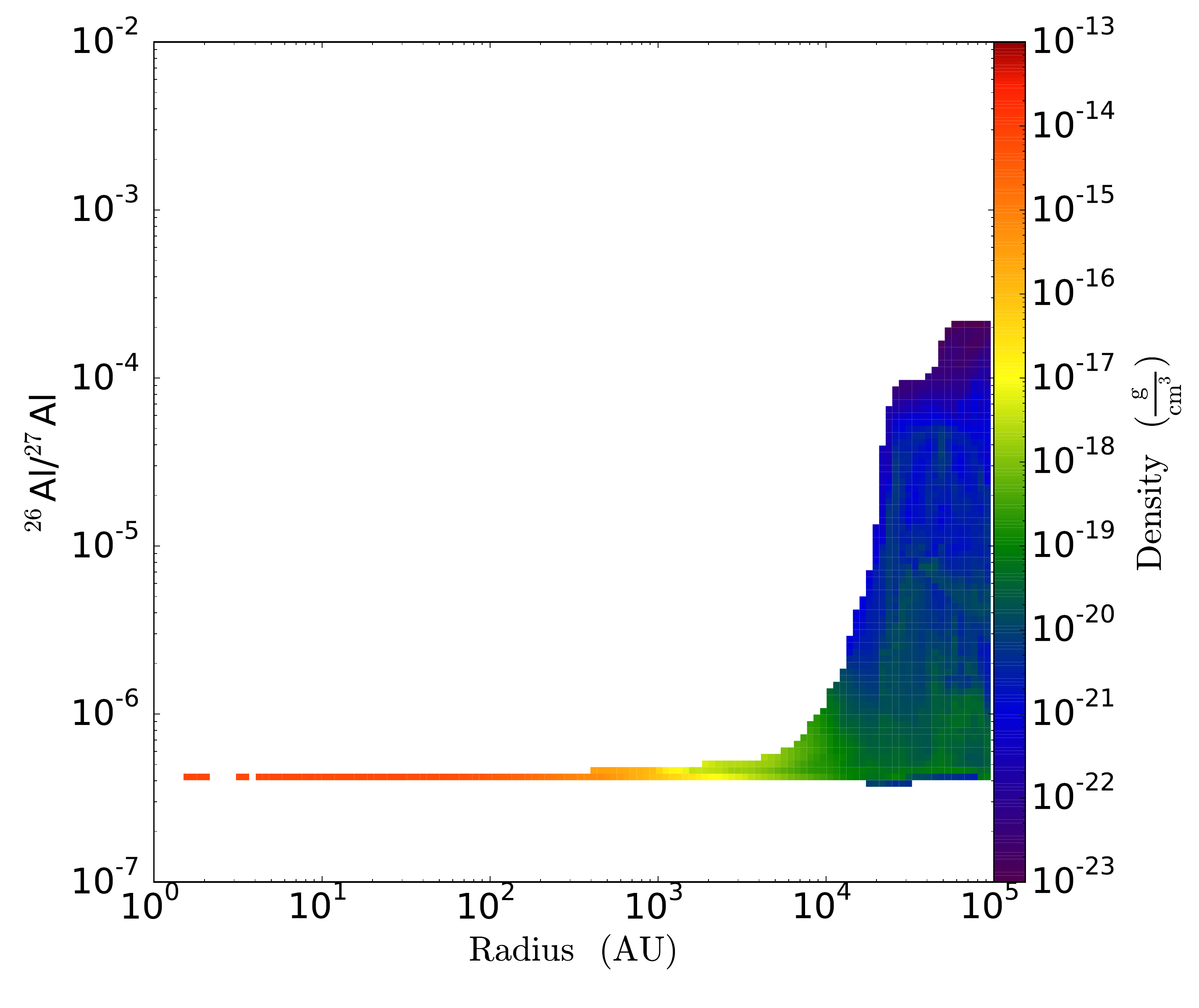} } \quad
\subfigure{\includegraphics[bb=0bp 250bp 612bp 550bp,clip,scale=0.4]{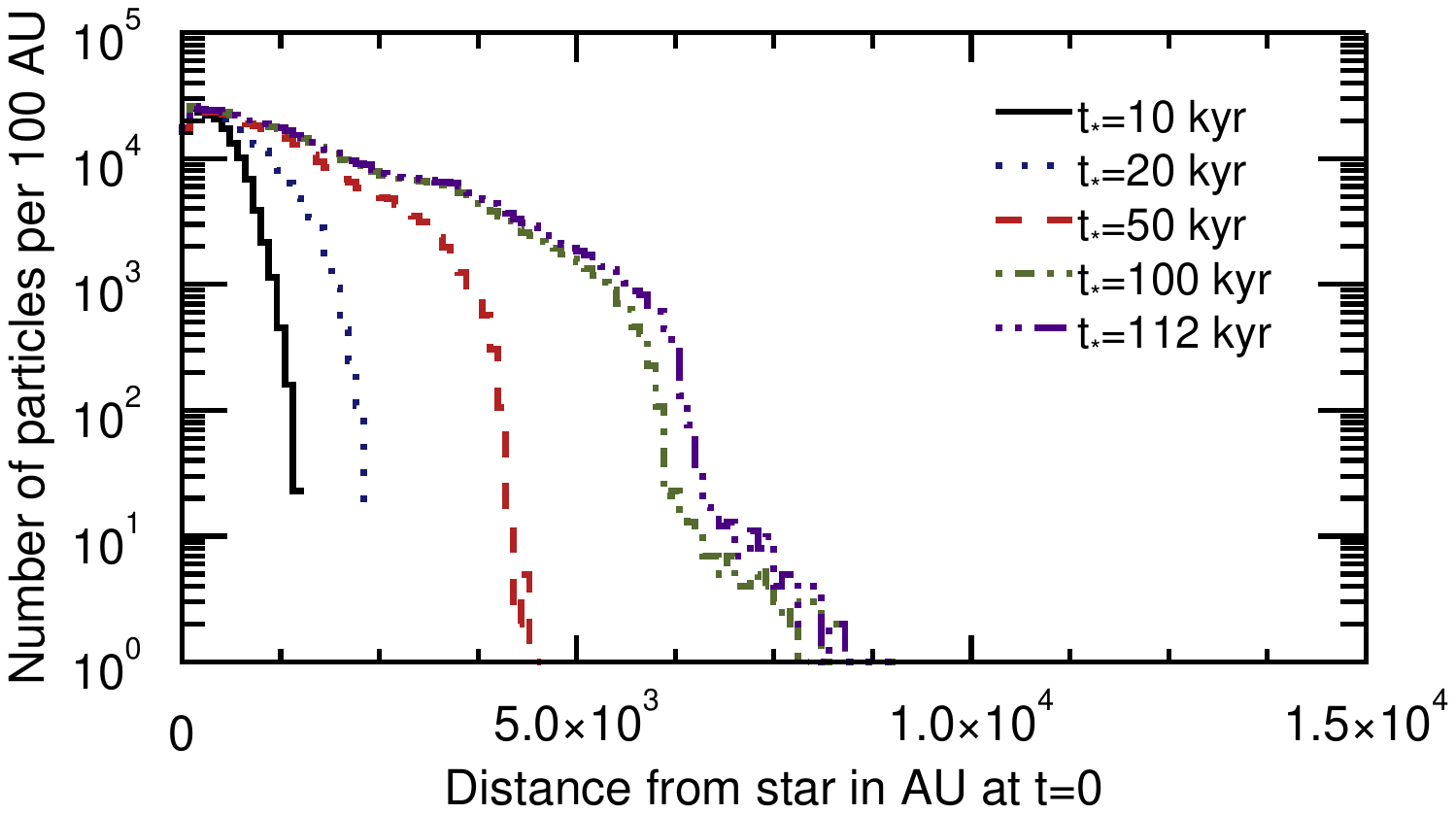} \includegraphics[scale=0.2]{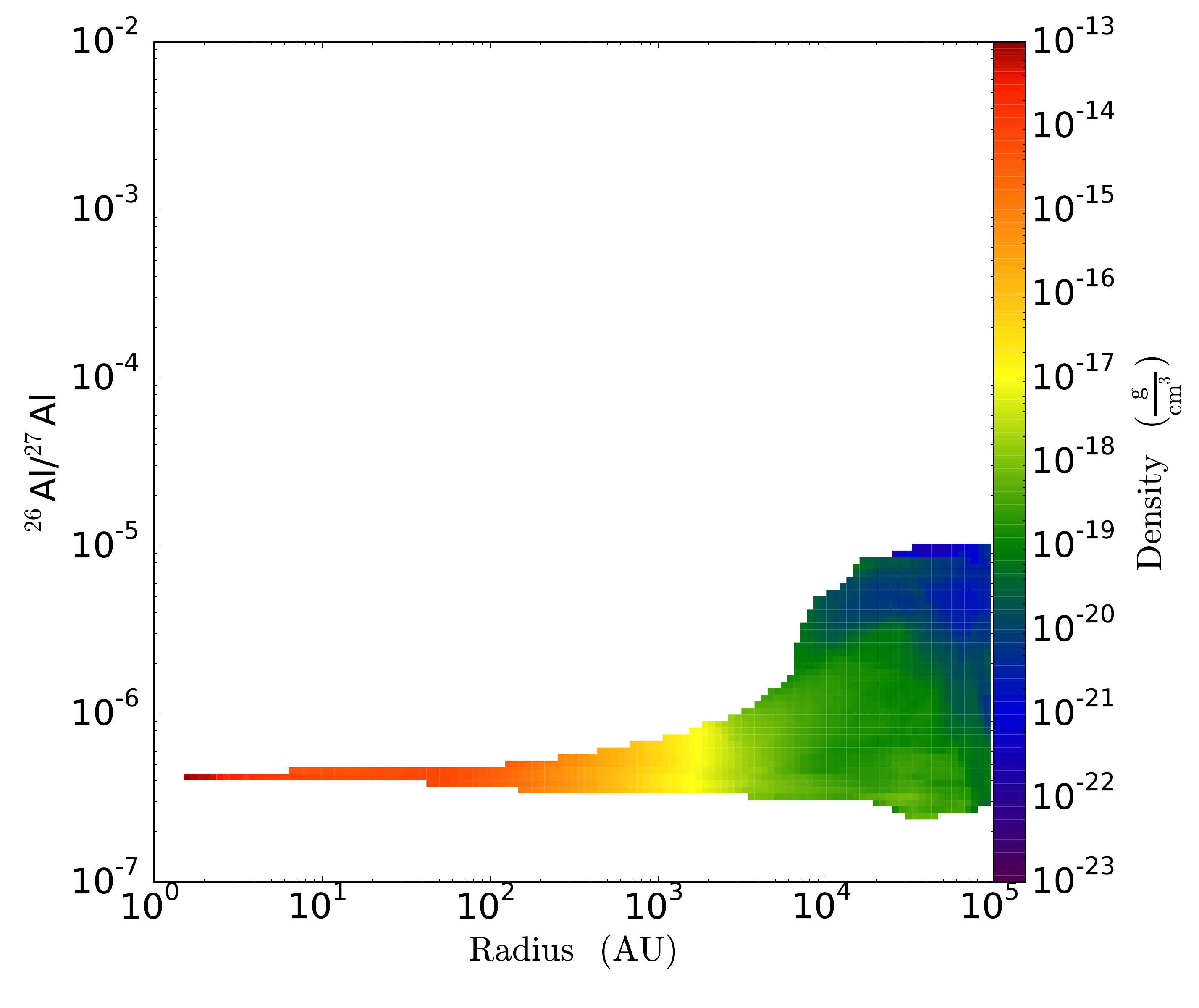} } \quad
\subfigure{\includegraphics[bb=0bp 250bp 612bp 550bp,clip,scale=0.4]{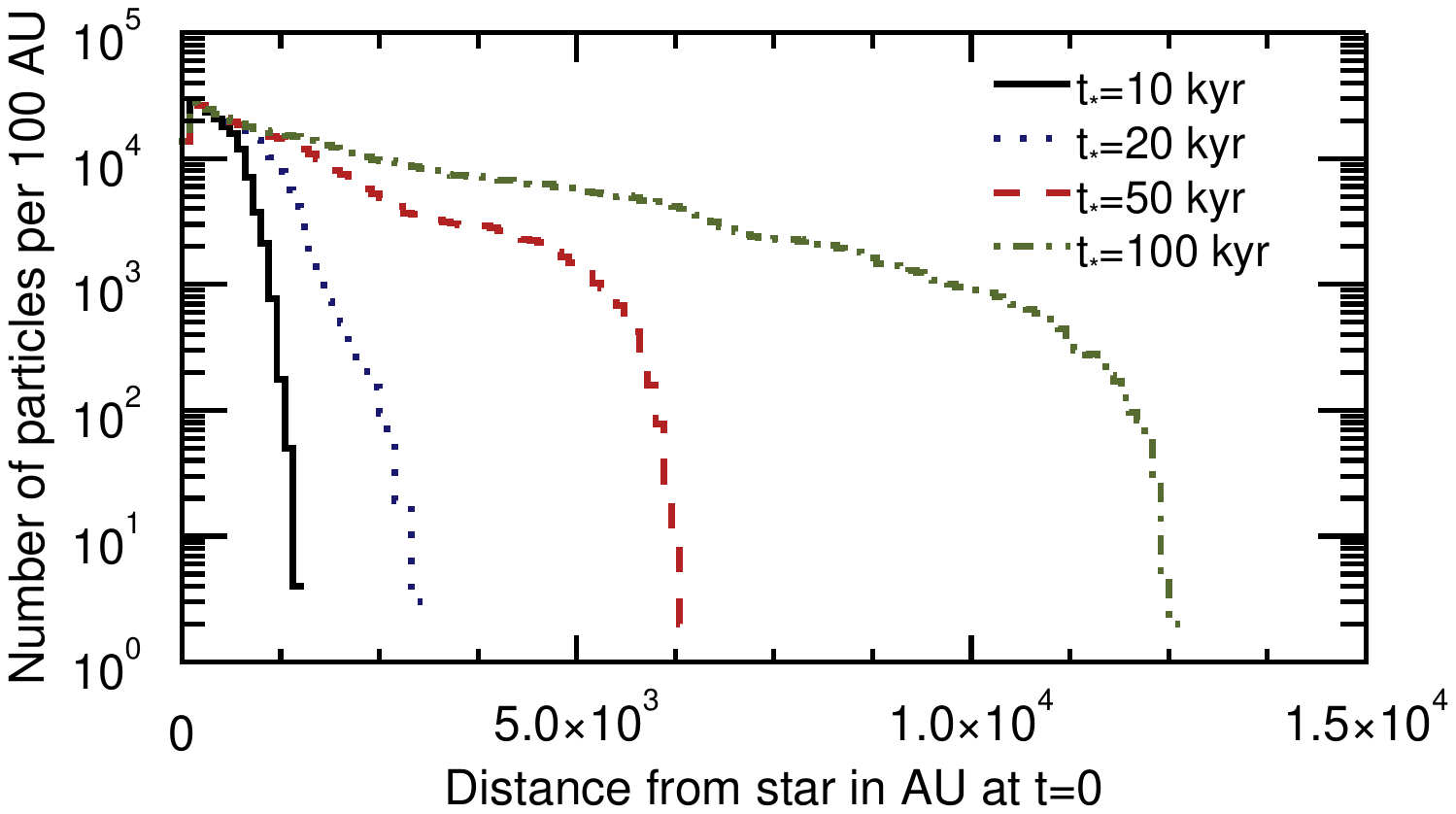} \includegraphics[scale=0.2]{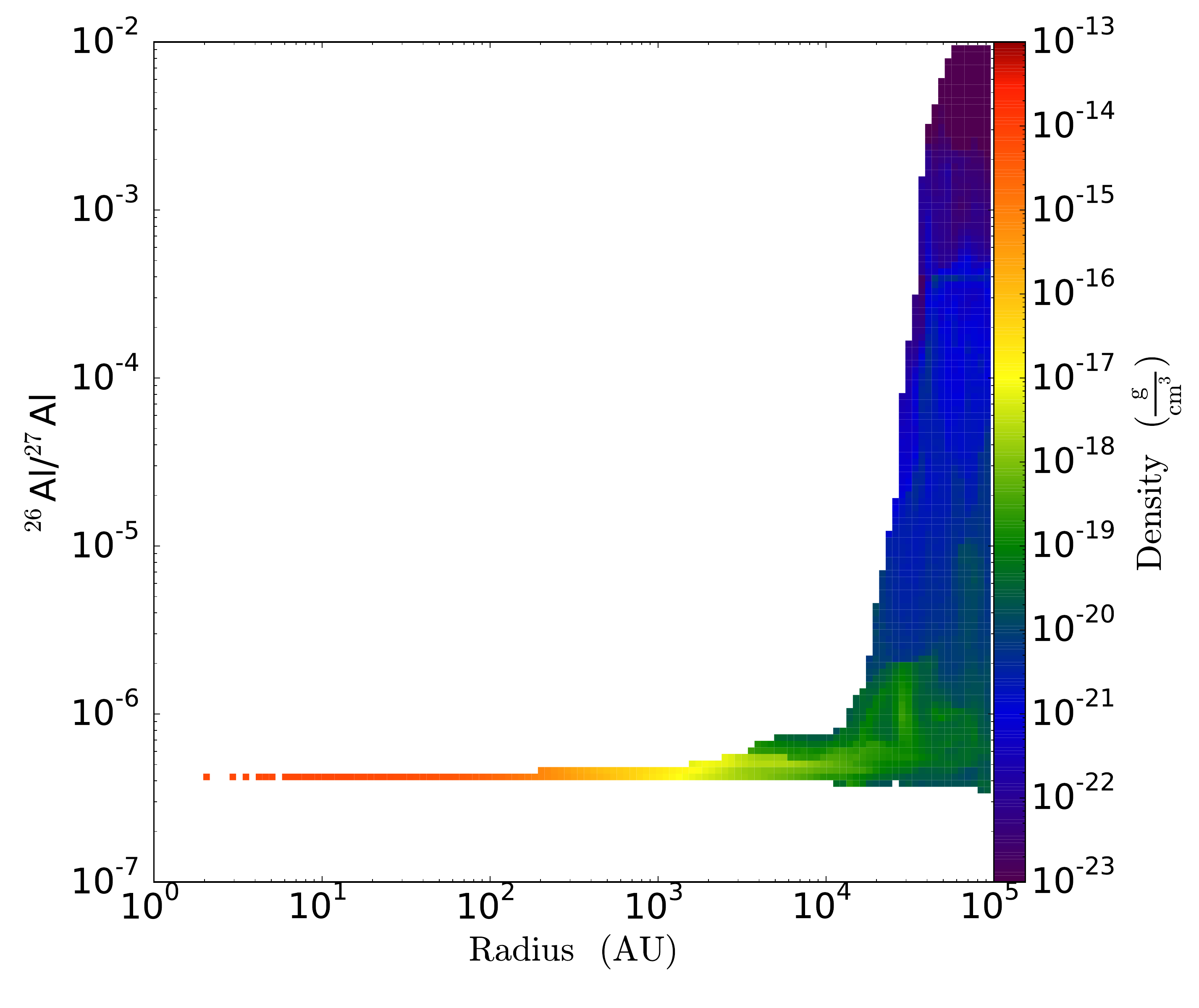} } \quad
\subfigure{\includegraphics[bb=0bp 250bp 612bp 550bp,clip,scale=0.4]{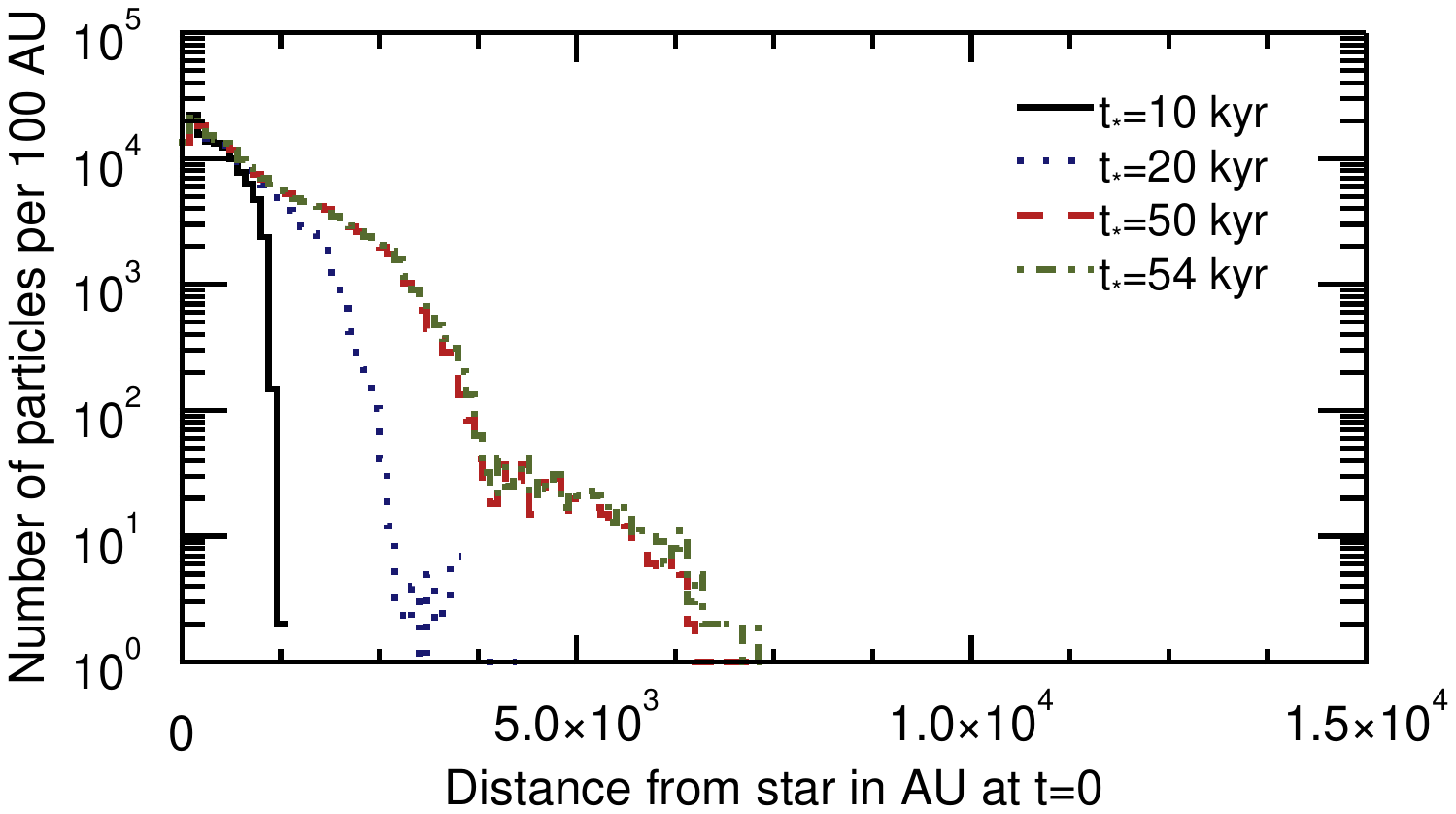} \includegraphics[scale=0.2]{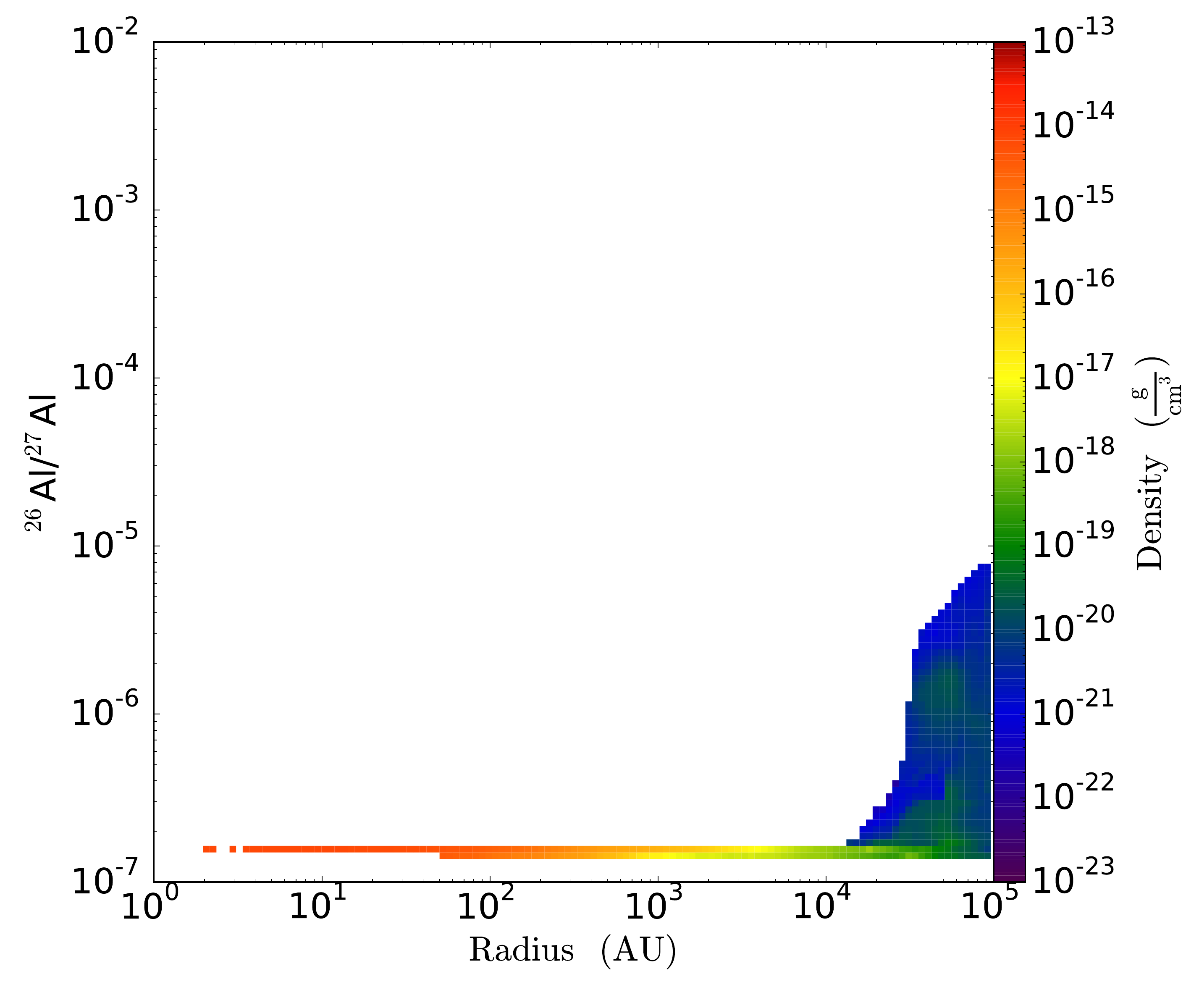} } \quad

\protect\caption{\label{fig:Origin-of-gas2} Same as \Fig{Origin-of-gas1}, but for stars 8,9,10 and 11}
\end{figure*}

\begin{figure*}
\subfigure{\includegraphics[scale=0.2]{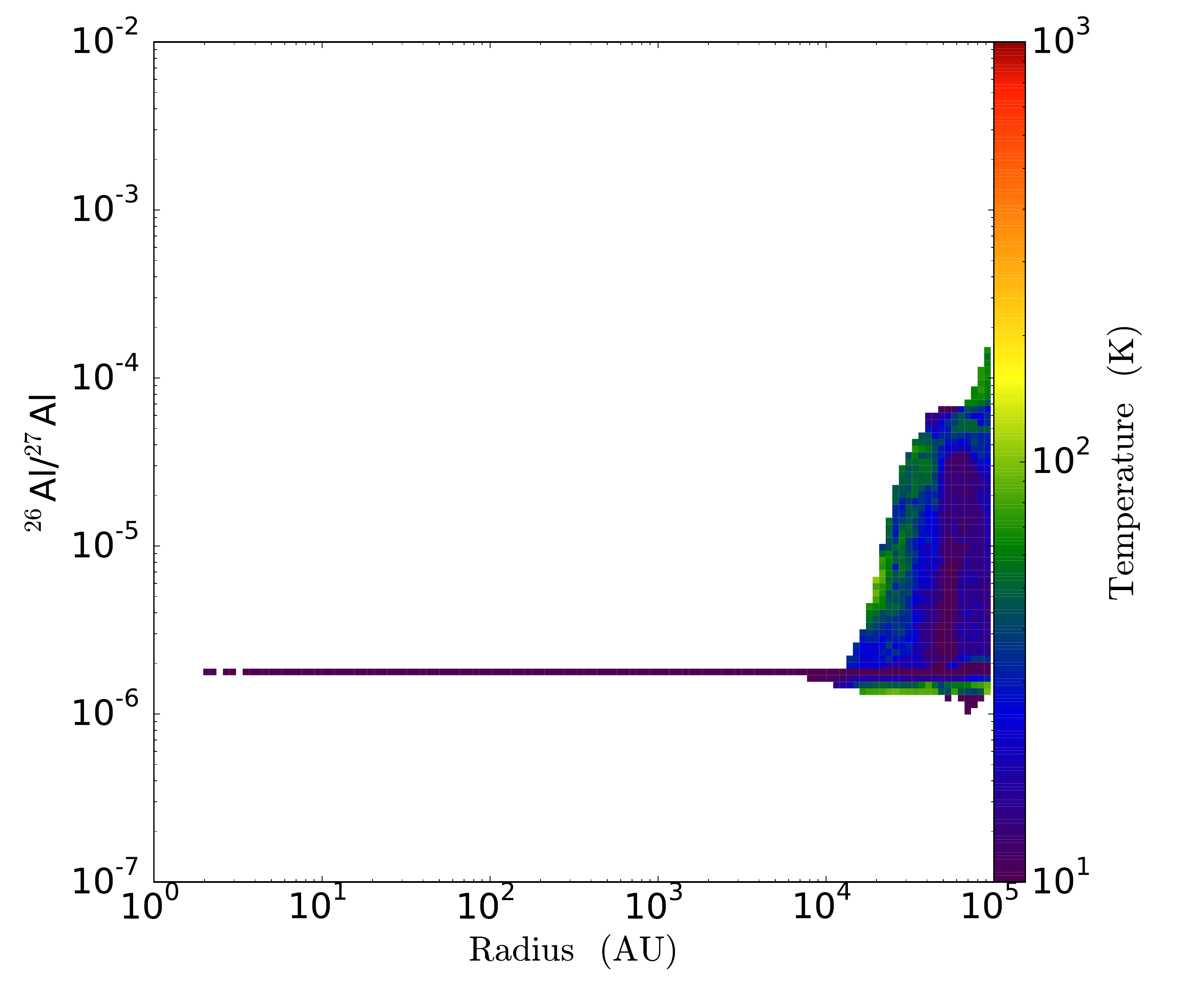}  \includegraphics[scale=0.2]{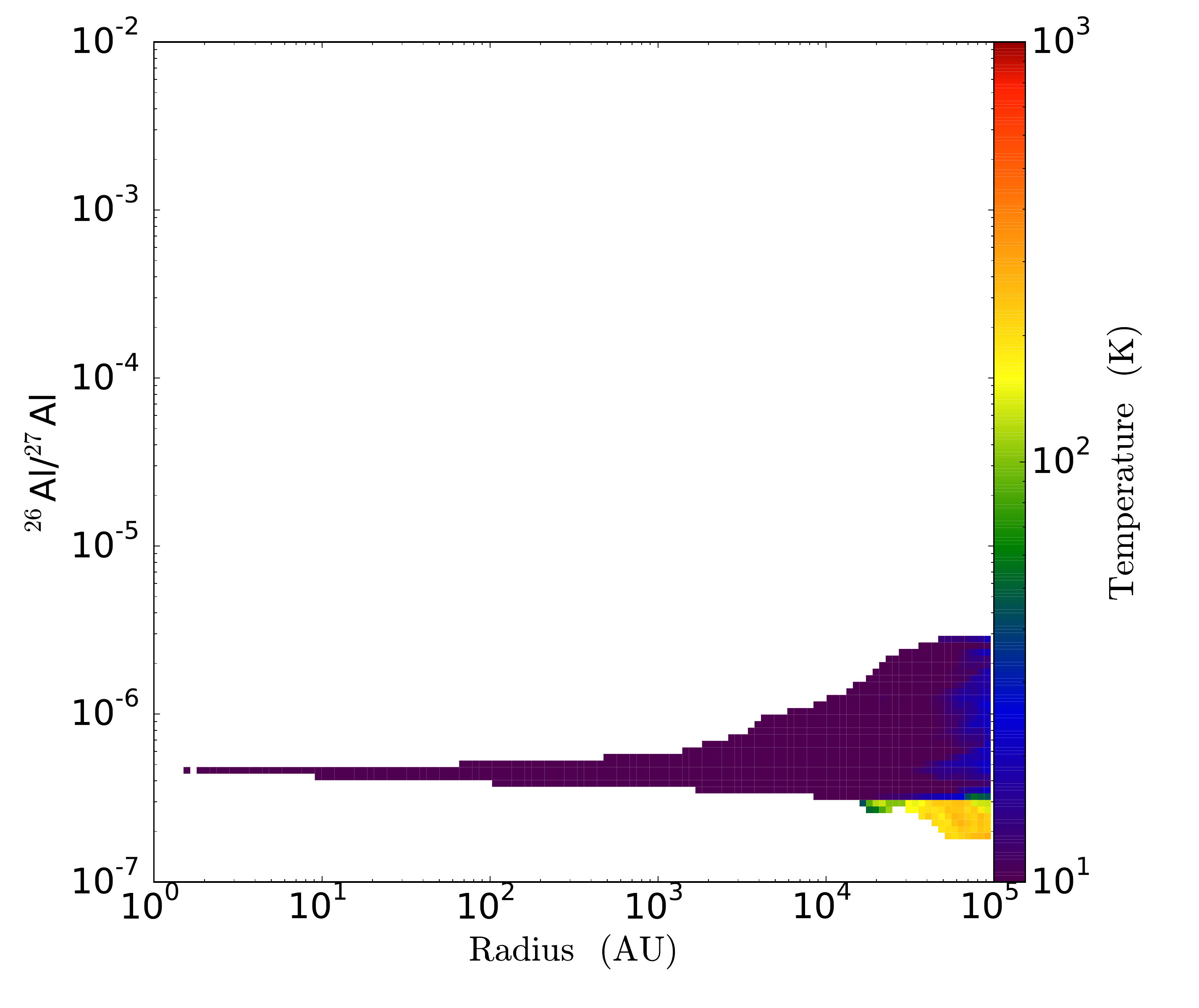}  \includegraphics[scale=0.2]{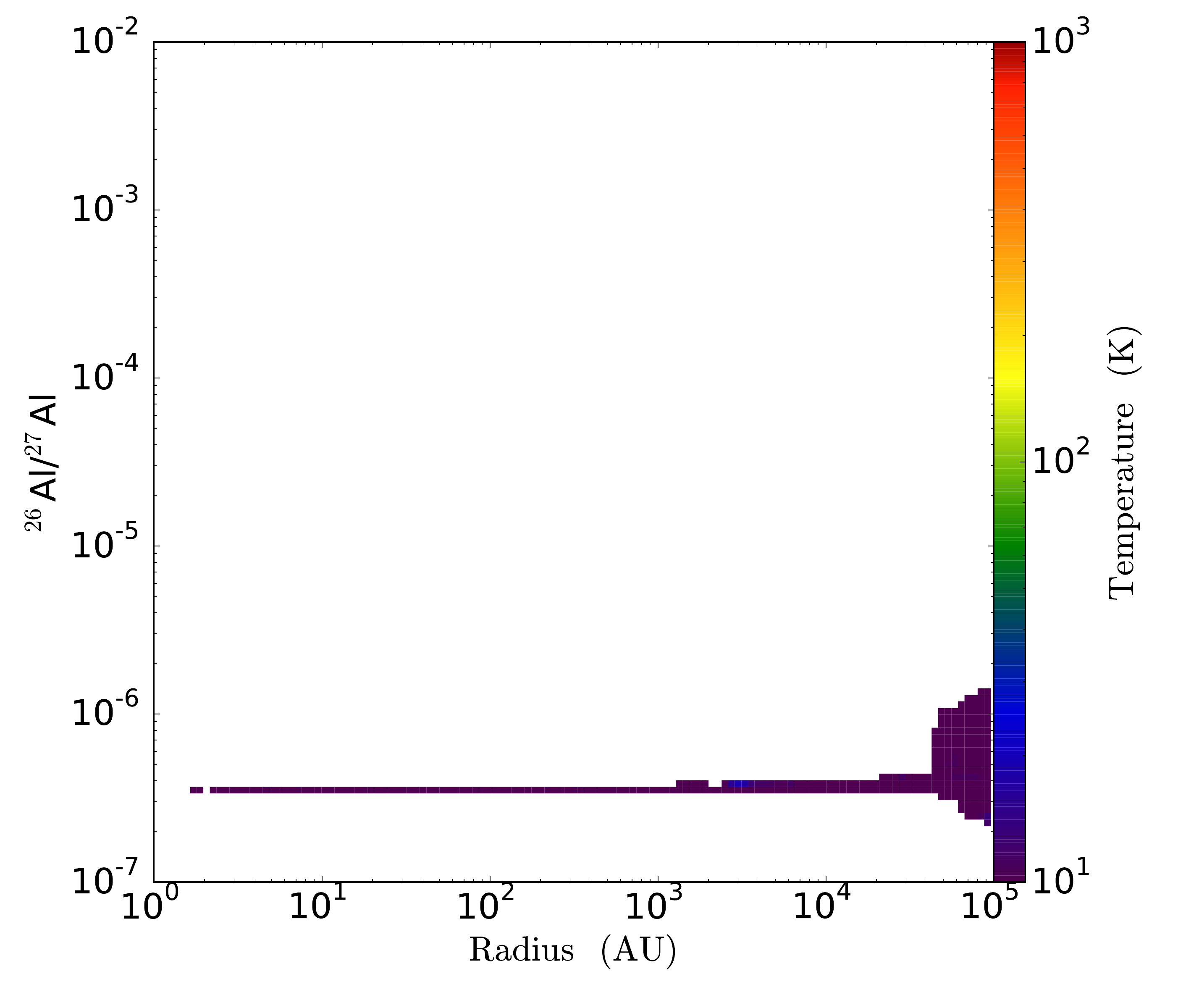} }  \\
\subfigure{\includegraphics[scale=0.2]{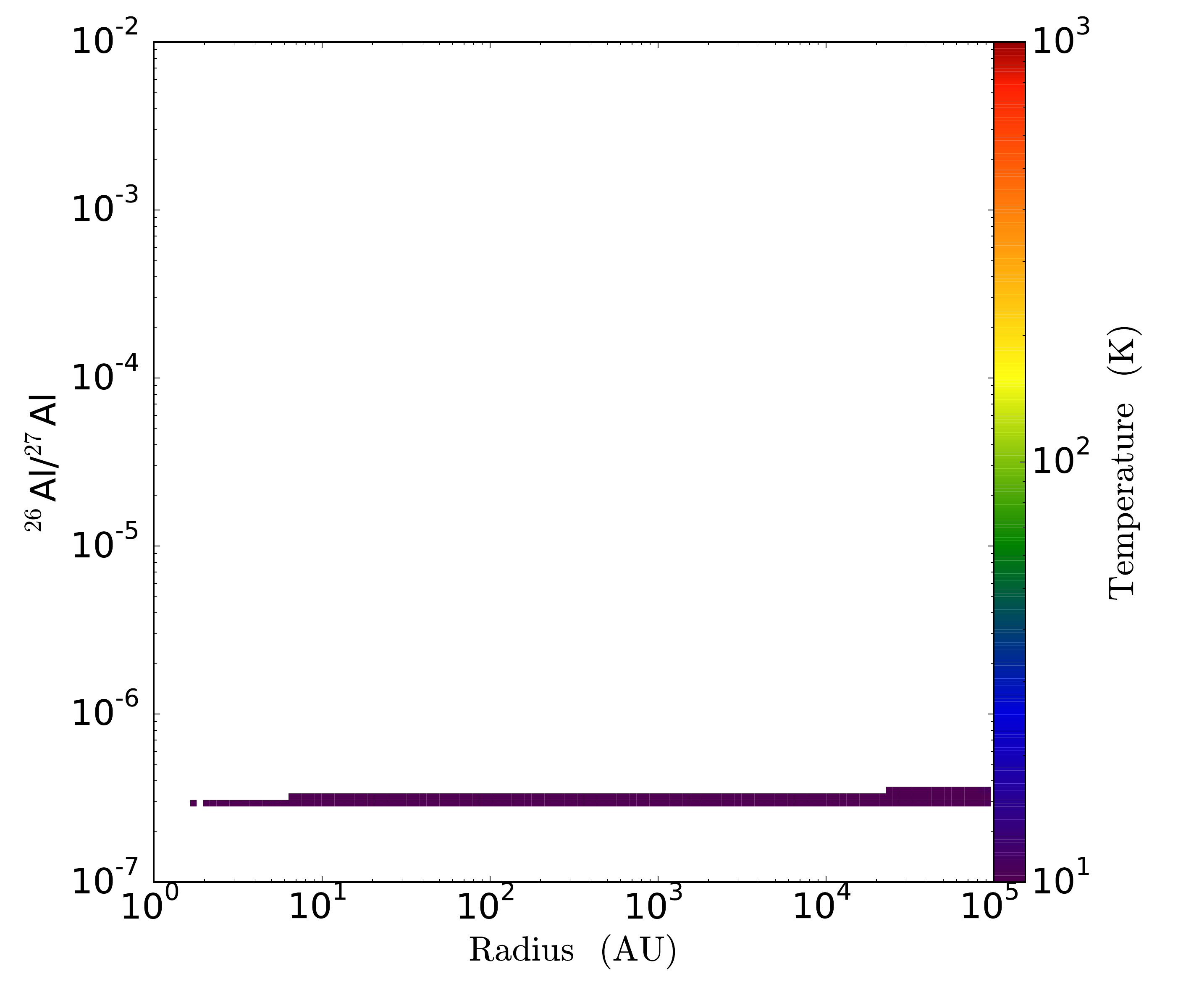} \includegraphics[scale=0.2]{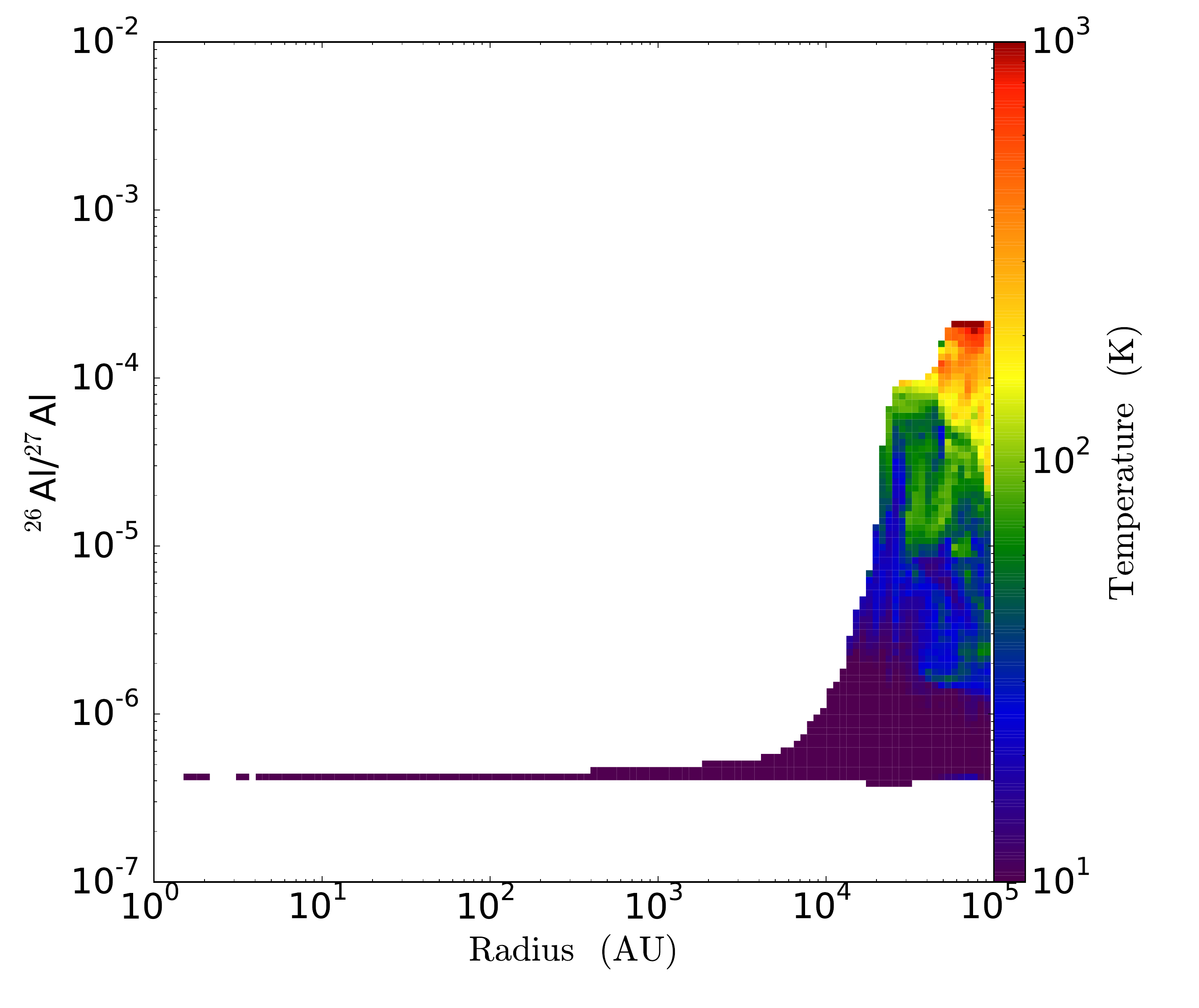}  \includegraphics[scale=0.2]{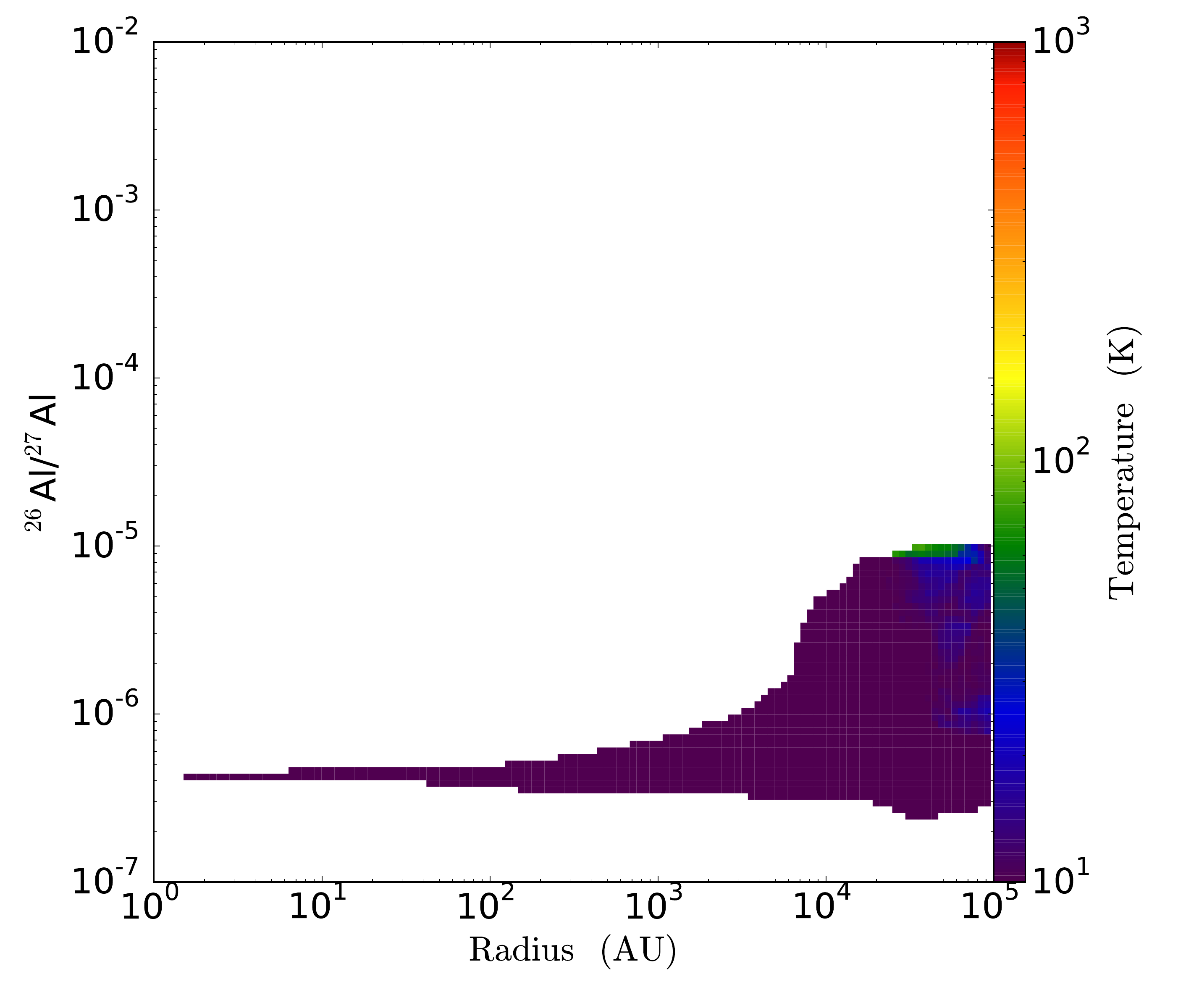} } \\
\subfigure{\includegraphics[scale=0.2]{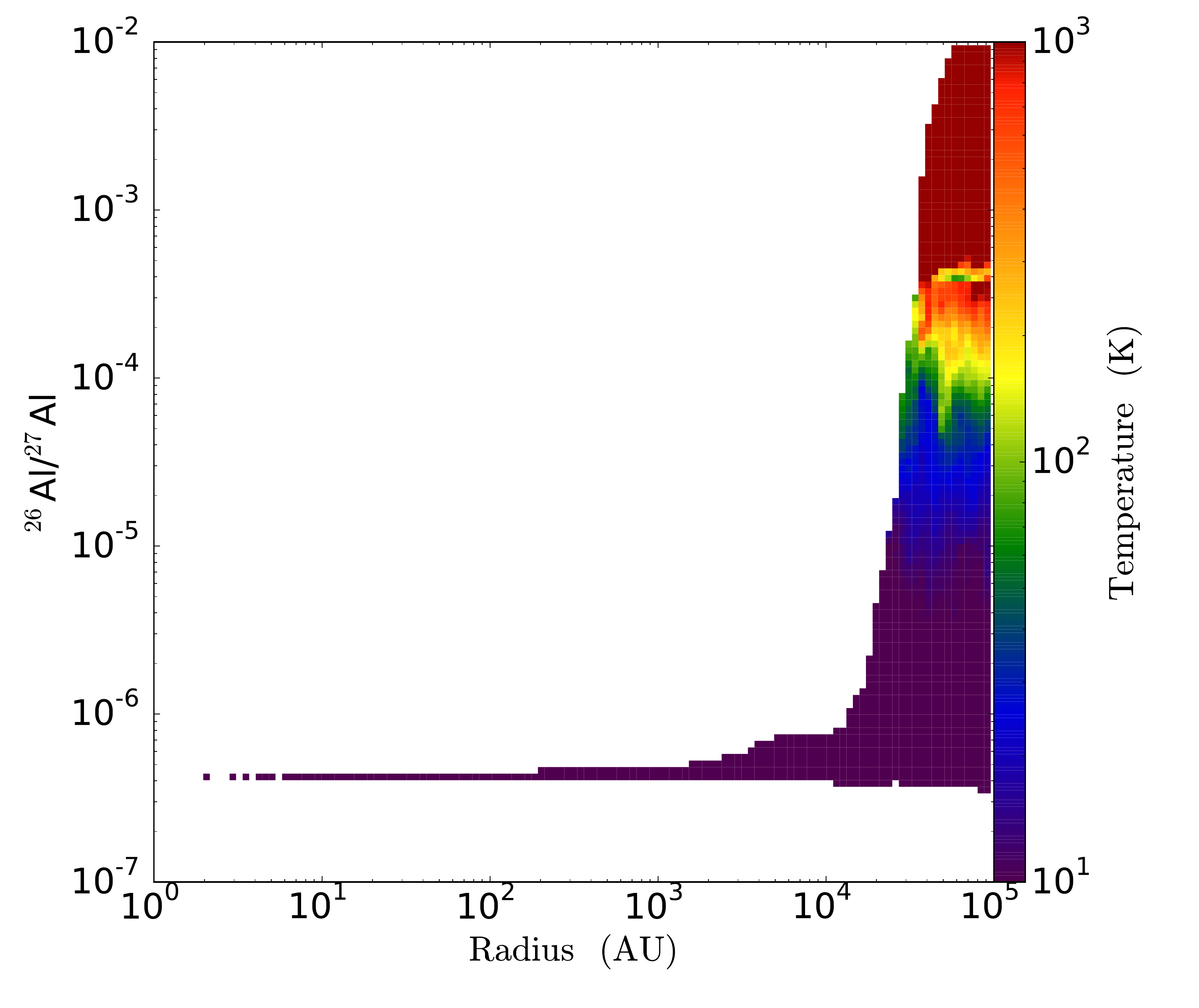}  \includegraphics[scale=0.2]{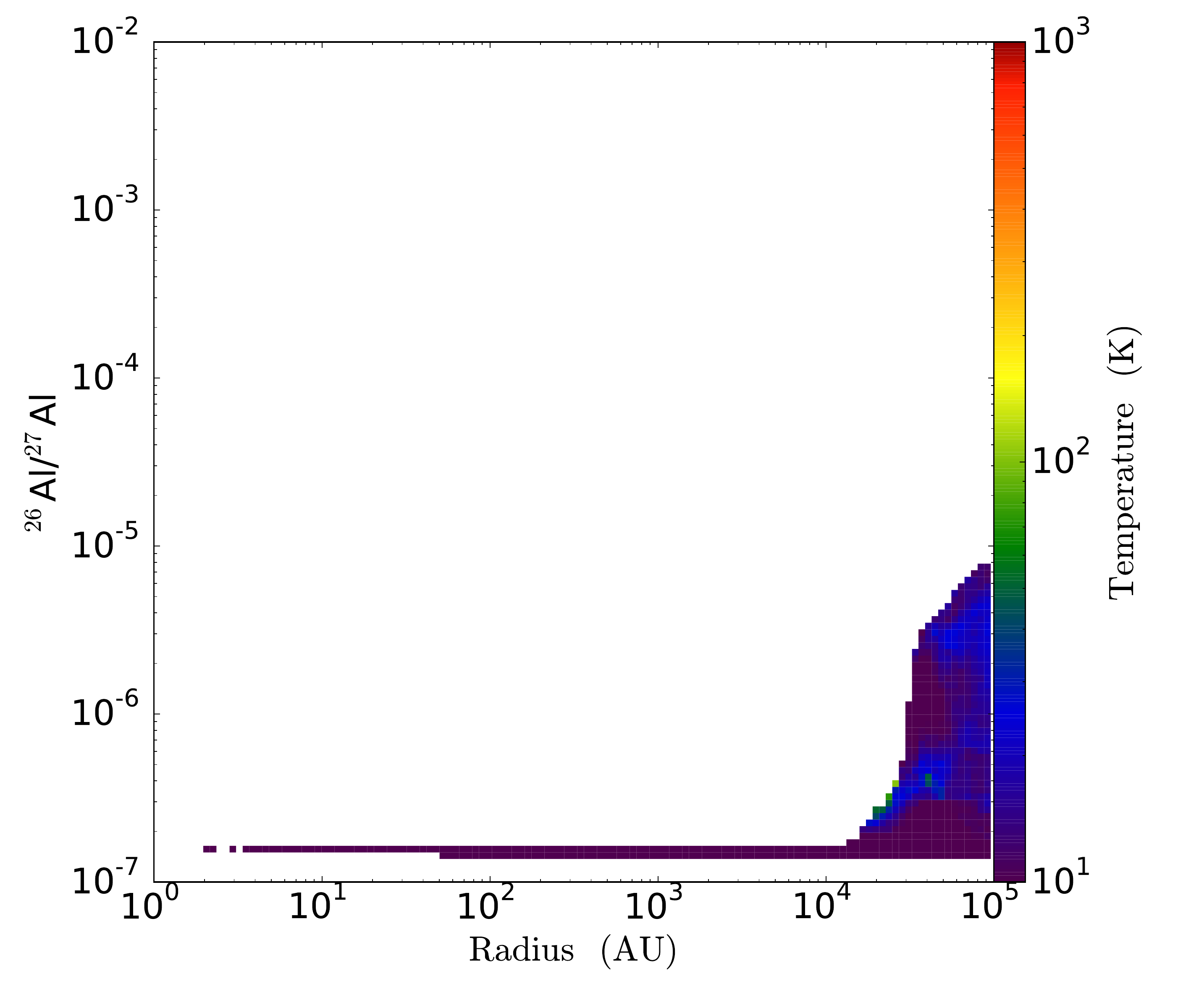} \includegraphics[scale=0.2]{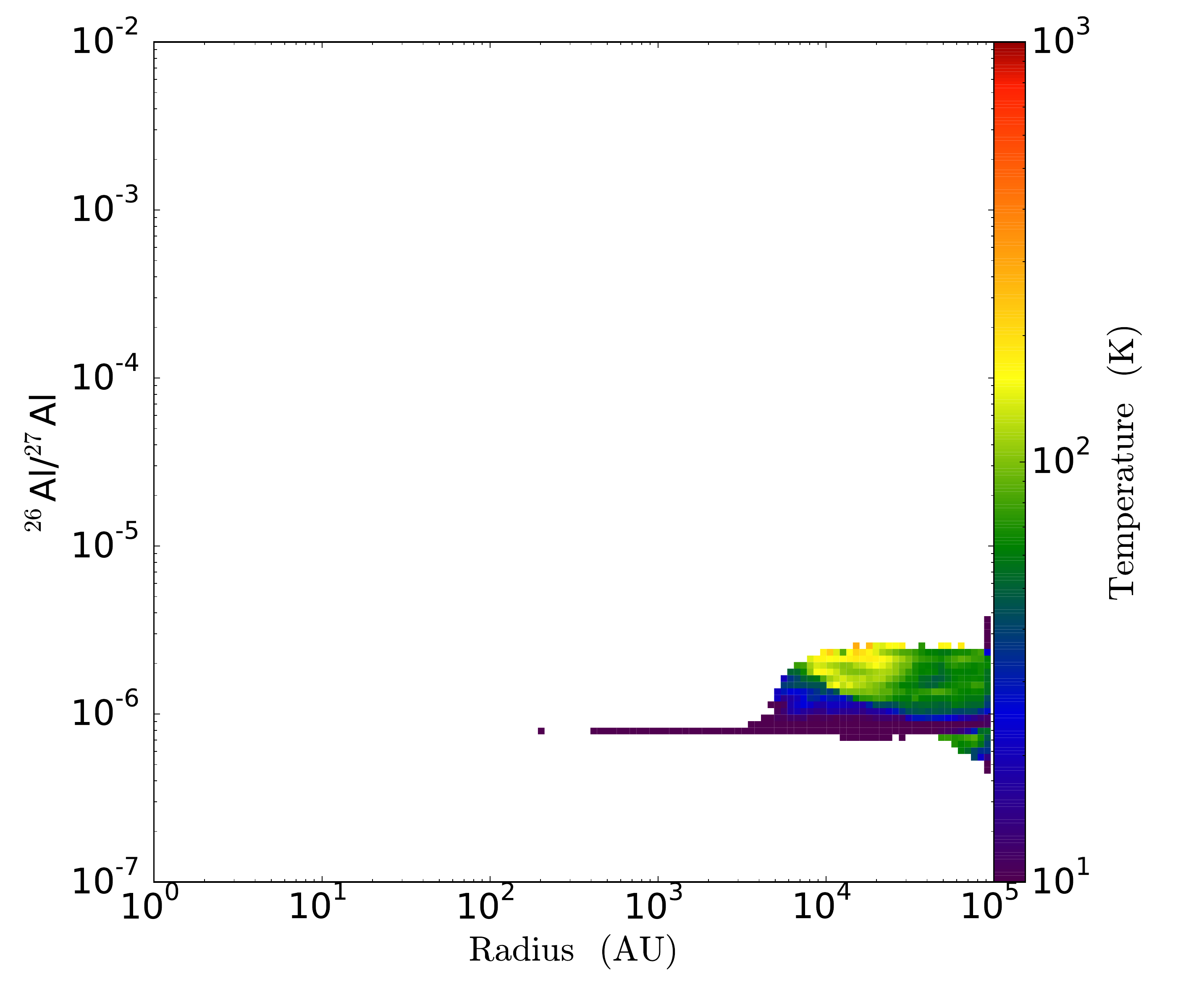} } \quad

\protect\caption{\label{fig:gasT} Same as right panel in \Fig{Origin-of-gas1} and \Fig{Origin-of-gas2}, but showing temperature instead of density from left to right and to bottom for stars 1, 5, 6, 7, 8, 9, 10, 11 and additionally 2. Only cells with $^{26}$Al/$^{27}$Al ratios below $10^{-2}$ are illustrated. The lower/upper cut-off value for the temperature is 10/1000 K and values below/above that value are colored purple/red. }
\end{figure*}

To better understand the reason for the spatial homogeneity of $^{26}$Al during the accretion, we investigated the origin of the gas and compared it with the SLR distribution at time $t=0$.
The right panels of \Figure{Origin-of-gas1} and \Fig{Origin-of-gas2} illustrate the $^{26}$Al/$^{27}$Al distribution of all cells within a distance of 100 kAU around the stars at time $t=0$ for the eight zoom runs including tracer particles. As indicated by \Fig{SLR_t_indi}, the gas does not show spreads of more than a factor of 5 in $^{26}$Al/$^{27}$Al abundance within about 10 kAU at the time of stellar birth. In contrast, the gas distribution beyond $\sim 10^4$ AU can be of very different abundance, but it is also of much lower density. When following the motion of the gas with tracer particles, we find that -- for at least the first 50 kyr -- all of the gas within the inner 100 AU was located less than 10 kAU away from the star at stellar birth. Considering the narrow spread in SLR abundances for the inner several thousand AU at the time of star formation, this explains why we only see small differences in SLR distribution during the early accretion process of the stars. In \Fig{gasT}, we show the same phase diagrams as in the right panels of \Fig{Origin-of-gas1} and \Fig{Origin-of-gas2}, but coloring the temperature instead of the density of the gas. Densities decrease with increasing distance from the star and moreover the gas close to the star is of low temperature. This in agreement with observations that stars form from cores of cold gas of about $\sim10$ kAU in size \citep{2007ApJ...666..982E}.

Our results indicate that heterogeneous accretion processes cannot account for the variability in the $^{26}$Al/$^{27}$Al ratios between canonical and FUN CAIs during the early stages of star formation, which may appear counterintuitive considering the variability of several orders of magnitude in SLR abundances present within the entire GMC.
We consider below the cascade of events leading to star formation in a GMC, to better understand our results.
Turbulent motions inside GMCs pre-dominantly cause the formation of filaments of pc-size inside GMCs.
Inside these filaments gas gets further compressed to form pre-stellar cores of sizes of 5 to 10 kAU
consistent with theoretical predictions of a Bonnor-Ebert sphere.
Eventually, the cores become dense enough to overcome the threshold value for gravitational collapse, and they collapse to stars.
Compared to typical sizes of GMCs, pre-stellar cores are about three orders of magnitude smaller,
and fill only a small part of the GMC (\Fig{phase_space_box}). In order to contaminate a pre-stellar core during formation, it has to be
near the boundary between two regions with different SLR-abundances.
Nevertheless, we consider the hypothetical case of a pre-stellar core that is
located close to such a large difference boundary. Then we are likely to have the following two regimes:
On the one hand, densities inside pre-stellar cores are very high compared to the average in the rest of the GMC, on the other hand,
SLR enriched gas is associated to recent supernova events and therefore located in regions of warm gas, and particularly
of low density (right panel of \Fig{Origin-of-gas1} and \Fig{Origin-of-gas2}, \Fig{gasT}). Thus, even if the gas that is in the vicinity of the pre-stellar core has a significantly different abundance, it
is difficult for that gas to penetrate the core, because of the large density contrast.

Although our simulations have not identified the existence of appreciable spatial and/or temporal heterogeneity in SLR abundances during the early accretion phases, we consider also the possibility that a pre-stellar core is contaminated by enriched gas at the beginning of its existence when the densities are still rather low. Similarly to our analysis in the previous section for the mixing on GMC scales, we compare the relevant timescales for the mixing at the scale of pre-stellar cores, which is the life time of a pre-stellar core with the crossing time of the gas. Observational constraints suggest that the life times of pre-stellar-cores range from 100 kyr up to 500 kyr \citep{2008ApJ...684.1240E}.
As stars form in regions of cold gas of mostly molecular hydrogen, we adopt a value of 10 K for the temperature of the sound-speed.
Considering radii of pre-stellar cores of about 5 to 10 kAU for a solar-mass star, we obtain crossing times of about 100 kyr to 250 kyr, similar to life-times of pre-stellar cores, which could in principle allow for insufficient mixing inside the core.
However, our results show only modest variations in SLR ratios and thus we conclude that the cold gas is already well mixed before the formation of the star forming cores. Even if a potential contamination occurs, it only contributes slightly to the abundance per mass and only penetrates the outer edge of the core, from where it takes often more than 100 kyr for the gas to fall in towards the star (left panel of \Fig{Origin-of-gas1} and \Fig{Origin-of-gas2}, \Fig{gasT}).

\subsection{Late phase: Potential heterogeneity of SLRs at later times}

\begin{figure}
\subfigure{\includegraphics[width=\linewidth]{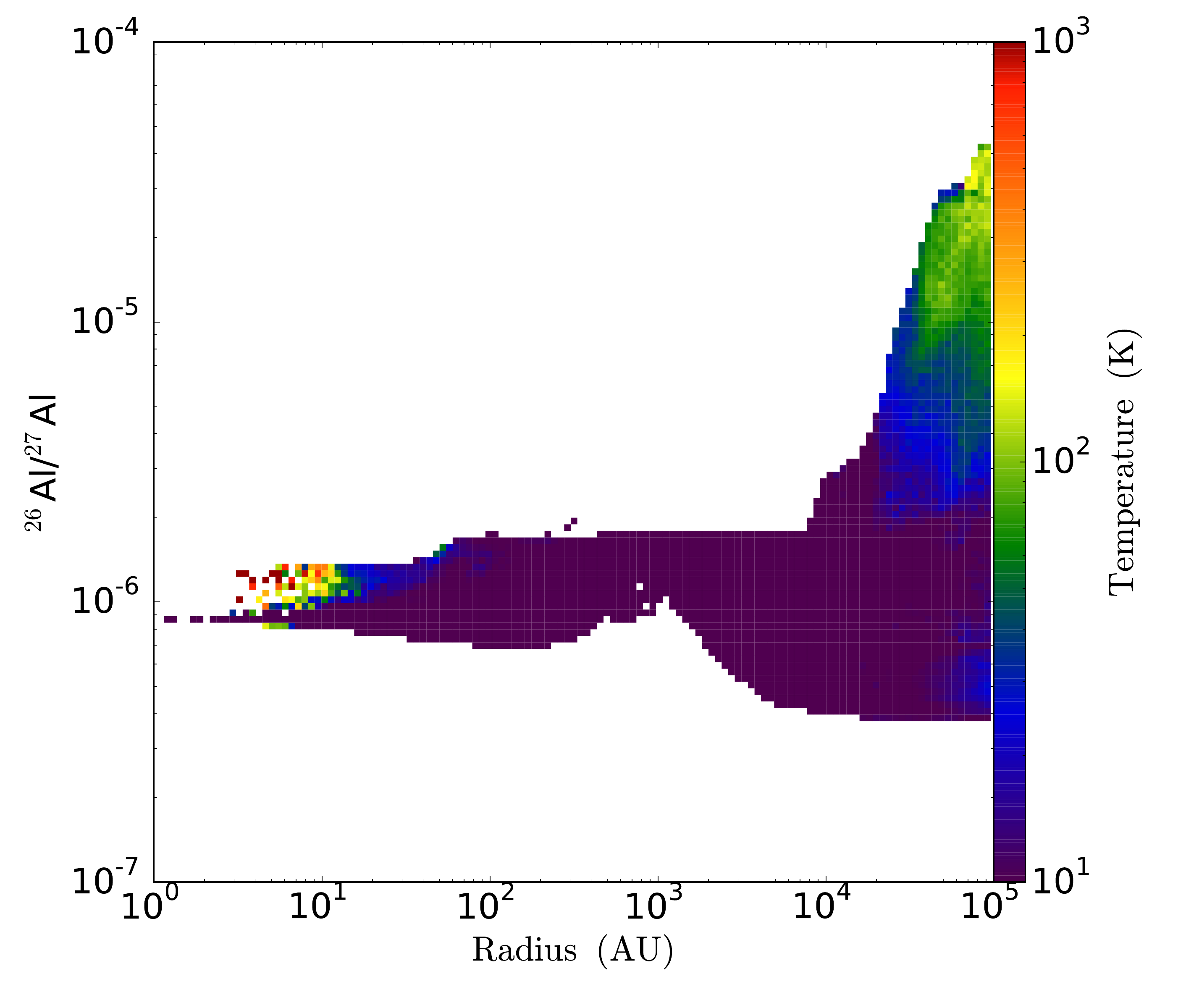} } \quad
\protect\caption{\label{fig:s74late} Phase-space diagram for star 2 at $t=160$ kyr after star formation.}
\end{figure}

Although we do not detect any significant differences in $^{26}$Al/$^{27}$Al at early times, the fact that gas from initial distances beyond 10 kAU accretes to the star over time scales $>$ 100 kyr allows for the possibility of contaminations at later times. Therefore, we consider the evolution of the stellar
surroundings, including the protoplanetary disks, at a phase later than the
initial $\sim$ 100 kyr, which we refer to as the `late phase' for simplicity.
Due to high computational costs, only data for three stars (star 2, 3 and 4) were acquired for the late phase, and none of these runs include tracer particles. One of these stars shows enhancements of up to a factor of 2 in $^{26}$Al/$^{27}$Al within a distance of only 10 AU from the star after about 160 kyr (\Fig{s74late}).
Such contaminations at later times are possible due to massive accretion of mass onto the young star during the early phase, which causes a decrease in density around the star.
Hence, it becomes possible for material that was not initially bound to the protostar to approach the vicinity of the star at later times.
Unfortunately, we do not have data from simulations including
tracer particles for this run, with which we could analyze the origin
of the gas causing the enrichments. We point out that this
late pollution is different from the idea of contaminating the collapsing pre-stellar core or a specific local
injection into the protoplanetary disk from a supernova \citep{2007ApJ...662.1268O}. Finally, we emphasize that such contaminations occurring
at late stages in our run probably cannot account for differences in $^{26}$Al abundance in CAIs,
considering that temperatures and pressures are too low to form CAIs directly out of the gas phase at this later stage.

\section{Origin of the variability in the $^{26}$Al/$^{27}$Al ratio between canonical and FUN CAIs}

As indicated by the radioactive decay of the SLRs, and the initial average abundance in the gas phase, most of the stars seemed to have formed from the initial gas reservoir present in the GMC.
Towards the end of the run more and more stars that potentially could have ended as solar mass stars formed from SLR enriched gas, and the spread in SLR abundances seems to be higher among the stars than at the beginning of our simulation (\Fig{SLR_sinks} and left/middle panel of \Fig{phase_space_box})).
In contrast, it appears that the average SLR abundance in the gas reservoir that contributes to star formation is generally enhanced after $3.7$ Myr and the spread becomes narrower again. This general enrichment of the reservoir stems from supernovae that occurred at the beginning of the simulation, in agreement with the increase of the average $^{26}$Al/$^{27}$Al in the gas phase, and consistent with the findings in \cite{2013ApJ...769L...8V} that the average abundance of $^{26}$Al/$^{27}$Al for the star forming gas increases at later times.
Thus, later times of GMC evolution are more favorable for larger variabilities in SLR ratios around stars. Nevertheless, we do not detect any significant contaminations that could account for differences in $^{26}$Al/$^{27}$Al of more than one order of magnitude for any of the more than 200 stars considered in this study. This suggests that the measured heterogeneities in $^{26}$Al/$^{27}$Al between canonical and FUN CAIs are caused by a different mechanism than supernova conatmination.

\subsection{An extrasolar origin for FUN CAIs?}
In contrast to the early phase, our results suggest the presence of gas with variable $^{26}$Al/$^{27}$Al ratios in the stellar surrounding during later phases of accretion, namely $>100$ kyr after star formation.
As such, we evaluate the possibility that FUN CAIs represent objects formed around other $^{26}$Al-poor solar mass stars located in the vicinity of the proto-Sun and thereafter transported to the Sun via stellar outflows \citep{2011PNAS..10819152M}.
Two lines of evidence are in apparent support of such a scenario.
First, most stars form in cluster often separated by less than 1000 AU \citep{2003ARA&A..41...57L}. Second, approximately 1 solar mass of material is lost to outflows during the accretion of solar mass type star \citep{2014prpl.conf...53O}. Thus, the cross contamination of nearby nascent systems through stellar outflows appears to be likely the outcome of star formation in clusters. We assess this possibility by investigating the spatial distribution of stars with contrasting $^{26}$Al/$^{27}$Al values in our simulations. We restrict our analysis to stars more massive than $0.2$ M$_{\odot}$ that formed within a timeframe of $1.6$ Myr, as this represent the maximum age difference between canonical and FUN CAIs inferred from the $^{182}$Hf-$^{182}$W system \citep{Holst28052013}. In \Fig{Rel_diff_abun}, we show that all pair of stars formed within a distance of 50 kAU have initial $^{26}$Al/$^{27}$Al values within one order of magnitude, although greater variability is observed at greater distances. Therefore, our analysis requires the transport of CAI material in the ISM over distance greater to 50 kAU to explain the contrasting initial $^{26}$Al/$^{27}$Al ratios observed between FUN and canonical CAIs, which appears unrealistic.

\begin{figure}
\includegraphics[bb=0bp 250bp 612bp 550bp,clip,width=\linewidth]{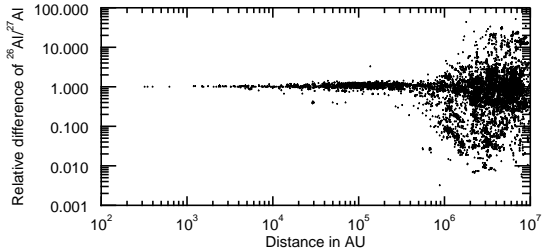}

\protect\caption{\label{fig:Rel_diff_abun}Relative difference in $^{26}$Al/$^{27}$Al abundance between all stars of mass higher than $0.2$ M$_{\odot}$ and all stars that have an age difference of at most $1.6$ Myr with respect to their distance at the time of formation of one of the stars.}
\end{figure}

One might argue that CAIs formed instead around stars lower than 0.2 M$_{\odot}$ in mass, which are closer to the star. Even with our state of the art zoom-run, we do not resolve the formation of these very low mass stars properly, and we do not have a reliable statistics for these stars. However, we are confident that the SLR to distance relation would not be very different for low mass stars, considering that these stars formed from the same gas reservoir as the higher mass stars. Taking additionally into account the uncertainty whether CAIs can travel through the ISM and accrete onto foreign star-disk system, we consider this scenario to be  unlikely.

\subsection{Thermal processing of dust grains}

Instead, we argue that measured differences in CAIs and chondrules are
most likely caused by physical processes neglected in our simulation. In our
model, we only considered the motion of the gas to evaluate the influence
from GMC scales down to protoplanetary disk scales. However, observations
show that GMCs and protoplanetary disks consist of about 1\% of dust.
Therefore, we suggest that the contrasting initial $^{26}$Al/$^{27}$Al values recorded by canonical and FUN CAIs reflects unmixing of the $^{26}$Al carrier by thermal processing during the early stage of solar system formation. This is supported by the observation that FUN CAIs plot on the solar system nucleosynthetic correlation line defined by inner solar system solids, asteroids and planets \citep{2015E&PSL.420...45S}. Although a detailed study of thermal processing is beyond
the scope of this paper, we discuss the basics of the process
here.

\begin{figure}
\includegraphics[width=\linewidth]{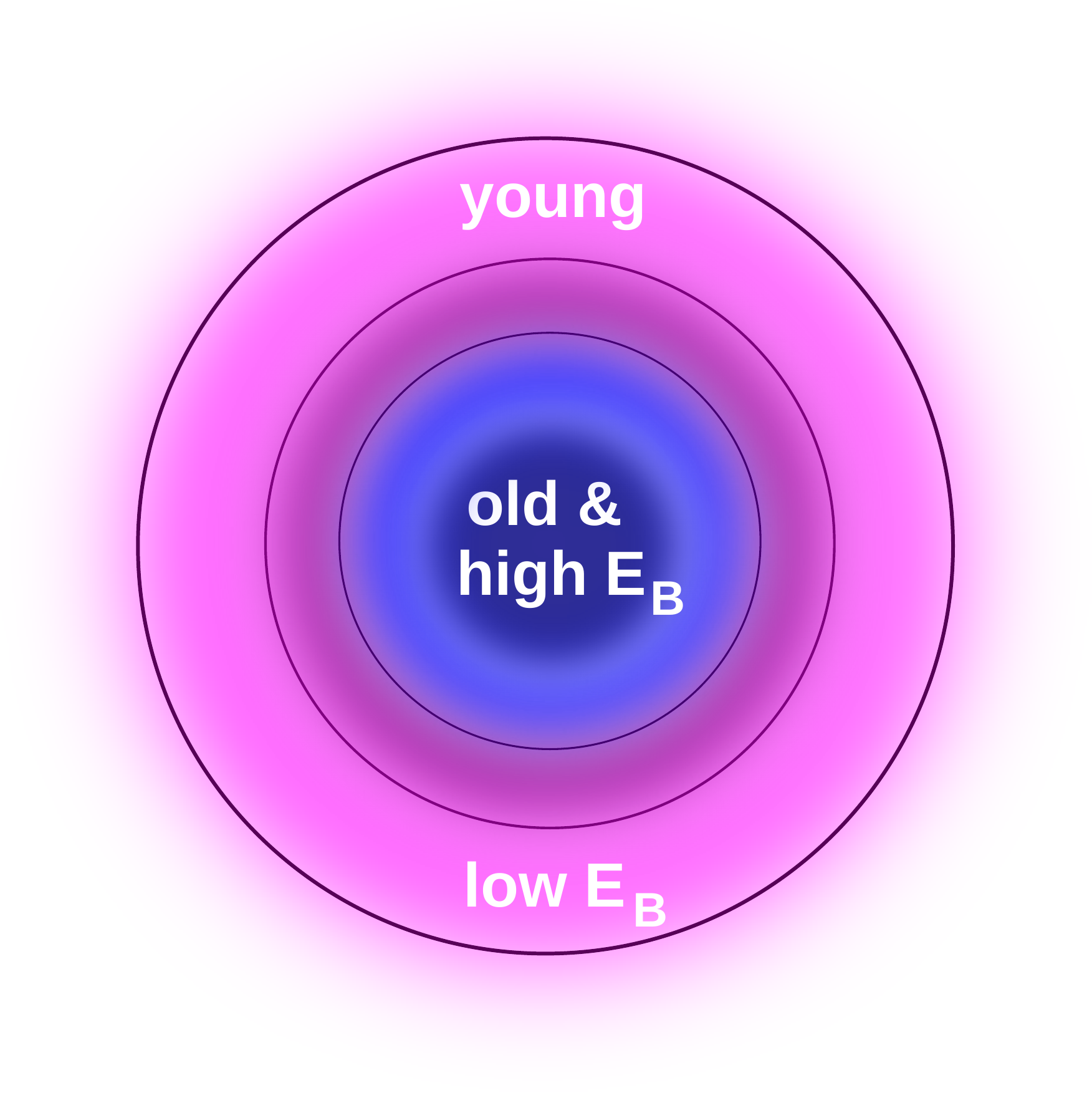}

\protect\caption{\label{fig:Rel_diff_abun}Sketch of a spherical dust grain. Old dust has high binding energy and settles at the center of a grain, while younger dust of lower binding energy is located at the surface layer. Due to radioactive decay older dust components are more likely to contain less $^{26}$Al than younger dust grains. }
\end{figure}

Consider the evaporation time
\begin{equation}
t_{{\rm ev}}\propto\nu^{-1}e^{\frac{E_{B}}{k_{B}T}}
\end{equation}
\citep{2015ARA&A..53..541B,1987ppic.proc..333T}, where $\nu$ is
the vibrational frequency, $E_{B}$ the binding energy of the dust
grain, $k_{B}$ the Boltzmann constant and $T$ the temperature. Approximating
dust grains to be perfectly spherical, the energy of one photon $h\nu$,
where $h$ is the Planck constant, is assumed to scale with the inverse
of the volume of the grains. Thus, we obtain

\begin{equation}
t_{{\rm ev}}\propto a^{3}e^{\frac{E_{b}}{k_{B}T}}.
\end{equation}

The dust composition in the ISM consists of grains of different size
(up to $\mu$m size) and different age. Due to radioactive decay,
older dust grains are considered to show lower SLR abundances than
younger grains. Taking additionally into account that older grains
were exposed to potentially destructive radiation for a longer time,
the surviving grains have higher binding energies and/or
larger grain sizes than the younger grains. Furthermore, older components
and thus SLR depleted components are more likely to be in the central
layers of the grains whereas younger components are more likely to
accumulate on the surface of the grains. Thus the younger grain components
generally
\begin{enumerate}
\item vaporize at lower temperatures,
\item are considered to be in smaller grains and
\item shield the older components from radiation.
\end{enumerate}
During the star-forming process, the well mixed dust grains in the
pre-stellar core fall in towards the star, where they are exposed
to stellar radiation. Since the younger, SLR enriched components vaporize
more easily, the gas phase can become locally enriched
in SLRs. Depending on the temporally changing strength of irradiation, more or less layers of the dust grains get vaporized, eventually causing a continuous distribution of different $^{26}$Al/$^{27}$Al ratios in the gas around the star.
Due to the short-timescales of this collapsing phase, of the order of kyr or less, the gas cannot mix to a homogeneous reservoir before CAIs are formed by condensation out of the gas phase.
In this way, CAIs inherit the thermally induced, local heterogeneities of $^{26}$Al/$^{27}$Al ratios in the early gas phase.

Progressive thermal processing of in-falling $^{26}$Al-rich molecular cloud material in the inner solar system has also been invoked to account for large scale heterogeneity in $^{26}$Al that existed at the time of accretion of most asteroidal bodies \citep{2009Sci...324..374T,2011ApJ...735L..37L,2013ApJ...763L..40P,2015E&PSL.420...45S,2015GeCoA.149...88S,2016vanKooten}. We note that the initial $^{60}$Fe abundance for stars in our simulations is much higher than that inferred for early solar system based on differentiated meteorites and chondritic components \citep{2012E&PSL.359..248T,2015ApJ...802...22T}. Similarly to \citet{2013ApJ...769L...8V}, we interpret this discrepancy as reflecting efficient removal of $^{60}$Fe from disk solids via thermal processing of their precursors, which requires that the carrier phase of $^{60}$Fe was significantly more volatile than the $^{26}$Al carrier. As such, inferred initial $^{60}$Fe estimates based on meteoritic material such as differentiated meteorites and/or chondrules may not be representative of that of the bulk solar system.

\section{Conclusions}

For the first time, we followed the dynamics
of GMC gas down to the small scales relevant for individual star and protoplanetary disk formation,
while we simultaneously accounted for the large-scale effects induced by supernovae, magnetic fields, and turbulence in the GMC.
In particular, we analyzed the abundance of the $^{26}$Al and $^{60}$Fe SLRs around newly formed stars in simulations of
a ($40$ pc)$^{3}$ GMC carried out with the adaptive-mesh-refinement code \ramses. First, we simulated the dynamics in the GMC including enrichments from supernovae with a highest resolution of 126 AU for about 4 Myr. During this time, more than 200 stars with masses in the range of $0.2$ M$_{\odot}$ to $8 $M$_{\odot}$ formed inside the GMC.
To model the gas dynamics in the early phases of star formation in further detail, we investigated the distribution of the $^{26}$Al/$^{27}$Al ratio
during the first $\sim$100 kyr for eleven of the stars, by rerunning their formation and early evolution phases with higher spatial resolution, using grid sizes down to $2 $ AU.
We conclude from
our simulations that huge variations in abundance ratios of $^{26}$Al/$^{27}$Al
and $^{60}$Fe/$^{56}$Fe generated by supernova explosions exist in GMCs. However, highly enhanced values only occur in the
hot, low density gas located close to recent supernova events. Over
time, the ejecta are cooled down and get incorporated into star-forming
gas. Here the gas gets mixed, such that the variations in the cold dense gas are modest.
None of the more than 200 stars showed abundances variations of a magnitude that could have accounted for the measured differences between canonical CAIs and FUN CAIs.
Considering that we only see marginal deviations
from the characteristic decay curve of the initial SLR abundance
for the gas around the stars selected for zoom-ins, we conclude that the gas in
pre-stellar cores is already well mixed before the formation of the stars.
We demonstrated that the gas forming the star-disk system
accretes from distances within about $10^4$ AU, which is in agreement with observations and theoretical predictions for the size of a Bonnor-Ebert sphere for a 1 M$_{\odot}$ star.
The collapsing gas in the core that forms the star is gravitationally
bound and by definition overdense compared to its surrounding.
Thus, hot gas from recent supernovae cannot easily pollute the stellar environment in the
early phases corresponding to stage 0/I, when the star still has a
massive envelope and is strongly accreting from its initial gas reservoir. 

However, we point out that the situation changes for times later than about 100 kyr,
when most of the surrounding gas has accreted onto the star-disk system
and the density of the envelope has dropped by a few orders of magnitude.
At these times, gas with different abundance can penetrate the environment around the star and might potentially
lead to significant variability of the SLR abundances in the protoplanetary
disk. As in previous models these variations are related to the production
of SLRs by supernovae, but the picture is different in the sense that the SLR enrichments already occur in cold gas.
In contrast to a specific injection into the star-disk system, the star moves through the interstellar medium
and eventually enters gas reservoirs of different SLR composition. Therefore, the traditional model of supernova injection is also misleading at later times in star formation.

Instead of being caused by early heterogeneities in the SLR distribution in the gas phase around young stars, we suggest thermal processing of the dust components as the main explanation for the differences in $^{26}$Al/$^{27}$Al ratios between canonical CAIs and FUN CAIs.
The main point is that new and old dust are likely to differ both in binding energies and distribution over grain size, with the fraction of new dust being larger in small dust grains, which also are expected to have on the average smaller binding energies.  The old dust component has been -- perhaps even repeatedly -- subjected to the harsh conditions in the interstellar medium, and what still remains intact is thus expected to be the fraction with the highest binding energies.

During the early phase of star formation, the temperature and pressure distributions near the star are likely to
populate the regime where refractory solids can form, but these conditions are likely changing on
time-scales of thousands of years or less. Moreover, the precursor mix of gas and dust will take different -- and
possibly complex -- paths through the protoplanetary disk, subjecting it to varying temperatures and pressures.
Thus, different fractions
of `new' and `old' dust will vaporize under varying conditions during the formation process, allowing variable
SLR abundance patterns in the gas out of which CAIs form. Such a mechanism can not only explain the
large differences in abundances between canonical and FUN CAIs, but also the continuous and broad distribution
of $^{26}$Al/$^{27}$Al measured among FUN CAIs \citep{2013M&PSA..76.5085P}.
Because the old dust component, with its high binding energy, is
expected to be most resilient to heating, the proposed mechanism is also
consistent with the findings by \citet{makide_oxygen_2009} that CAIs
that contain high temperature minerals such as grossite and hibonite
present in CR chondrites formed with a low initial abundance of
$^{26}$Al.

\acknowledgements
This research was supported by a grant from the Danish Council for Independent Research to {\AA}N,
a Sapere Aude Starting Grant from the Danish Council for Independent Research to TH,
and by the European Research Council (ERC) under the EU Horizon 2020 research and innovation
programme (grant agreement No 616027) through ERC Consolidator Grant ``STARDUST2ASTEROIDS'' to MB.
Research at Centre for Star and Planet Formation is funded by the Danish National Research Foundation (DNRF97).
We acknowledge PRACE for awarding us access to the computing resource CURIE based in France at CEA for
carrying out part of the simulations.
Archival storage and computing nodes at the University of Copenhagen HPC center, funded with a research
grant (VKR023406) from Villum Fonden, were used for carrying out part of the simulations and the post-processing. Finally, we acknowledge the developers of the python-based analysing tool yt (http://yt-project.org/) \citep{2011ApJS..192....9T} that simplified our analysis. 

\bibliographystyle{apj}

\clearpage{}
\end{document}